\documentclass[a4paper,11pt]{article}
\usepackage{jheppub} 
\usepackage{tabularx}
\usepackage{array}
\usepackage{tikz}
\usetikzlibrary{positioning}
\usepackage{algorithm}       
\usepackage{algorithmic}     
\usepackage{amsthm}  
\usetikzlibrary{calc}

\newcommand{\CALL}[2]{\textbf{call}~\textsc{#1}(#2)}
\newtheorem{theorem}{Theorem}[section]
\newtheorem{lemma}[theorem]{Lemma}

\theoremstyle{definition}
\title{\boldmath Exact CHY Integrand Construction Using Combinatorial Neural Networks and Discrete Optimization}

\author[a]{Simeng Li,}
\author[a]{Yaobo Zhang}

\affiliation[a]{School of Physics, Ningxia University, Yinchuan, 750021, China}

\emailAdd{lisimeng1@stu.nxu.edu.cn}
\emailAdd{yaobozhang@nxu.edu.cn}

\abstract{
Constructing a rational CHY integrand that realizes prescribed physical pole constraints is a discrete inverse problem whose combinatorial complexity grows with multiplicity.
We encode the pole hierarchy through generalized pole degrees $K(A)$ (channels $s_A$), defined as signed internal-edge counts associated with particle subsets in a colored integrand graph.
Additivity under integrand multiplication together with the elementary face recursion on the subset lattice expresses all higher-channel $K(A)$ as linear functions of the two-particle data $\{K(s_{ij})\}$ and reduces the inverse step to a mixed-integer linear feasibility problem.
The subset lattice provides a fixed dependency graph for deterministic message passing with forward evaluation and backward residual propagation; this computation is parameter-free and involves no training.
In factorial-rescaled variables $\widetilde K(A)=(|A|-2)!\,K(A)$, every local update is integral, so propagation is exact in the rescaled recursion variables and does not rely on numerical reconstruction.
We further organize generalized integrand graphs by an $n$-regular grading under multiplication, where degree-zero (0-regular) factors act as M\"obius-invariant insertions that can be decomposed into four-point cross ratios.
We illustrate the construction at six and eight points, including pick-pole selection and higher-order pole reduction.
}


\begin{document} 
\maketitle
\flushbottom

\section{Introduction}
\label{sec:intro}

The Cachazo--He--Yuan (CHY) formalism \cite{Cachazo:2013gna,Cachazo:2013iea,Cachazo:2013hca}
expresses many tree-level scattering amplitudes in massless theories (e.g.\ Yang--Mills, gravity, and scalar models) as integrals over the moduli space of $n$ marked points on the Riemann sphere.
The scattering equations localize these integrals to a finite set of solutions fixed by the external kinematics \cite{Cachazo:2013gna,Cachazo:2013iea,Cachazo:2013hca}.

Given a CHY integrand, evaluating the corresponding CHY integral (the \emph{forward CHY problem})
can be approached by combinatorial integration rules and diagrammatic correspondences for simple poles \cite{Baadsgaard:2015voa,Baadsgaard:2015aha},
as well as by contour-integral algorithms on $\mathcal{M}_{0,n}$ \cite{Cachazo:2015nsa}.
Reduction schemes for higher-order poles based on cross-ratio identities provide a further systematic handle \cite{Cardona:2016gon,Huang:2016FeynmanRules,Zhou:2017CrossRatio}.
The complementary \emph{inverse CHY problem} asks for a rational CHY integrand whose physical singularities match a prescribed set of poles \cite{Huang:2018pbu}.
Existing inverse constructions are mostly limited to special settings (for example, reconstructing Parke--Taylor data from sets of cubic diagrams in bi-adjoint scalar theory \cite{Huang:2018pbu}),
and the inverse direction is discrete and becomes rapidly combinatorial at higher multiplicity and with nontrivial numerators.

In this paper we present an algorithmic approach to the inverse CHY problem that makes the subset structure of pole data explicit.
We organize pole information using a generalized pole degree $K(s_A)$, defined as a signed internal-edge count for each subset $A\subset\{1,\ldots,n\}$
in the standard 4-regular graph representation of CHY integrands \cite{Baadsgaard:2015voa,Baadsgaard:2015aha,Cardona:2016gon}.
Additivity under integrand multiplication and local face-recursion constraints on the subset lattice imply that all higher-channel $K(s_A)$ are linear functions of the two-particle data $\{K(s_{ij})\}$,
so the inverse step can be formulated as a mixed-integer linear feasibility problem \cite{Schrijver:1998,Wolsey:1998}.
The same subset lattice also provides a fixed dependency graph for deterministic constraint propagation.
After a factorial rescaling, every local update can be implemented using only integer additions, so the propagation remains exact in the rescaled variables.
This avoids workflows where exact expressions are recovered \emph{a posteriori} from numerical evaluations \cite{DeLaurentis:2019seampy,Peraro:2016FiniteFields,Peraro:2019FiniteFlow,Klappert:2019FireFly,Ferguson:1999PSLQ,Laporta:2000DifferenceEquations}.

\subsection{Background: The CHY Formalism and Computational Motivation}

At tree level, the CHY representation writes an $n$-particle scattering amplitude $\mathcal{A}_n$ as an integral over puncture positions $\{z_i\}$ on the Riemann sphere \cite{Cachazo:2013hca,Cachazo:2014xea}:
\begin{equation}
\mathcal{A}_{n}=\int\frac{\prod_{i=1}^{n}dz_{i}}{\text{vol}(\mathrm{SL}(2,\mathbb{C}))}\; \Omega(\mathcal{E})\; \mathcal{I}(z).
\end{equation}
The measure $\Omega(\mathcal{E})$ imposes the scattering equations
\begin{equation}
\mathcal{E}_{a} \equiv \sum_{b \ne a} \frac{s_{ab}}{z_a - z_b} = 0,
\end{equation}
which implement momentum conservation and on-shell kinematics in this representation \cite{Cachazo:2013hca}.
Different quantum field theories correspond to different choices of the integrand $\mathcal{I}(z)$.
The formalism was introduced for tree-level amplitudes in a range of massless theories \cite{Cachazo:2013hca,Cachazo:2014xea},
and later extended to polynomial forms \cite{Dolan:2014dia}.

Beyond the original tree-level setting, CHY-inspired structures have been explored in several directions \cite{Zhang:2025OnNewFactorizations,Zhang:2025,He:2015yua,Mason:2013sva,Geyer:2015bja}.
For example, factorization properties of Yang--Mills amplitudes have been studied within the CHY framework \cite{Zhang:2025OnNewFactorizations,Zhang:2025}.
Loop-level generalizations and related worldsheet formulations have also been explored, including forward-limit constructions \cite{He:2015yua}
and the connection to ambitwistor string theory \cite{Mason:2013sva,Geyer:2015bja}.
We will not review these developments here; our focus is the inverse problem at tree level formulated in terms of pole data.

On the computational side, a variety of tools exist for manipulating and evaluating CHY integrals \cite{Baadsgaard:2015voa,Baadsgaard:2015aha,Cachazo:2015nsa,DeLaurentis:2019seampy}.
The ``pick pole'' algorithm isolates contributions containing specified physical singularities $1/s_A$ and has been implemented for bi-adjoint scalar theory \cite{Feng:2016nsv,Feng:2019winn}.
Extensions to more general integrands, including labelled tree-graph representations \cite{Gao:2017dek} and the appearance of higher-order poles \cite{Cardona:2016gon},
motivate systematic methods that organize pole constraints and help manage the associated combinatorics.

More broadly, the CHY formalism has stimulated interactions with tropical geometry \cite{Cachazo:2019JHEP06},
positive geometries \cite{Arkani-Hamed:2017tmz}, and twisted (co)homology / intersection-theoretic formulations
of worldsheet integrals on $\mathcal{M}_{0,n}$ \cite{Mizera:2017KLT,Mizera:2018Intersection,Mizera:2019,BrownDupont:2021}. Motivated by these connections, we focus on discrete optimization and combinatorial/topological structures as a way to organize the pole hierarchy and to formulate inverse-construction tasks.

Machine-learning methods have been applied in several areas of theoretical and computational physics,
including scans of Calabi--Yau data \cite{he2017,he2021,ashmore2020} and amplitude-adjacent tasks such as phase-space integration and multi-loop integral evaluation \cite{Bothmann:2020,Winterhalder:2022a},
as well as symbolic or diagnostic applications \cite{Dersy:2023,Cai:2024wjd,Alnuqaydan:2023symba,carleo2019}.
Here, however, we introduce no learnable parameters and perform no data-driven training.
Instead, the computation is fixed by discrete CHY constraints on integer edge data.

Exact results in amplitude calculations are frequently obtained by solving rational constraints, for example in integration-by-parts (IBP) reductions over $\mathbb{Q}$ \cite{vonManteuffel:2014IBP}.
Such workflows can be practical, but the exact dependence is typically recovered \emph{a posteriori} from numerical data \cite{Peraro:2016FiniteFields,Peraro:2019FiniteFlow,Klappert:2019FireFly,Ferguson:1999PSLQ,Laporta:2000DifferenceEquations}.
In our setting, the integrand data are integer edge multiplicities and the pole hierarchy comes with local recursion constraints.
We exploit this structure to implement a constraint-preserving algorithm on a fixed dependency graph, with updates that remain discrete and integral after a natural rescaling.

\subsection{From Graphs to Combinatorial Message Passing}

The pole hierarchy of a CHY integrand is indexed by particle subsets $A \subset \{1,\ldots,n\}$, ordered by
inclusion. This inclusion poset can be viewed as the face poset of the full simplex on $n$ labels,
and it provides a fixed graph on which one can run local message-passing updates \cite{ebli2020,Bodnar:2021,morris2019}.
We use neural-network terminology only as an analogy for a fixed computation graph with forward/backward sweeps; the procedure is fully specified by discrete constraints and involves no training and no learnable parameters.
Two structural properties then give local constraints on this poset:
\begin{itemize}
\item \textbf{Additivity.} Under integrand multiplication, pole data add:
$K_{\alpha\cdot\beta}(A)=K_\alpha(A)+K_\beta(A)$.
\item \textbf{Local face recursions.} For $|A|\ge3$,
\[
K(A)=\frac{1}{|A|-2}\sum_{\substack{B\subset A\\|B|=|A|-1}}K(B).
\]
\end{itemize}

These relations define a deterministic upward propagation (a ``forward pass'') on the subset lattice.
To keep the propagation integral, we introduce the rescaling
\[
\widetilde K(A):=(|A|-2)!\,K(A)\qquad(|A|\ge2),
\]
for which the recursion becomes an unnormalized face-sum,
\[
\widetilde K(A)=\sum_{\substack{B\subset A\\|B|=|A|-1}}\widetilde K(B),\qquad |A|\ge3.
\]
In this representation, each local update is an integer operation.

With these ingredients, inverse CHY construction and pole selection reduce to a discrete constraint-satisfaction problem on a fixed dependency graph:
the unknowns are the integer two-particle data $\{K(s_{ij})\}$, the recursion fixes all higher-subset values, and pole requirements appear as hard constraints on selected nodes.
There is no training stage and no learnable parameters.
We analyze six- and eight-point examples to illustrate the resulting constraint propagation and MILP-based construction.

Beyond the familiar 4-regular graph representation that encodes many simple-pole CHY integrands
(e.g.\ products of two Parke--Taylor factors) \cite{Baadsgaard:2015voa,Baadsgaard:2015aha},
we will work in a slightly more general colored-graph language in which each integrand carries an integer
\emph{regularity} (net degree) and integrand multiplication is graded (Sec.~\ref{sec:generalized-chy}).
In particular, the product of an $m$-regular and an $n$-regular graph is $(m+n)$-regular, while the
degree-zero sector (0-regular graphs) consists of M\"obius-invariant multiplicative factors.
This graded multiplication is naturally suggested by the twisted (co)homology viewpoint of worldsheet
integrals on $\mathcal{M}_{0,n}$, where logarithmic singularities, cross-ratio coordinates, and intersection
pairings provide a canonical bookkeeping of divisor data
\cite{Mizera:2017KLT,Mizera:2018Intersection,BrownDupont:2021}.
Practically, 0-regular factors give a constructive handle to generate families of integrands within a fixed
regularity class and to engineer the cross-ratio insertions used to reduce higher-order poles
\cite{Cardona:2016gon,Huang:2016FeynmanRules,Zhou:2017CrossRatio}.

\paragraph{Summary of contributions.}
(i) We formulate the inverse CHY problem as a mixed-integer linear feasibility problem in the two-particle
data $\{K(s_{ij})\}$ using additivity and face-recursion constraints.
(ii) We implement deterministic constraint propagation on the subset lattice using factorial-rescaled
variables $\widetilde K(A)$, so that updates can be kept integral.
(iii) We introduce a regularity grading for generalized colored graphs and identify the 0-regular
(degree-zero) sector of M\"obius-invariant factors; we use 0-regular factorizations to construct cross-ratio to organize pick-pole selection and higher-order pole reduction on explicit low-point examples.

\paragraph{Organization.} Section~\ref{sec:graph-structure} reviews the graph-theoretic representation of CHY integrands and the
associated pole data. Section~\ref{sec:topological-perspective} develops the simplicial/topological viewpoint and the resulting fixed
message-passing architecture. Section~\ref{sec:applications} presents applications to pick-pole selection and higher-order pole
expansions. We conclude in Section~\ref{sec:conclusion}.

\section{Graph-Theoretic Structure of CHY Integrands}
\label{sec:graph-structure}

\subsection{Integrand as 4-Regular Graph and Integration Rules}
The CHY formalism allows for a direct translation from the algebraic form of an integrand to a graph-theoretic object.
For an $n$-particle process, the punctures $z_i$ are treated as vertices of a graph.
For each factor of $(z_i - z_j)^{-k}$ in the denominator of the integrand, $k$ edges are drawn between vertices $i$ and $j$.
For the BAS integrand, $\mathcal{I}_{\text{BAS}} = \text{PT}(\alpha) \text{PT}(\beta)$, each PT factor contributes edges between cyclically adjacent vertices in its respective ordering.
Since each vertex $k$ is connected to two neighbors in the $\alpha$ ordering and two neighbors in the $\beta$ ordering, every vertex has exactly four edges connected to it.
Such a graph, where every vertex has a degree of 4, is known as a 4-regular (or quartic) graph \cite{Baadsgaard:2015voa, Baadsgaard:2015aha, Cachazo:2014nsa, Mason:2013sva, Adamo:2013tsa}.

\begin{figure}[htbp]
  \centering
  \includegraphics[width=0.4\textwidth]{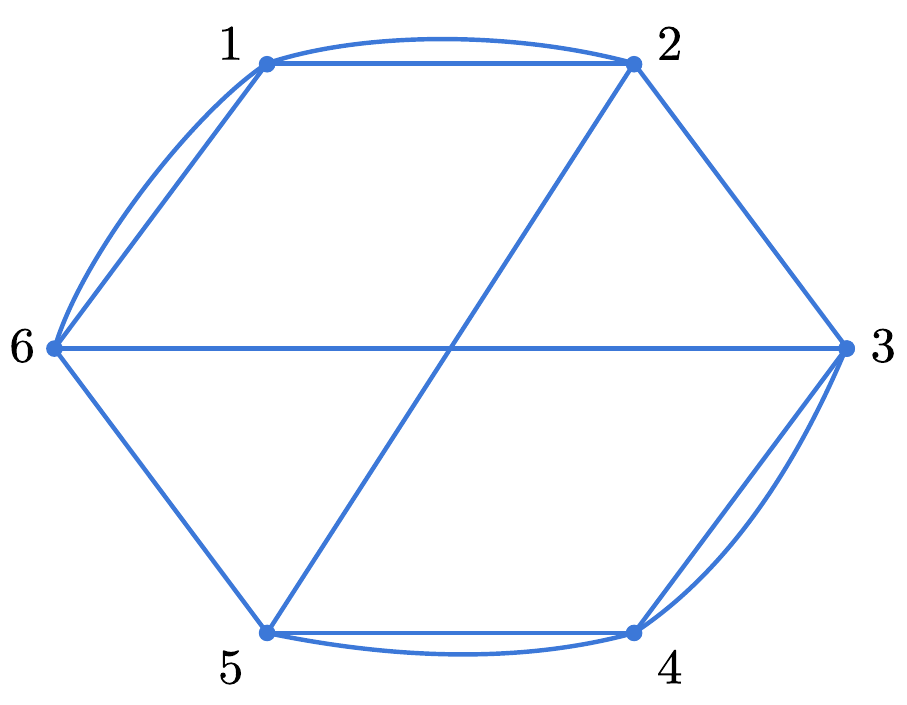}
  \caption{The 4-regular graph of integrand $\mathcal{I} = \text{PT}(123456)\text{PT}(125436)$}
  \label{fig:sample}
\end{figure}

This graphical representation leads to a set of purely combinatorial integration rules, which provide a practical algorithm for evaluating the CHY integral without explicitly solving the scattering equations \cite{Baadsgaard:2015voa, Baadsgaard:2015aha, Cachazo:2014nsa, Cardona:2016gon}.
The final amplitude is a sum over terms, each being a product of inverse Mandelstam variables, $\prod 1/s_{A_i}$.
The rules determine which sets of Mandelstam variables appear.

\begin{enumerate}
    \item \textbf{Identify Potential Poles:} For any subset of external labels $A \subset \{1, \dots, n\}$, a "pole index" $\chi(A)$ is calculated:
    \begin{equation}
    \chi(A) := L[A] - 2(|A| - 1)
    \end{equation}
    where $L[A]$ is the number of edges in the 4-regular graph with both endpoints in $A$, and $|A|$ is the number of particles in $A$.
A Mandelstam variable $s_A$ corresponds to a physical pole of order $\chi(A)+1$ only if $\chi(A) \ge 0$ \cite{Baadsgaard:2015voa, Adamo:2013tsa}.
For BAS theory, all physical poles are simple poles, corresponding to $\chi(A) = 0$.
\item \textbf{Find Compatible Sets:} A single term in the final amplitude corresponds to a valid Feynman diagram with $n-3$ propagators.
The algorithm requires finding sets of $n-3$ distinct subsets $\{A_1, \dots, A_{n-3}\}$, each satisfying $\chi(A_i)=0$, that are "compatible."
Compatibility is a graph-theoretic condition ensuring that the corresponding propagators can coexist in a single cubic Feynman diagram \cite{Geyer:2015bja}.
\item \textbf{Sum Contributions:} Each compatible set contributes a term $1/(s_{A_1} \cdots s_{A_{n-3}})$ to the amplitude.
The full amplitude is the sum over all such contributions \cite{Adamo:2013tsa}.
\end{enumerate}

For example, for a 6-point BAS amplitude with integrand $\mathcal{I} = \text{PT}(123456)\text{PT}(125436)$, one would begin by drawing the corresponding 12-edge, 6-vertex, 4-regular graph.
Then, one would identify all subsets $A$ with $\chi(A)=0$ (e.g., $A=\{1,2,6\}$) and find all compatible triplets of such subsets.
The sum of the inverse products of the corresponding Mandelstam variables gives the final amplitude, matching the standard cubic-diagram expansion in bi-adjoint scalar theory \cite{Baadsgaard:2015aha, Baadsgaard:2015voa}.

\subsection{Combinatorial Structure of Generalized CHY Graphs}
\label{sec:generalized-chy}
The graph-theoretic methods outlined for the bi-adjoint scalar (BAS) theory can be extended into a more general framework. This allows for a unified combinatorial analysis of any theory described by a rational CHY integrand.

To begin, we generalize the graph representation itself. To accommodate any rational function of differences $(z_i-z_j)$, we introduce two types of edges:
\begin{itemize}
    \item A factor of $(z_i - z_j)^{-1}$ in the denominator corresponds to a \textbf{solid edge}.
    \item A factor of $(z_i - z_j)^{+1}$ in the numerator corresponds to a \textbf{dashed edge}.
\end{itemize}
This establishes a direct map between any rational integrand $I(z)$ and a corresponding graph with two edge types.

With this representation, we can define a local property for each vertex $v$, its \textbf{Net Degree}, which counts the net power of factors involving $z_v$ in the integrand:
\begin{equation}
\text{deg}_{\text{net}}(v) := \text{deg}_{\text{solid}}(v) - \text{deg}_{\text{dashed}}(v)
\end{equation}
This concept of a local net count can be conceptually extended to a "regional" property for any subset of vertices $A \subset \{1, \dots, n\}$. This gives rise to the \textbf{Generalized Pole Degree, K(A)}, which measures the net number of internal edges within the region defined by $A$:
\begin{equation}
K(A) := L_{\text{solid}}[A] - L_{\text{dashed}}[A]
\end{equation}
where $L_{\text{solid}}[A]$ and $L_{\text{dashed}}[A]$ are the number of solid and dashed edges, respectively, with both endpoints in the set $A$. This new degree directly relates to the original pole index $\chi(A)$ found in the literature via the simple formula:
\begin{equation}
\chi(A) = K(A) - 2(|A| - 1)
\label{eq:chi-K-relation}
\end{equation}
This relation elegantly separates the theory-dependent contribution to the pole structure, captured by $K(s_A)$, from the standard geometric contribution of the worldsheet, $-2(|A|-1)$.

This framework becomes useful when we consider the multiplication of integrands, as in the "double copy" construction. The multiplication of two graphs, $G_\alpha$ and $G_\beta$, corresponding to integrands $I_\alpha(z)$ and $I_\beta(z)$, is defined as a new graph $G_\gamma = G_\alpha * G_\beta$ corresponding to the product integrand $I_\alpha(z)I_\beta(z)$. The edge set of the product graph is the union of the edge sets of the original graphs. The index $K(s_A)$ behaves very simply under this operation.

\textbf{Theorem:} The pole degree $K(s_A)$ is additive under graph multiplication.
\textit{Proof:} For a product graph $G_\gamma = G_\alpha * G_\beta$, the number of internal edges in a subset $A$ is the sum of internal edges from the constituent graphs. Thus,
\begin{align*}
K_\gamma(A) &:= L_{\text{solid, }\gamma}[A] - L_{\text{dashed, }\gamma}[A] \\
&= (L_{\text{solid, }\alpha}[A] + L_{\text{solid, }\beta}[A]) - (L_{\text{dashed, }\alpha}[A] + L_{\text{dashed, }\beta}[A]) \\
&= (L_{\text{solid, }\alpha}[A] - L_{\text{dashed, }\alpha}[A]) + (L_{\text{solid, }\beta}[A] - L_{\text{dashed, }\beta}[A]) \\
&= K_\alpha(A) + K_\beta(A). \quad \square
\end{align*}
This additivity provides a simple algebraic tool to compute the pole structure of complicated theories by summing the known indices of their simpler constituents.

\begin{figure}[h]
    \centering
    \includegraphics[width=0.92\linewidth]{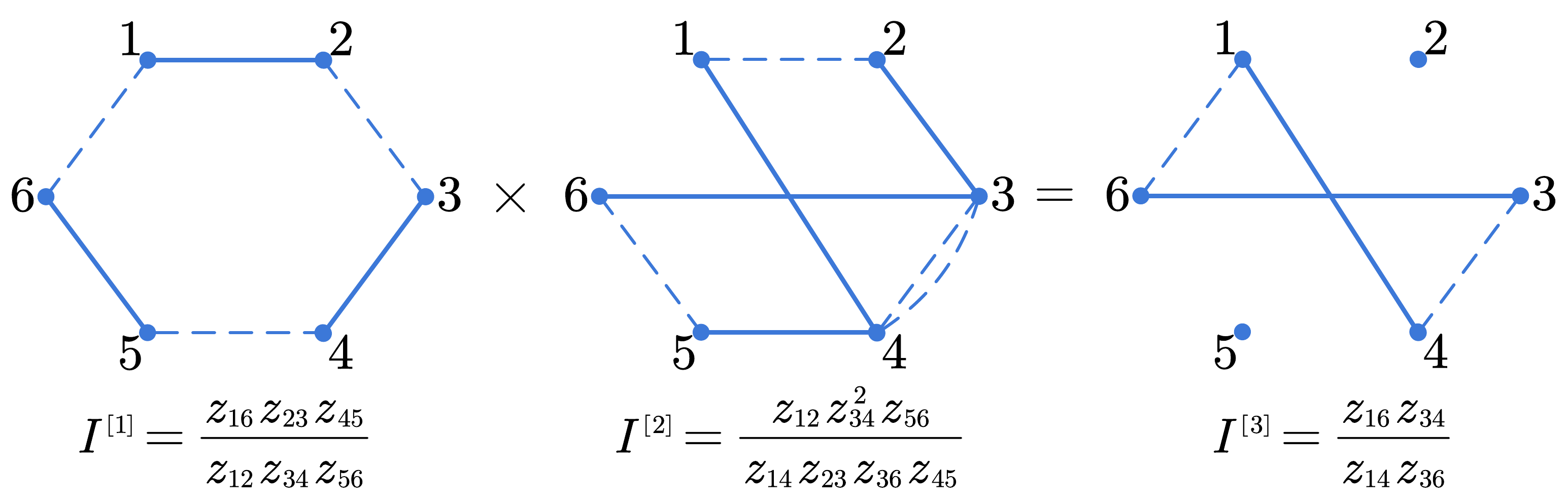}
    \caption{The set of 0-regular graphs is closed under multiplication.}
    \label{0-regular}
\end{figure}

\begin{figure}[h]
\begin{center}
  \begin{tabularx}{0.92\textwidth} {   
  !{\vrule width 1.2pt}  >{\centering\arraybackslash}X     
  | >{\centering\arraybackslash}X   
  | >{\centering\arraybackslash}X  
  | >{\centering\arraybackslash}X !{\vrule width 1pt} 
    >{\centering\arraybackslash}X     
  | >{\centering\arraybackslash}X   
  | >{\centering\arraybackslash}X  
  | >{\centering\arraybackslash}X!{\vrule width 1.2pt}  } 
  \hline   
  &\(I^{[1]}\) & \(I^{[2]}\) & \(I^{[3]}\) &  &\(I^{[1]}\) & \(I^{[2]}\) & \(I^{[3]}\) \\ 
  \hline 
  \(deg(s_{1})\)&0  & 0  & 0 &\(K(s_{35})\)&1  & -1  & 0 \\ 
   \hline 
  \(deg(s_{2})\)&0  & 0  & 0 &\(K(s_{36})\)&0  & 0  & 0 \\ 
  \hline 
  \(deg(s_{3})\)&0  & 0  & 0  &\(K(s_{45})\)&0  & 1  & 1 \\
  \hline 
  \(deg(s_{4})\)&0  & 0  & 0  &\(K(s_{46})\)&0  & 0  & 0 \\
  \hline 
  \(deg(s_{5})\)&0  & 0  & 0  &\(K(s_{56})\)&-1  & 0  & -1 \\
  \hline 
  \(deg(s_{6})\)&0  & 0  & 0  &\(K(s_{123})\)&0  & 0  & 0\\
  \hline 
  \(K(s_{12})\)&1  & -1  & 0 &\(K(s_{124})\)&1  & 0  & 1 \\ 
  \hline 
  \(K(s_{13})\)&0  & 0  & 0  &\(K(s_{125})\)&1  & -1 & 0 \\
  \hline 
  \(K(s_{14})\)&0  & 1  & 1  &\(K(s_{126})\)&0  & -1  & -1\\
  \hline 
  \(K(s_{15})\)&0  & 0  & 0  & \(K(s_{134})\)&1  & -1  & 0 \\
  \hline 
  \(K(s_{16})\)&-1  & 0  & -1  &\(K(s_{135})\)&0  & 0  & 0\\
  \hline 
  \(K(s_{23})\)&-1  & 1  & 0  &\(K(s_{136})\)&-1  & 1  & 0\\
  \hline 
  \(K(s_{24})\)&0  & 0  & 0  &\(K(s_{145})\)&-1  & 2  & 1 \\
  \hline 
  \(K(s_{25})\)&0  & 0  & 0  &\(K(s_{146})\)&-1  & 1  & 0  \\
  \hline 
  \(K(s_{26})\)&0  & 0  & 0  &\(K(s_{156})\)&0  & -1  & -1 \\
  \hline 
  \(K(s_{34})\)&1  & -2  & -1   & & & & \\
  \hline
  \end{tabularx}
\end{center}

 \caption{The $K_{I^{[3]}}(s_A)=K_{I^{[1]}}(s_A)+K_{I^{[2]}}(s_A)$ for all poles}
\end{figure}

\begin{figure}[h]
    \centering
    \includegraphics[width=0.92\linewidth]{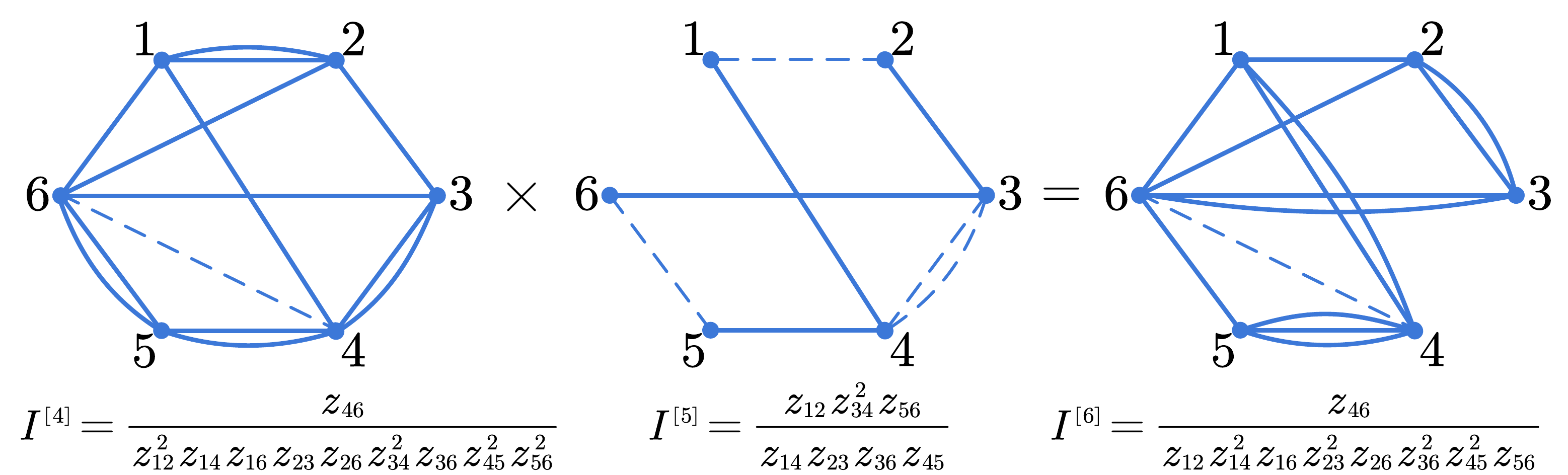}
    \caption{The "0-regular group" acts as a set of transformations on other graphs}
    \label{4--regular}
\end{figure}

\begin{figure}[h]
\begin{center}
  \begin{tabularx}{0.92\textwidth} {   
  !{\vrule width 1.2pt} >{\centering\arraybackslash}X     
  | >{\centering\arraybackslash}X   
  | >{\centering\arraybackslash}X  
  | >{\centering\arraybackslash}X  !{\vrule width 1pt} 
    >{\centering\arraybackslash}X     
  | >{\centering\arraybackslash}X   
  | >{\centering\arraybackslash}X  
  | >{\centering\arraybackslash}X!{\vrule width 1.2pt} } 
  \hline   
  &\(I^{[4]}\) & \(I^{[5]}\) & \(I^{[6]}\)&  &\(I^{[4]}\) & \(I^{[5]}\) & \(I^{[6]}\) \\ 
   \hline 
  \(deg(s_{1})\)&4  & 0  & 4 &\(K(s_{35})\)&0  & -1  & -1 \\ 
   \hline 
  \(deg(s_{2})\)&4  & 0  & 4 & \(K(s_{36})\)&1  & 0  & 1  \\ 
  \hline 
  \(deg(s_{3})\)&4  & 0  & 4  & \(K(s_{45})\)&2  & 1  & 3  \\
  \hline 
  \(deg(s_{4})\)&4  & 0  & 4 &\(K(s_{46})\)&-1  & 0  & -1 \\
  \hline 
  \(deg(s_{5})\)&4  & 0  & 4 &\(K(s_{56})\)&2  & 0  & 2  \\
  \hline 
  \(deg(s_{6})\)&4  & 0  & 4  &\(K(s_{123})\)&3  & 0  & 3 \\
  \hline 
  \(K(s_{12})\)&2  & -1  & 1 &\(K(s_{124})\)&3  & 0  & 3 \\ 
  \hline 
  \(K(s_{13})\)&0  & 0  & 0  &\(K(s_{125})\)&2  & -1 & 1  \\
  \hline 
  \(K(s_{14})\)&1  & 1  & 2  &\(K(s_{126})\)&4  & -1  & 3\\
  \hline 
  \(K(s_{15})\)&0  & 0  & 0  &\(K(s_{134})\)&3  & -1  & 2 \\
  \hline 
  \(K(s_{16})\)&1  & 0  & 1  &\(K(s_{135})\)&0  & 0  & 0\\
  \hline 
  \(K(s_{23})\)&1  & 1  & 2  &\(K(s_{136})\)&2 & 1  & 3 \\
  \hline 
  \(K(s_{24})\)&0  & 0  & 0  &\(K(s_{145})\)&3  & 2  & 5 \\
  \hline 
  \(K(s_{25})\)&0  & 0  & 0  &\(K(s_{146})\)&1  & 1  & 2 \\
  \hline 
  \(K(s_{26})\)&1  & 0  & 1  &\(K(s_{156})\)&3  & -1  & 2 \\
  \hline 
  \(K(s_{34})\)&2  & -2  & 0  & & & & \\
  \hline
  \end{tabularx}
\end{center}
 \caption{The $K_{I^{[6]}}(s_A)=K_{I^{[4]}}(s_A)+K_{I^{[5]}}(s_A)$ for all poles}
\end{figure}

Finally, we return to the local property of net degree to classify the global structure of graphs. A graph is called "n-regular" if all its vertices have a net degree of $n$. The additivity of the net degree under multiplication implies that the product of an $m$-regular and an $n$-regular graph is an $(m+n)$-regular graph. This implies that the set of "0-regular graphs" is closed under multiplication. Furthermore, one can show that this set forms a commutative group, with the empty graph ($I(z)=1$) as the identity and graph inversion (swapping solid and dashed edges) as the inverse operation. This "0-regular group" acts as a set of transformations on other graphs; multiplying an $n$-regular graph by any 0-regular graph yields another $n$-regular graph, preserving the graph's regularity class while modifying its detailed pole structure, as tracked by $K(s_A)$.

\subsection{Recursion Relation for Generalized Pole Degree $K(s_A)$}
\label{sec:recursion-K}

The generalized pole degree $K(A)$ obeys a simple recursion on the subset lattice.
For any $A\subseteq\{1,\dots,n\}$ with $|A|\ge 3$,
\begin{equation}
K(A)=\frac1{|A|-2}\sum_{\substack{B\subset A\\|B|=|A|-1}}K(B).
\label{eq:Krec}
\end{equation}
We refer to \eqref{eq:Krec} as the \emph{elementary face recursion}. It is an edge-counting identity:
any internal edge $(i,j)$ with $i,j\in A$ is counted in $K(B)$ for exactly $|A|-2$ of the
$(|A|-1)$-subsets $B=A\setminus\{k\}$ (namely all $k\neq i,j$). We record \eqref{eq:Krec} because it makes
all higher-channel values $K(A)$ \emph{linear functions} of the two-particle data $\{K(ij)\}$,
which is the key input for the mixed-integer feasibility formulation in Sec.~\ref{sec:bootstrap} and Appendix~\ref{app:ilp}.

Illustrative cases include
$K(123)=K(12)+K(13)+K(23)$ and
$K(1234)=\tfrac12\bigl(K(123)+K(124)+K(134)+K(234)\bigr)$.

More generally, averaging over codimension-$d$ faces gives, for $1\le d\le |A|-2$,
\begin{equation}
K(A)=\frac1{\binom{|A|-2}{d}}\sum_{\substack{B\subset A\\|B|=|A|-d}}K(B).
\label{eq:general-rec}
\end{equation}
Different values of $d$ provide redundant evaluations of the same $K(A)$ and serve as
useful consistency checks in implementations.

\subsection{Bootstrap Equations for Integrands}
\label{sec:bootstrap}

The relationship between the generalized pole degree $K(s_A)$ and the pole index $\chi(A)$ established in equation \eqref{eq:chi-K-relation} provides a useful framework for solving the inverse CHY problem. This inverse problem asks: given a desired set of poles in a scattering amplitude, can we construct a CHY integrand that produces exactly those poles?

From equation \eqref{eq:chi-K-relation}, we know that the pole index is determined by \eqref{eq:chi-K-relation}. A basic CHY integration rule tells us that only subsets $A$ with $\chi(A) = 0$ contribute simple poles $1/s_A$ to the final amplitude \cite{Baadsgaard:2015voa, Baadsgaard:2015aha}. This immediately gives us a classification scheme for any subset $A \subset \{1, \dots, n\}$:
\begin{itemize}
\item \textbf{Physical poles}: $\chi(A) = 0$ $\rightarrow$ simple poles $1/s_A$ appear in the amplitude
\item \textbf{Absent poles}: $\chi(A) < 0$ $\rightarrow$ no poles appear for these channels  
\item \textbf{Higher-order poles}: $\chi(A) > 0$ $\rightarrow$ would correspond to multiple poles, but these typically do not appear in tree-level amplitudes for generic theories
\end{itemize}

This classification leads to a key insight: for any desired amplitude with a specific pole structure, the pattern of which poles appear and which are absent constrains the possible values of $K(s_A)$. Consider a target amplitude that has simple poles for a set $\mathcal{P} = \{A_1, A_2, \ldots, A_k\}$ and no poles for the complementary set $\overline{\mathcal{P}} := \mathcal{S}_n \setminus \mathcal{P}$, where $\mathcal{S}_n$ denotes the set of all subsets of $\{1,\ldots,n\}$. The bootstrap equations are:
\begin{align}
\text{For } A \in \mathcal{P}: \quad &K(A) = 2(|A| - 1) \label{eq:bootstrap-eq}\\
\text{For } A \in \overline{\mathcal{P}}: \quad &K(A) < 2(|A| - 1) \label{eq:bootstrap-ineq}
\end{align}
The equality constraints ensure that the desired poles appear with the correct order, while the inequality constraints ensure that unwanted poles are absent.

However, the values $K(A)$ cannot be chosen arbitrarily—they must satisfy the recursion relations derived in Section \ref{sec:recursion-K}. This creates a highly constrained system where the unknowns are the two-particle pole degrees $K(s_{ij})$ for all pairs $\{i,j\}$. Using the recursion relation \eqref{eq:Krec}, any higher-order pole degree can be expressed in terms of lower-order ones. For example:
\begin{align}
K(s_{ijk}) &= K(s_{ij}) + K(s_{ik}) + K(s_{jk})\\
K(s_{ijkl}) &= \frac{K(s_{ijk}) + K(s_{ijl}) + K(s_{ikl}) + K(s_{jkl})}{2}
\end{align}
Continuing this process, we can express all pole degrees $K(A)$ with $|A| \geq 3$ as linear combinations of the two-particle pole degrees $K(s_{ij})$. This reduces our bootstrap system to a set of linear constraints on the $\binom{n}{2}$ variables $\{K(s_{ij})\}_{1 \leq i < j \leq n}$.

A crucial observation is that for rational CHY integrands constructed from products of Parke-Taylor factors and their generalizations considered here, all pole degrees $K(s_A)$ are integers. This follows from the fact that each factor $(z_i - z_j)^{\pm 1}$ contributes $\pm 1$ to the edge count. Therefore, the bootstrap equations become a mixed system of linear Diophantine equations and inequalities:
\begin{align}
\mathbf{M}_{\text{eq}} \vec{K}_2 &= \vec{b}_{\text{eq}} \label{eq:bootstrap-linear-eq}\\
\mathbf{M}_{\text{ineq}} \vec{K}_2 &< \vec{b}_{\text{ineq}} \label{eq:bootstrap-linear-ineq}
\end{align}
where $\vec{K}_2 = (K(s_{12}), K(s_{13}), \ldots, K(s_{n-1,n}))^T$ is the vector of two-particle pole degrees, $\mathbf{M}_{\text{eq}}$ and $\mathbf{M}_{\text{ineq}}$ encode the recursion relations for equality and inequality constraints respectively, while $\vec{b}_{\text{eq}}$ and $\vec{b}_{\text{ineq}}$ encode the corresponding target pole structures.

The bootstrap approach yields a constructive algorithm for the inverse CHY problem. The basic idea is to transform the desired pole structure into a system of linear constraints on the two-particle variables $K(s_{ij})$, solve this constrained system, and then use the additivity property under graph multiplication to construct the corresponding integrand. The algorithm proceeds in the following steps:

\begin{algorithm}[H]
\caption{CHY Integrand Bootstrap Construction}
\begin{algorithmic}[1]
\REQUIRE Target pole structure $\mathcal{P}$ (set of desired simple poles)
\ENSURE CHY integrand $I(z)$ producing amplitude with pole structure $\mathcal{P}$
\STATE \textbf{Input:} Specify desired pole structure $\mathcal{P} \subset \mathcal{S}_n$ where $\mathcal{S}_n$ is the set of all possible subsets
\STATE \textbf{Constraint Setup:} For each $A \in \mathcal{P}$, impose $K(A) = 2(|A| - 1)$
\STATE \textbf{Constraint Setup:} For each $A \in \overline{\mathcal{P}}$, impose $K(A) < 2(|A| - 1)$
\STATE \textbf{Recursion Matrix:} Construct matrices $\mathbf{M}_{\text{eq}}, \mathbf{M}_{\text{ineq}}$ encoding relations between $K(A)$ and $K(s_{ij})$
\STATE \textbf{Mixed Integer System:} Formulate $\mathbf{M}_{\text{eq}} \vec{K}_2 = \vec{b}_{\text{eq}}$ and $\mathbf{M}_{\text{ineq}} \vec{K}_2 < \vec{b}_{\text{ineq}}$
\STATE \textbf{Integer Solution:} Solve for integer vector $\vec{K}_2$ satisfying the mixed system
\IF{no integer solution exists}
    \RETURN "No CHY integrand exists for this pole structure"
\ENDIF
\STATE \textbf{Decomposition:} Use additivity $K_{\alpha \star \beta} = K_\alpha + K_\beta$ to factor solution
\STATE \textbf{Construction:} Build integrand $I(z) = \prod_i I_i(z)$ from elementary factors
\RETURN CHY integrand $I(z)$
\end{algorithmic}
\end{algorithm}

For illustration, consider a six-particle amplitude where we want simple poles for $s_{12}$, $s_{34}$, and $s_{56}$, but no other two-particle poles. The bootstrap equations become $K(s_{12}) = K(s_{34}) = K(s_{56}) = 2$ and $K(s_{ij}) < 2$ for all other pairs $\{i,j\}$. Combined with the recursion relations for higher-order poles, this system constrains the possible CHY integrands. One solution corresponds to the integrand $I = \frac{1}{z_{12}^2 z_{14} z_{15} z_{23}z_{26}z_{34}^2z_{36} z_{45} z_{56}^2}$, which indeed produces poles only in the appropriate channels.

This bootstrap framework thus provides a systematic method to reverse-engineer CHY integrands from desired amplitude structures, supporting amplitude construction and the exploration of additional theories within the CHY formalism. The method is particularly useful because it leverages the additive structure of the generalized pole degree $K(s_A)$, allowing complex integrands to be built systematically from simpler components.

\subsection{Examples of Bootstrap Construction}
\label{sec:examples}

To illustrate the utility of the bootstrap framework developed in Section \ref{sec:bootstrap}, we present concrete examples demonstrating how to reverse-engineer CHY integrands from desired pole structures. These examples showcase both the computational implementation of the bootstrap algorithm and the interplay between the algebraic constraints and the resulting graph structures.

Consider a six-particle amplitude where we desire simple poles for the channels $s_{12}$, $s_{23}$, $s_{34}$, $s_{56}$, $s_{123}$, $s_{156}$, $s_{234}$, and $s_{456}$. This represents a highly constrained scenario with eight simple poles, approaching the maximum possible for a six-particle amplitude.

Following the bootstrap algorithm, we begin by translating these requirements into constraints on the K-values. The equality constraints \eqref{eq:bootstrap-eq} yield:
\begin{align}
K(s_{12}) = K(s_{23}) = K(s_{34}) = K(s_{56}) &= 2\\
K(s_{123}) = K(s_{156}) = K(s_{234}) = K(s_{456}) &= 4
\end{align}

For all other channels, the inequality constraints \eqref{eq:bootstrap-ineq} require $K(s_A) < 2(|A| - 1)$. Additionally, the single-particle K-values must satisfy the regularity condition:
\begin{equation}
K_i = \sum_{j \neq i} K(s_{ij}) = 4 \quad \text{for all } i \in \{1,\ldots,6\}
\end{equation}

This yields a mixed integer linear programming problem with 15 two-particle variables $K(s_{ij})$, subject to 14 equality constraints (8 from desired poles, 6 from single-particle conditions) and 27 inequality constraints from unwanted channels. The constraint matrix $\mathbf{M}_{\text{eq}}$ encodes how each higher-order K-value depends on the two-particle degrees through the recursion relations. For instance, the constraint for $K(s_{123}) = 4$ translates to:
\begin{equation}
K(s_{12}) + K(s_{13}) + K(s_{23}) = 4
\end{equation}

The complete system of inequalities includes 27 inequality constraints (from unwanted two-particle and three-particle channels):
\begin{align*}
&K_{1,3}<2,K_{1,4}<2,K_{1,5}<2,K_{1,6}<2,K_{2,4}<2,K_{2,5}<2,K_{2,6}<2,\\
&K_{3,5}<2,K_{3,6}<2,K_{4,5}<2,K_{4,6}<2,K_{1,2}+K_{1,4}+K_{2,4}<4,\\
&K_{1,2}+K_{1,5}+K_{2,5}<4,K_{1,2}+K_{1,6}+K_{2,6}<4,K_{1,3}+K_{1,4}+K_{3,4}<4,\\
&K_{1,3}+K_{1,5}+K_{3,5}<4,K_{1,3}+K_{1,6}+K_{3,6}<4,K_{1,4}+K_{1,5}+K_{4,5}<4,\\
&K_{1,4}+K_{1,6}+K_{4,6}<4,K_{2,3}+K_{2,5}+K_{3,5}<4,K_{2,3}+K_{2,6}+K_{3,6}<4,\\
&K_{2,4}+K_{2,5}+K_{4,5}<4,K_{2,4}+K_{2,6}+K_{4,6}<4,K_{2,5}+K_{2,6}+K_{5,6}<4,\\
&K_{3,4}+K_{3,5}+K_{4,5}<4,K_{3,4}+K_{3,6}+K_{4,6}<4,K_{3,5}+K_{3,6}+K_{5,6}<4,
\end{align*}
and the complete system of 14 equality constraints (8 from desired poles, 6 from single-particle conditions):
\begin{align*}
&K_{1,2}=2,K_{2,3}=2,K_{3,4}=2,K_{5,6}=2,K_{1,2}+K_{1,3}+K_{2,3}=4,\\
&K_{1,5}+K_{1,6}+K_{5,6}=4,K_{4,5}+K_{4,6}+K_{5,6}=4,K_{2,3}+K_{2,4}+K_{3,4}=4,\\
&K_1=K_{1,2}+K_{1,3}+K_{1,4}+K_{1,5}+K_{1,6}=4,\\
&K_2=K_{1,2}+K_{2,3}+K_{2,4}+K_{2,5}+K_{2,6}=4,\\
&K_3=K_{1,3}+K_{2,3}+K_{3,4}+K_{3,5}+K_{3,6}=4,\\
&K_4=K_{1,4}+K_{2,4}+K_{3,4}+K_{4,5}+K_{4,6}=4,\\
&K_5=K_{1,5}+K_{2,5}+K_{3,5}+K_{4,5}+K_{5,6}=4,\\
&K_6=K_{1,6}+K_{2,6}+K_{3,6}+K_{4,6}+K_{5,6}=4.
\end{align*}

Solving this system yields the unique integer solution:
\begin{equation}
\begin{aligned}
K_{1,2} &\to 2, & K_{1,3} &\to 0, & K_{1,4} &\to 0,  \\
K_{2,3} &\to 2, & K_{2,4} &\to 0, & K_{2,5} &\to 0, \\
K_{2,6} &\to 0, & K_{3,4} &\to 2, & K_{3,5} &\to 0, \\
K_{3,6} &\to 0, & K_{4,5} &\to 1, & K_{4,6} &\to 1, \\
K_{5,6} &\to 2, & K_{1,5} &\to 1, & K_{1,6} &\to 1
\end{aligned}
\end{equation}

The resulting CHY integrand:
\begin{equation}
I_6 = \frac{1}{z_{12}^2 z_{15} z_{16} z_{23}^2 z_{34}^2 z_{45} z_{46} z_{56}^2}
\end{equation}
shows that the bootstrap construction produces an integrand with the desired pole structure in this example. The resulting integrand $I_6$ generates an amplitude whose only simple poles are $s_{12}$, $s_{23}$, $s_{34}$, $s_{56}$, $s_{123}$, $s_{156}$, $s_{234}$, and $s_{456}$, exactly as specified in our requirements. The K-values directly translate to the powers in the integrand: each $K(s_{ij}) = 2$ yields a factor $z_{ij}^{-2}$ in the denominator, while $K(s_{ij}) = 1$ gives $z_{ij}^{-1}$, and $K(s_{ij}) = 0$ means the corresponding factor is absent.

For this particular example, the mixed integer linear programming problem yields a unique solution, making it an ideal illustration of the bootstrap method. The constraint system can be solved using standard integer programming packages available in Mathematica (through functions like \texttt{LinearProgramming} with integer constraints) or Python (using libraries such as \texttt{cvxpy}, \texttt{PuLP}, or \texttt{scipy.optimize.milp}). These computational tools efficiently handle the combinatorial nature of finding integer solutions to our linear constraint system. More complex examples with non-unique solutions and detailed computational methods for solving the bootstrap equations are presented in Appendix \ref{app:ilp}. 

\paragraph{Summary.}
Representing a rational CHY integrand as a graph with solid/dashed edges turns the generalized pole degree
$K(A)$ into a signed internal-edge count on any subset $A$. Two structural properties organize the inverse
construction: (i) additivity under integrand multiplication, and (ii) the elementary face recursion \eqref{eq:Krec},
which expresses every higher-channel $K(A)$ as a linear combination of lower-layer values. As a result,
all pole constraints reduce to linear equalities/inequalities in the two-particle variables $\{K(ij)\}$, so the
inverse step can be posed as a mixed-integer feasibility problem (Appendix~\ref{app:ilp}). In Sec.~\ref{sec:conn-autodiff} we repackage the
same subset lattice as a face poset/Hasse diagram to organize deterministic constraint propagation used by
the pick-pole and pole-reduction algorithms.

\clearpage
\section{From Poles to Simplices: A Topological Perspective}
\label{sec:topological-perspective}
\subsection{Simplicial Structure in CHY Pole Hierarchies}
The pole data of a CHY integrand are indexed by particle subsets $A\subset\{1,\ldots,n\}$, partially ordered by inclusion. For each $A$ we attach the generalized pole degree $K(A)$ and the pole index $\chi(A)=K(A)-2(|A|-1)$. The elementary face recursion \eqref{eq:Krec} expresses $K(A)$ for $|A|\ge 3$ as an average over its codimension-one subsets, so the hierarchy is determined by the two-particle layer $\{K(ij)\}$.

It is convenient to identify each subset $A$ with a simplex $\sigma_A$ of dimension $|A|-1$ in the full $(n-1)$-simplex. Then $B\subset A$ is the face relation $\sigma_B\prec\sigma_A$, and \eqref{eq:Krec} can be written as a local averaging rule on the face poset (Hasse diagram):
\begin{equation}
K(\sigma_A)=\frac{1}{|A|-2}\sum_{\substack{B\subset A\\ |B|=|A|-1}} K(\sigma_B),\qquad |A|\ge 3.
\label{eq:simplicial-averaging}
\end{equation}

We use this simplicial viewpoint mainly as a bookkeeping device for the subset lattice and recursion constraints. The channels with $\chi(A)=0$ are treated as a marked subset of faces (together with inequality constraints for $\chi(A)<0$ and higher-order poles), rather than as a subcomplex closed under taking subfaces. What we will use in practice is the induced fixed dependency graph (the inclusion lattice with marked/fixed nodes), which supports deterministic message passing and constraint propagation in Section~\ref{sec:conn-autodiff}. We do not rely on homological invariants in what follows.

\subsection{Combinatorial Neural Networks and Discrete Automatic Differentiation}
\label{sec:conn-autodiff}

\paragraph{Scope and terminology.}
We borrow the language of Combinatorial Neural Networks (CoNNs) to describe message passing on the fixed face lattice induced by the recursion relations. There are no learnable parameters, no training, and no loss minimization. In this paper, ``forward pass'' means deterministic evaluation of higher-layer pole data from lower layers via the recursion, while the ``backward pass''/``backpropagation'' refers to the discrete propagation of constraint residuals triggered when an update attempts to enter a fixed node.

\paragraph{Integer message passing via factorial rescaling.}
Although the recursion for $K(A)$ contains the normalization factor $1/(|A|-2)$, the propagation can be implemented using strictly integer updates after a layer-dependent rescaling. Define
\[
\widetilde K(A):=(|A|-2)!\,K(A)\qquad(|A|\ge2).
\]
Then
\[
\widetilde K(A)=\sum_{\substack{B\subset A\\|B|=|A|-1}}\widetilde K(B),\qquad |A|\ge3,
\]
so every local update along a face relation carries integer weight. This is exact within the integer-rescaled recursion variables: we only perform integer additions/subtractions on a fixed dependency graph.

Mainstream autodiff frameworks target differentiable real-valued programs; our integer, constraint-preserving propagation is outside their typical design scope, so we implement it explicitly.

\paragraph{Local influence weights (discrete sensitivities).}
For adjacent faces $B\subset A$ with $|A|=|B|+1$, a unit change in $\widetilde K(B)$ contributes a unit change to $\widetilde K(A)$. We use derivative notation as a compact bookkeeping device for these local influence weights:
\begin{equation}
\frac{\partial \widetilde K(A)}{\partial \widetilde K(B)} = 1,
\qquad B\subset A,\quad |A|=|B|+1.
\end{equation}
Multi-step influence is given by counting directed paths in the Hasse diagram. In the original $K$-normalization the corresponding coefficients are rational due to the $1/(|A|-2)$ factors. Equivalently, any update can be written as a linear combination of bottom-layer updates,
\begin{equation}
\Delta K(s_A) = \sum_{\substack{i,j \in A \\ i < j}} c_{A,ij} \Delta K(s_{ij}),\quad c_{A,ij} \in \mathbb{Q},
\qquad
\Delta \widetilde K(s_A) = \sum_{\substack{i,j \in A \\ i < j}} \widetilde c_{A,ij} \Delta \widetilde K(s_{ij}),\quad \widetilde c_{A,ij} \in \mathbb{Z}.
\end{equation}

\paragraph{Deterministic constraint propagation (CoNN-style message passing).}
The recursion constraints define a fixed computation graph on faces. CoNN-style message passing on simplicial complexes is commonly written as updates that aggregate information from boundary and coboundary relations; for example,
\begin{equation}
h_\sigma^{(t+1)} = \phi\left(h_\sigma^{(t)}, \bigoplus_{\tau \in \mathcal{B}(\sigma)} m_{\tau \to \sigma}^{(t)}, \bigoplus_{\tau \in \mathcal{C}(\sigma)} m_{\tau \to \sigma}^{(t)}\right)
\end{equation}
where $\mathcal{B}(\sigma)$ and $\mathcal{C}(\sigma)$ denote boundary and coboundary neighborhoods. In our setting, the ``messages'' are integer updates $\Delta \widetilde K$ propagated on the inclusion lattice, with fixed nodes encoding physical constraints.

\begin{algorithm}[H]
\caption{PropagateUpdate: Discrete constraint propagation}
\label{alg:propagate}
\begin{algorithmic}[1]
\REQUIRE Graph $G$, modified pole $s_A$, change $\Delta \widetilde K$, fixed poles $\mathcal{F}$
\ENSURE Updated $\widetilde K$-values satisfying all constraints
\STATE \textbf{Queue} $Q \leftarrow \{s_A\}$, \textbf{Visited} $V \leftarrow \emptyset$
\WHILE{$Q \neq \varnothing$}
    \STATE $s_B \leftarrow \textsc{Dequeue}(Q)$
    \IF{$s_B \in V$}
        \STATE \textbf{continue}
    \ENDIF
    \STATE $V \leftarrow V \cup \{s_B\}$
    
    \STATE \COMMENT{Upward propagation to supersets (integer update in $\widetilde K$)}
    \FOR{each superset $s_C \supset s_B$ with $|s_C| = |s_B| + 1$}
        \IF{$s_C \notin \mathcal{F}$}
            \STATE $\Delta \widetilde K_C \leftarrow \Delta \widetilde K$
            \STATE $\widetilde K^*(s_C) \leftarrow \widetilde K^*(s_C) + \Delta \widetilde K_C$
            \STATE \textsc{Enqueue}(Q, $s_C$)
        \ELSE
            \STATE \COMMENT{Fixed node encountered: trigger residual redistribution (backpropagation)}
            \STATE \textsc{Backpropagate}($G$, $s_B$, $s_C$, $\Delta \widetilde K$, $\mathcal{F}$)
        \ENDIF
    \ENDFOR
    
    \STATE \COMMENT{Downward step (only if needed when updating from higher layer)}
    \IF{$|s_B| > 2$}
        \STATE Select $s_D \subset s_B$ with $|s_D| = |s_B| - 1$ using a deterministic policy (see note)
        \IF{$s_D \notin \mathcal{F}$}
            \STATE $\widetilde K^*(s_D) \leftarrow \widetilde K^*(s_D) + \Delta \widetilde K$
            \STATE \textsc{Enqueue}(Q, $s_D$)
        \ENDIF
    \ENDIF
\ENDWHILE
\end{algorithmic}
\end{algorithm}

\noindent\textbf{Implementation notes.}
(1) We implement Algorithm~\ref{alg:propagate} using $\widetilde K(A)=(|A|-2)!\,K(A)$, so the upward update along every face relation is integer and requires no division. In particular, for $B\subset C$ with $|C|=|B|+1$ we use $\Delta \widetilde K(C)\leftarrow \Delta \widetilde K(B)$, which is equivalent to $\Delta K(C)\leftarrow \Delta K(B)/( |B|-1)$ in the original normalization.

(2) \emph{Selection policy (deterministic).} When multiple admissible subsets $s_D\subset s_B$ are available in the downward step, we choose the one that maximizes the remaining slack to inequality constraints; ties are broken lexicographically. This is for reproducibility and does not imply optimality.

\paragraph{Two phases.}
Operationally, the update alternates between: (i) upward propagation, which pushes $\Delta \widetilde K$ to immediate supersets to maintain the recursion, and (ii) residual redistribution (backpropagation), which is triggered by fixed nodes and reroutes the update through unfixed alternatives. At convergence, the two-particle data $\{K(s_{ij})\}$ determine the integrand via
\begin{equation}
I(z) = \prod_{1 \leq i < j \leq n} (z_i - z_j)^{-K(s_{ij})}
\end{equation}

\subsection{Neural-Inspired Pick-Pole Algorithm}
\label{sec:pick-pole}

The computational framework developed in Section \ref{sec:conn-autodiff}, combining Combinatorial Neural Networks with discrete automatic differentiation, provides the essential infrastructure for implementing the pole selection algorithm. We now turn to the central computational challenge in the CHY formalism: isolating specific physical singularities from the complete amplitude structure.

The pick-pole algorithm addresses a basic practical task in applying the CHY formalism \cite{Feng:2016nsv,Feng:2019winn}. Given a CHY integrand $I(z)$ that produces an amplitude with multiple poles, we often need to isolate contributions containing a specific pole $1/s_A$ where $s_A = (\sum_{i \in A} k_i)^2$. This capability is useful for studying factorization limits and related analytic structures, and it also interfaces naturally with forward-limit constructions of loop-level objects \cite{He:2015yua}.

For the bi-adjoint scalar theory, where the integrand takes the form $I[\pi|\rho] = \text{PT}(\pi) \times \text{PT}(\rho)$, the algorithm is elegant and provably correct \cite{Baadsgaard:2015voa,Huang:2018pbu}. The key insight is that a pole $s_A$ appears if and only if the set $A$ forms a contiguous block in both orderings $\pi$ and $\rho$. One can then multiply by an appropriate cross-ratio factor to eliminate unwanted contributions:
\begin{equation}
\mathcal{P}_{bd}^{ac} = \frac{z_{ac}z_{bd}}{z_{ad}z_{cb}}
\end{equation}
where $a, b$ are the endpoints of the contiguous set $A$ and $c, d$ are their nearest neighbors in the complement $\bar{A}$ \cite{Feng:2020lgt}.

For more general CHY integrands, particularly those associated with labelled tree graphs \cite{Gao:2017dek}, the notion of ``contiguity'' loses meaning, and the algorithm becomes noticeably more complex. As noted in \cite{Feng:2020lgt}, the current approach is heuristic, relying on a generate-and-test search through combinations of cross-ratio factors. This heuristic method, while sufficient to construct a broadly applicable one-loop CHY integrand for BAS theory \cite{Feng:2019winn}, suffers from exponential computational complexity and lacks a rigorous proof of correctness.

Our approach reformulates this challenge within the graph-theoretic and topological framework developed in previous sections. The key insight is that pole selection corresponds to a systematic adjustment of K-values while maintaining all recursion relations and fixed-point constraints. Rather than searching through combinations of cross-ratio factors, we directly manipulate the underlying graph structure through controlled modifications of the generalized pole degrees.

\begin{algorithm}[H]
\caption{Pick-Pole Algorithm for CHY Integrands}
\label{alg:pick-pole}
\begin{algorithmic}[1]
\REQUIRE CHY integrand $I(z)$ with pole structure, target pole $s_T$ to retain
\ENSURE Modified integrand $I^*(z)$ containing only terms with pole $s_T$
\STATE \textbf{Initialize:} Construct pole hierarchy graph $G$ with K-values from $I(z)$
\STATE \textbf{Identify:} Mark poles to retain $\mathcal{P}_{\text{retain}} = \{s_T\} \cup \text{CompatiblePoles}(s_T)$
\STATE \textbf{Identify:} Mark poles to remove $\mathcal{P}_{\text{remove}} = \text{AllPoles}(I) \setminus \mathcal{P}_{\text{retain}}$
\FOR{each pole $s_A \in \mathcal{P}_{\text{remove}}$}
    \STATE $K^*(s_A) \leftarrow K(s_A) - 1$ \COMMENT{Decrement K-value}
    \STATE \CALL{PropagateUpdate}{$G$, $s_A$, $-1$, $\mathcal{P}_{\text{retain}}$}
\ENDFOR
\RETURN Integrand $I^*(z)$ constructed from modified K-values
\end{algorithmic}
\end{algorithm}

The algorithm's correctness relies on several key properties established in previous sections. The additivity of K-values (Theorem in Section \ref{sec:generalized-chy}) ensures that modifications propagate correctly through the graph structure. The recursion relations \eqref{eq:Krec}-\eqref{eq:general-rec} maintain consistency across different levels of the pole hierarchy. Most critically, the discrete automatic differentiation framework embodied in the PropagateUpdate subroutine (Algorithm \ref{alg:propagate}) ensures that all constraints are satisfied throughout the modification process.

The connection to the topological perspective is particularly illuminating. Pole selection can be viewed as a projection operation on the simplicial complex, where we retain only those simplices (poles) compatible with our target. This geometric interpretation provides both intuition and a correctness criterion: the resulting integrand must correspond to a valid marked subset of the pole structure. The discrete nature of our approach--working with integer-valued integrand data and exact constraint satisfaction (integral in $\widetilde K$)--distinguishes it from continuous optimization methods and is designed to yield a valid CHY integrand within this framework.

The following sections illustrate the algorithm through explicit examples, showing how it isolates desired pole structures in both six-point and eight-point amplitudes. These examples serve not only as validation of the method but also as pedagogical illustrations of how the discrete automatic differentiation operates on the hierarchical pole structure of CHY integrands.

\subsection{Examples of 6-Point CHY Integrands}

We demonstrate the neural-inspired pick-pole algorithm through a concrete 6-point example. The algorithm operates on discrete structures where the bottom-layer $K(s_{ij})$ values are integers and propagation is implemented integrally in $\widetilde K$; the pole hierarchy forms a two-layer network structure that enables systematic pole selection through discrete automatic differentiation.

For any subset $A \subset \{1, 2, \cdots, n\}$, we define the K-value as:
$$K(s_A) := L[A]$$
In this example there are no dashed edges, so $K(s_A)=L[A]$.

Consider a six-point CHY integrand:
$$I_6 = \frac{1}{z_{12}^2 z_{23}^2 z_{34}^2 z_{56}^2 z_{16} z_{45} z_{46} z_{15}}$$

which produces five cubic Feynman diagrams:
$$\frac{1}{s_{12}s_{34}s_{56}} + \frac{1}{s_{123}s_{12}s_{56}} + \frac{1}{s_{123}s_{23}s_{56}} + \frac{1}{s_{156}s_{23}s_{56}} + \frac{1}{s_{156}s_{34}s_{56}}$$

To pick pole $s_{12}$, we need to retain poles $s_{12}$, $s_{34}$, $s_{56}$ and $s_{123}$ while removing poles $s_{23}$ and $s_{156}$.

\begin{figure}[H]
    \centering
    \includegraphics[width=1\linewidth]{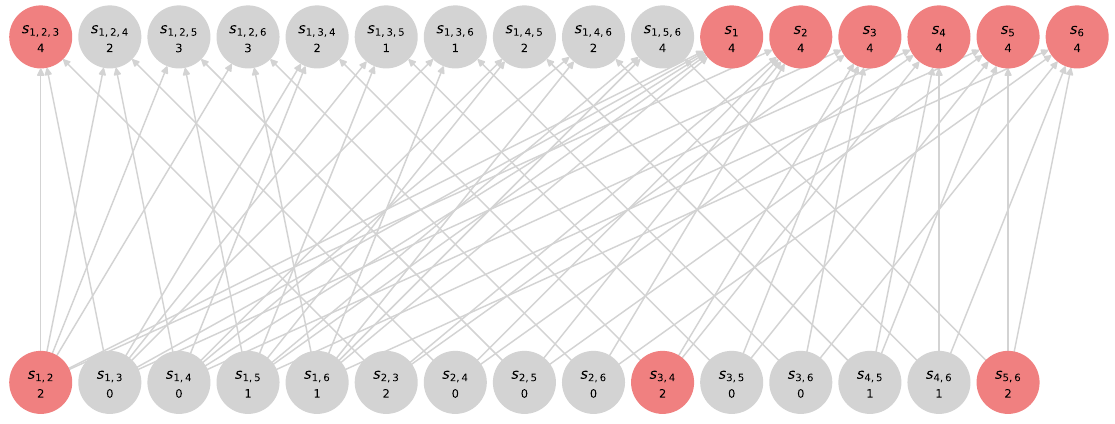}
    \caption{Initial pole hierarchy for a 6-point CHY integrand. The two-layer network has two-particle poles in the initial layer (bottom) and single-particle and three-particle poles in the second layer (top). Gray edges represent dependency relationships from the recursion relations $K(s_{abc}) = K(s_{ab}) + K(s_{ac}) + K(s_{bc})$ and $deg(s_{a}) = K(s_{ab}) + K(s_{ac}) + K(s_{ad}) + K(s_{ae}) + K(s_{af})$. Red nodes indicate fixed poles that must be preserved during the algorithm.}
    \label{eg-6-01}
\end{figure}

The recursion relations defining the pole hierarchy are:
$$K(s_{abc}) = K(s_{ab}) + K(s_{ac}) + K(s_{bc})$$
$$deg(s_{a}) = K(s_{ab}) + K(s_{ac}) + K(s_{ad}) + K(s_{ae}) + K(s_{af})$$

We fix the K-values of desired poles:
$$K(s_{12}) = 2,\quad K(s_{34}) = 2,\quad K(s_{56}) = 2,\quad K(s_{123}) = 4$$

\textbf{Step 1: Remove pole $s_{23}$}

We decrement the K-value of the unwanted pole $s_{23}$:
$$K^*(s_{23}) = K(s_{23}) - 1 = 1$$

\begin{figure}[H]
    \centering
    \includegraphics[width=1\linewidth]{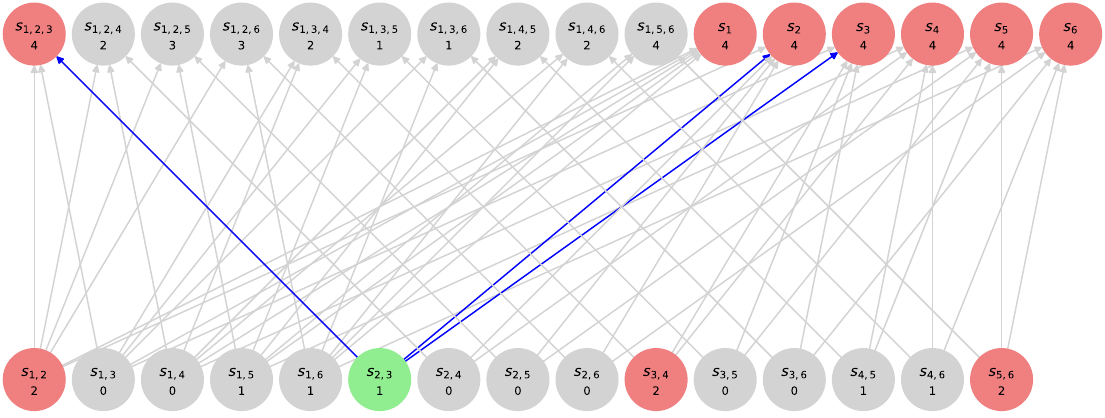}
    \caption{Initial upward propagation: The modification $\Delta K = -1$ at pole $s_{23}$ (green node) propagates upward through the network hierarchy. Blue arrows show the upward propagation paths to single-particle subsets and three-particle supersets. The algorithm encounters fixed constraints at $s_{123}$, $s_2$, and $s_3$ (red nodes), triggering the backward propagation mechanism.}
    \label{eg-6-02}
\end{figure}

The upward propagation encounters fixed points $s_{123}$, $s_2$, and $s_3$, triggering backward propagation.

\begin{figure}[H]
    \centering
    \includegraphics[width=1\linewidth]{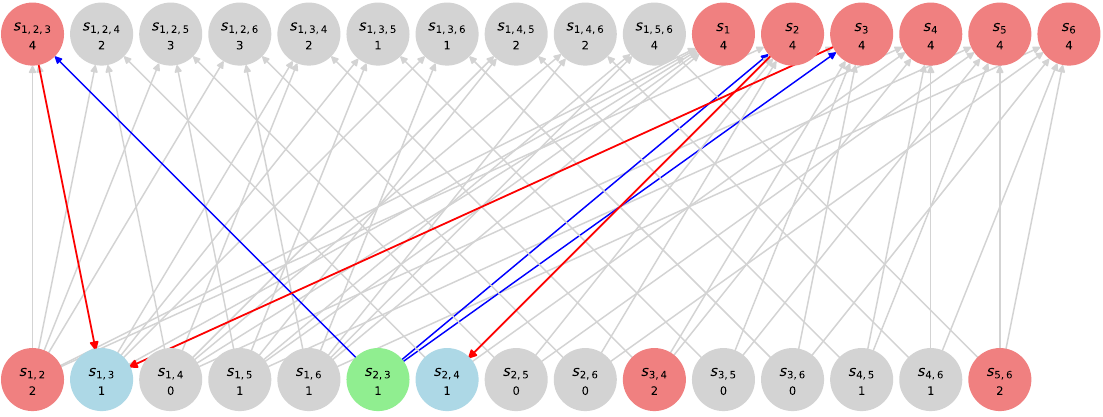}
    \caption{Initial backward propagation: When upward propagation hits fixed constraints, the discrete automatic differentiation algorithm redistributes the changes through backward propagation (red arrows). The algorithm selects intersection $s_{13}$ and superset $s_{24}$, incrementing their K-values by +1 to maintain all recursion relations while respecting fixed constraints.}
    \label{eg-6-03}
\end{figure}

The algorithm compensates by selecting appropriate poles:
$$K^*(s_{13}) = K(s_{13}) + 1= 1,\quad K^*(s_{24}) =K(s_{24}) + 1= 1$$

\begin{figure}[H]
    \centering
    \includegraphics[width=1\linewidth]{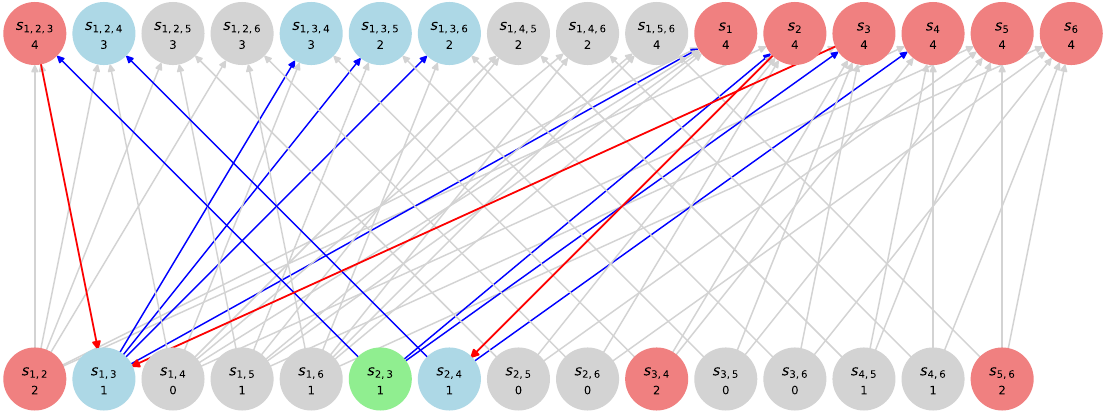}
    \caption{Second upward propagation: The changes at $s_{13}$ and $s_{24}$ propagate upward (blue arrows), updating their three-particle supersets $s_{134}$, $s_{135}$, $s_{136}$ and $s_{124}$ by +1. The propagation again encounters fixed points $s_1$ and $s_4$, necessitating another round of backward propagation.}
    \label{eg-6-04}
\end{figure}

The upward propagation again encounters fixed points $s_1$ and $s_4$.

\begin{figure}[H]
    \centering
    \includegraphics[width=1\linewidth]{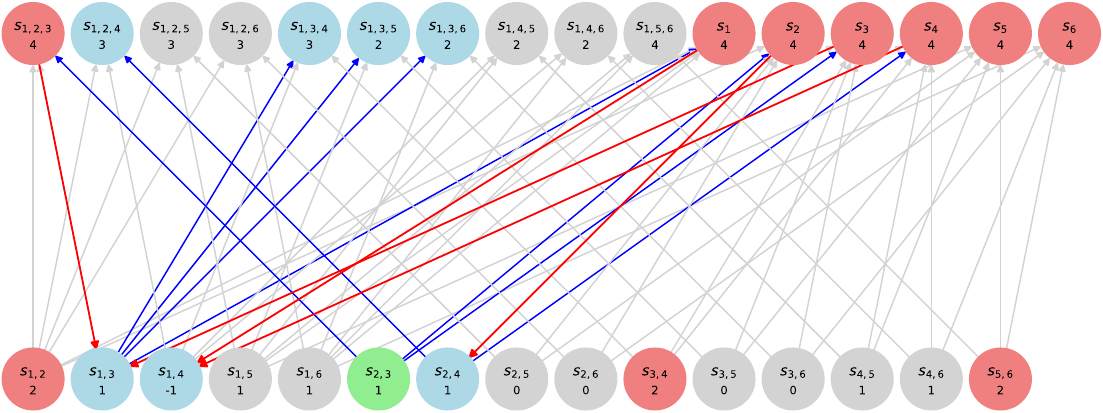}
    \caption{Second backward propagation: To maintain consistency with fixed single-particle constraints, the algorithm performs backward propagation from the second layer to the initial layer (red arrows). The intersection $s_{14}$ of dependent supersets from fixed nodes $s_1$ and $s_4$ is decremented by -1, ensuring all recursion relations remain satisfied.}
    \label{eg-6-05}
\end{figure}

\begin{figure}[H]
    \centering
    \includegraphics[width=1\linewidth]{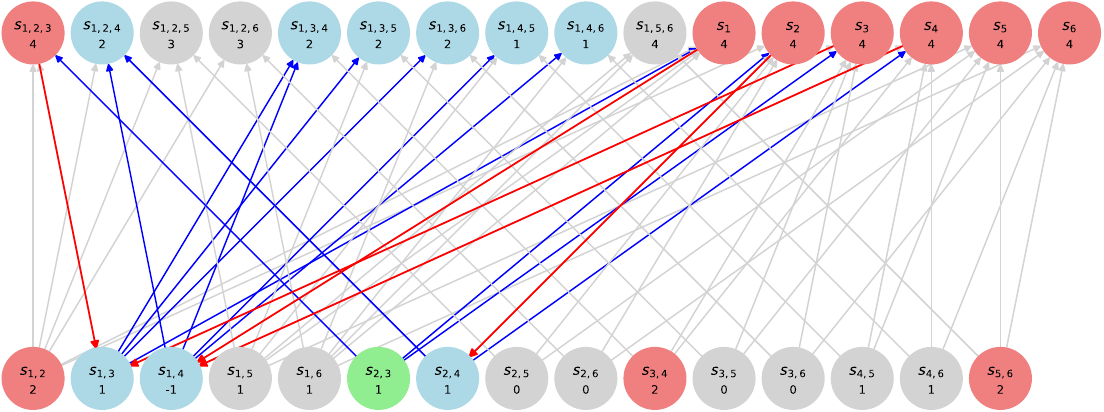}
    \caption{Final upward propagation for $s_{23}$ removal: The third upward propagation completes without encountering additional fixed points, indicating successful convergence of the discrete automatic differentiation algorithm. The removal of pole $s_{23}$ is now complete with all constraints satisfied.}
    \label{eg-6-06}
\end{figure}

The algorithm terminates when no fixed points are encountered, completing the removal of pole $s_{23}$.

\textbf{Step 2: Remove pole $s_{156}$}

We now repeat the procedure to remove pole $s_{156}$.

\begin{figure}[H]
    \centering
    \includegraphics[width=1\linewidth]{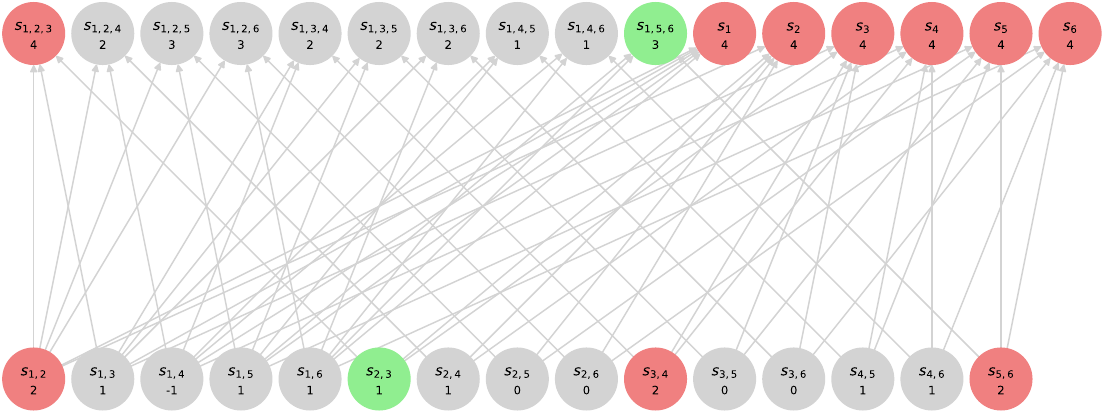}
    \caption{Updated pole hierarchy for $s_{156}$ removal: The network structure after removing $s_{23}$, showing updated K-values throughout the hierarchy. Fixed points (red) and the target pole $s_{156}$ for removal (green) are clearly marked. The algorithm now proceeds to remove the second unwanted pole while maintaining all established constraints.}
    \label{eg-6-07}
\end{figure}

We decrement the K-value of pole $s_{156}$:
$$K^*(s_{156}) = K(s_{156}) - 1 = 3$$

\begin{figure}[H]
    \centering
    \includegraphics[width=1\linewidth]{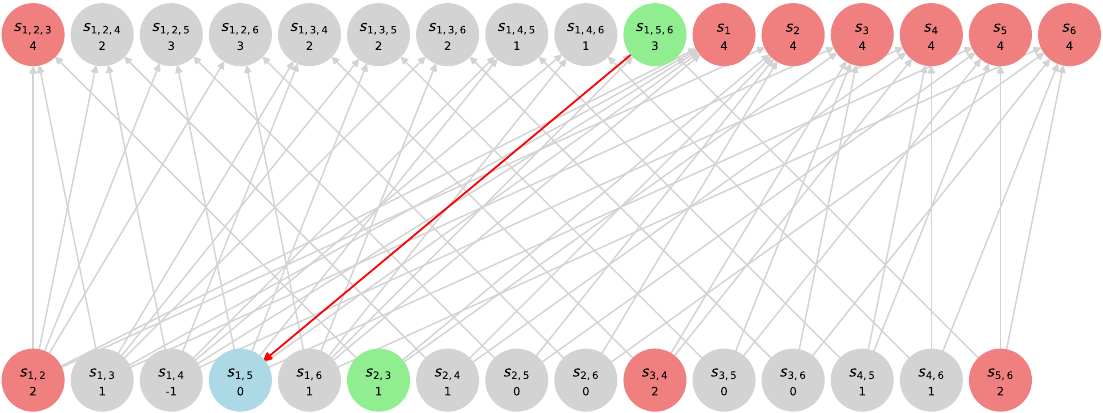}
    \caption{Downward propagation: Unlike the previous case, the modification at three-particle pole $s_{156}$ requires downward propagation to the initial layer (red arrows). The algorithm selects subset $s_{15}$ and decrements its K-value by -1, demonstrating the bidirectional nature of the discrete automatic differentiation framework.}
    \label{eg-6-08}
\end{figure}

The downward propagation selects subset $s_{15}$ for modification.

\begin{figure}[H]
    \centering
    \includegraphics[width=1\linewidth]{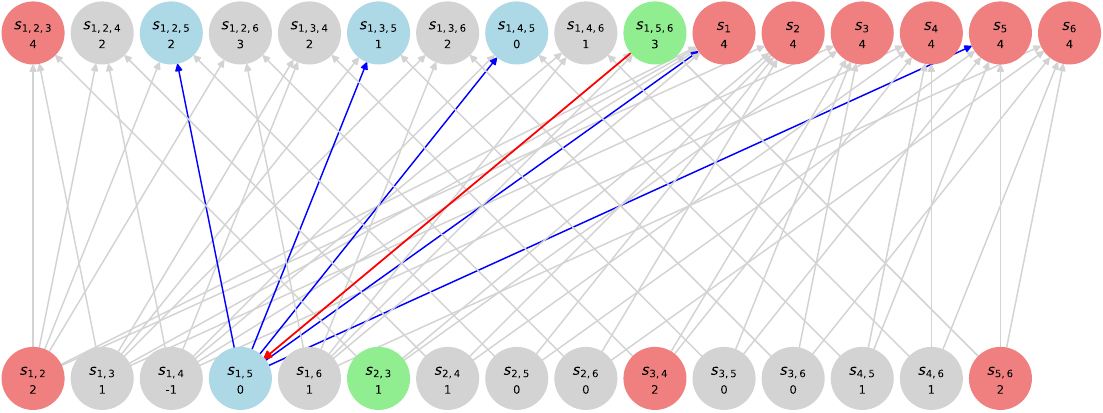}
    \caption{Upward propagation from $s_{15}$: The change at $s_{15}$ propagates upward (blue arrows) to its three-particle supersets $s_{125}$, $s_{135}$, and $s_{145}$, each decremented by -1. Fixed points $s_1$ and $s_5$ are encountered, triggering backward propagation to maintain network consistency.}
    \label{eg-6-09}
\end{figure}

\begin{figure}[H]
    \centering
    \includegraphics[width=1\linewidth]{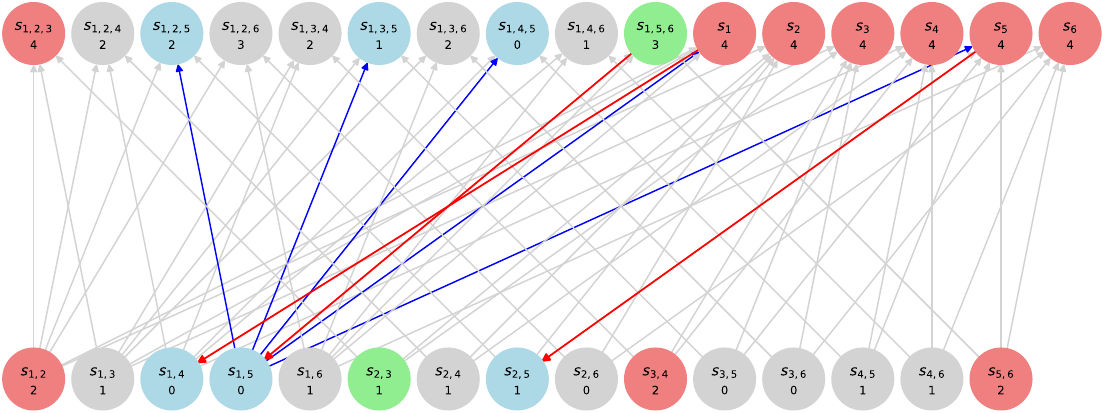}
    \caption{Backward propagation for fixed constraints: The algorithm redistributes changes through backward propagation (red arrows), selecting supersets $s_{14}$ and $s_{25}$ of fixed points $s_1$ and $s_5$ respectively. Each is incremented by +1 to compensate for the inability to modify the fixed single-particle poles.}
    \label{eg-6-10}
\end{figure}

\begin{figure}[H]
    \centering
    \includegraphics[width=1\linewidth]{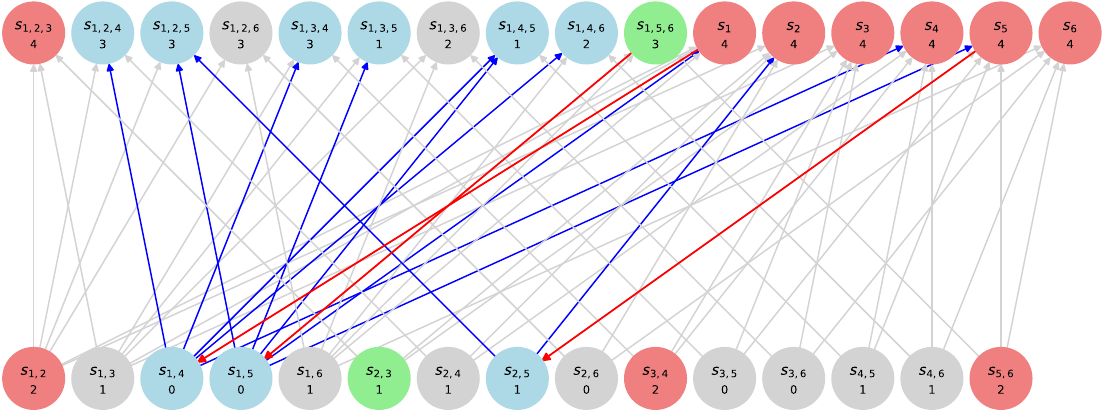}
    \caption{Continued upward propagation: Changes at $s_{14}$ and $s_{25}$ propagate upward, updating their three-particle supersets by +1. The algorithm encounters additional fixed points $s_2$ and $s_4$, requiring another round of backward propagation to complete the constraint satisfaction process.}
    \label{eg-6-11}
\end{figure}

\begin{figure}[H]
    \centering
    \includegraphics[width=1\linewidth]{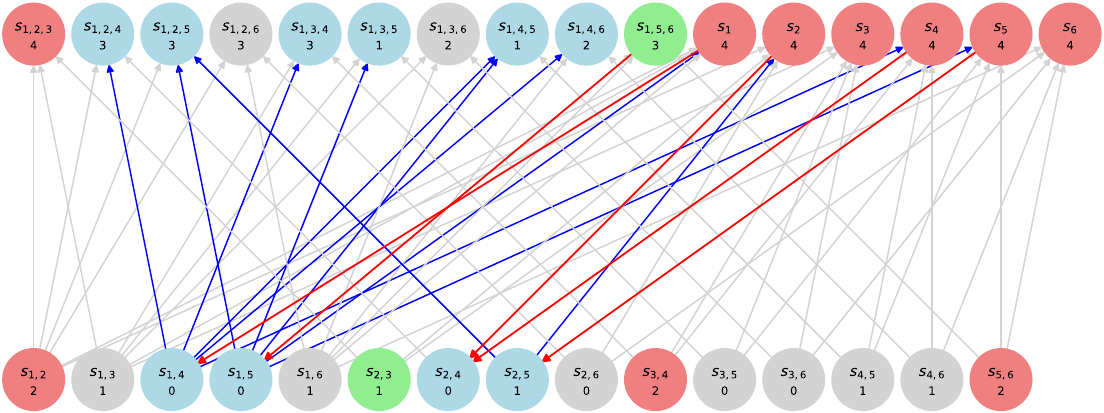}
    \caption{Final backward propagation: The last backward propagation step decrements the intersection $s_{24}$ of supersets from fixed nodes $s_2$ and $s_4$ by -1. This completes the constraint redistribution necessary to maintain all recursion relations while respecting fixed point constraints.}
    \label{eg-6-12}
\end{figure}

\begin{figure}[H]
    \centering
    \includegraphics[width=1\linewidth]{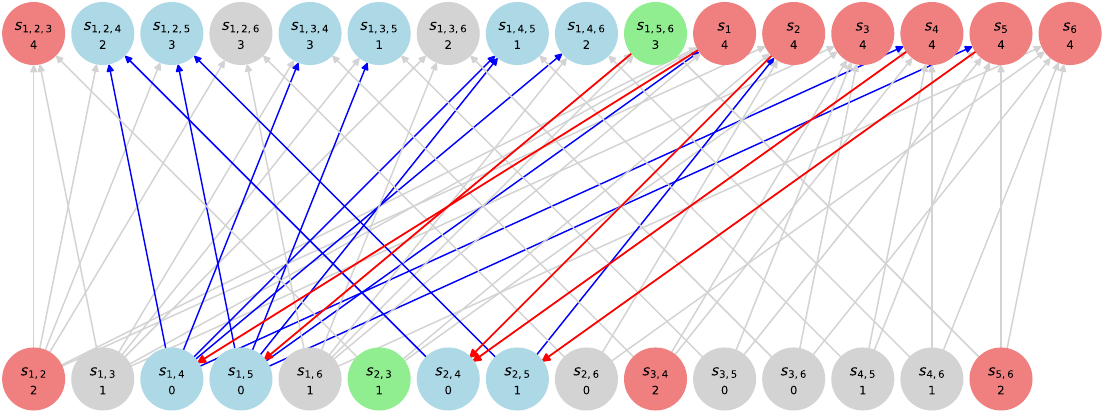}
    \caption{Algorithm convergence: The final upward propagation from $s_{24}$ completes without encountering fixed constraints, indicating successful convergence of the discrete automatic differentiation algorithm. The removal of both unwanted poles $s_{23}$ and $s_{156}$ is now complete, yielding the desired CHY integrand.}
    \label{eg-6-13}
\end{figure}

The algorithm terminates successfully, yielding the modified CHY integrand:
$$I_6^* = \frac{1}{z_{12}^2 z_{13}z_{16}z_{23}z_{25} z_{34}^2 z_{45} z_{46}z_{56}^2}$$

This produces the amplitudes:
$$\frac{1}{s_{12}s_{34}s_{56}} + \frac{1}{s_{123}s_{12}s_{56}} $$

which successfully picks up all terms containing the desired pole $s_{12}$.

\paragraph{Summary (6-point).}
This example illustrates the pick-pole pipeline as deterministic constraint propagation on a fixed
(two-layer) dependency graph. Two-particle channels form the base layer, while the single-particle and
three-particle nodes form the next layer.

Starting from an integer modification of an unwanted pole (implemented integrally in $\widetilde K$), the procedure:
(i) propagates the induced update upward along face relations to maintain the recursion constraints; and
(ii) upon encountering a fixed node, triggers \emph{backpropagation}, i.e. discrete residual redistribution
to nearby unfixed nodes so that all constraints remain satisfied. The same update rules extend to higher
multiplicity, where additional layers (e.g. four-particle nodes at eight points) appear in the dependency
graph.

\clearpage
\section{Applications to Higher-Order Pole Expansion}
\label{sec:applications}

\subsection{Cross-Ratio Identities and Higher-Order Poles}
\label{sec:cross-ratio-identities}

Physical tree-level amplitudes must contain only simple poles corresponding to the propagation of on-shell intermediate particles—a requirement dictated by unitarity and locality. Yet when we expand the Pfaffian factors in Yang-Mills or gravity integrands, individual terms routinely exhibit poles of the form $1/s_A^{\alpha}$ with $\alpha > 1$. These unphysical singularities must precisely cancel when summing over all $(n-3)!$ solutions to the scattering equations, but verifying such cancellations quickly becomes computationally expensive even for moderate multiplicities \cite{Cardona:2016gon}.

For a given $n$-point CHY integrand, the poles and the order of the poles can be determined by the pole index $\chi(A) := L[A] - 2(|A| - 1)$. Explicitly, each subset $A$ gives a nonzero pole contribution if and only if $\chi(A) \geq 0$. From the previous algorithm for Feynman diagrams we can obtain all Feynman diagrams for the CHY integrand with only simple poles. But for higher-order poles, the problem becomes complicated.

For a CHY integrand with a pole of higher order $\frac{1}{s_A^{\alpha}}$, we can reduce the order of the CHY integrand by multiplying an identity containing that pole $s_A$. The construction of the identity for general higher-order poles is similar to the cross-ratio. In \cite{Cardona:2016gon}, the cross-ratio identity for pole $s_A$ with selected index $j \in A$ and $p \in \bar{A}$ is
\begin{equation}
I_n[A,j,p] \equiv -\sum_{i \in A \setminus \{j\}} \sum_{b \in \bar{A} \setminus \{p\}} \frac{s_{ib}}{s_A} \frac{z_{bp}z_{ij}}{z_{ib}z_{jp}} = 1
\label{eq:cross-ratio-identity}
\end{equation}

This identity holds only when the worldsheet coordinates $\{z_i\}$ satisfy the scattering equations, making it an on-shell relation that can be freely inserted into the CHY integral. The appearance of the factor $s_{ib}/s_A$ is crucial—when multiplied with an integrand containing a higher-order pole $1/s_A^{\alpha}$, it effectively reduces the pole order by one. The cross-ratio structure encoded in the factors $z_{bp}z_{ij}/(z_{ib}z_{jp})$ ensures that the resulting terms maintain the correct transformation properties under Möbius transformations.

Multiplying by this cross-ratio identity reduces the pole order and produces a linear combination of CHY integrands with simpler poles. The systematic algorithm iteratively reduces the higher-order poles until every resulting integrand is composed of simple poles. The algorithm proceeds as follows: for each subset $A$ with $\chi(A) > 0$, we apply the appropriate cross-ratio identity, generating a linear combination of new integrands. This process continues until all resulting integrands satisfy $\chi(B) \leq 0$ for all subsets $B$, ensuring only physical simple poles remain.

However, during the iteration process we have no control over whether the decomposition increases or decreases the higher-order poles of a CHY integrand, which can lead to redundant terms. Our goal is to minimize the number of CHY integrands that contain only simple poles. In the following, we propose a decomposition strategy based on the systematic algorithm that reduces redundant terms. In this section, we explore how the methods developed in previous sections can be applied to this problem.

\subsection{Decomposing 0-Regular Graphs into Cross-Ratios}
\label{Decompose the 0-regular}
As established in Section~\ref{sec:generalized-chy}, the 0-regular graphs form a group under multiplication and act as transformations on CHY integrands while preserving their regularity class. A crucial insight for decomposing higher-order poles is that multiplying a CHY integrand by an appropriate 0-regular graph can systematically reduce the pole orders. Specifically, when a CHY integrand contains a higher-order pole $1/s_A^{\alpha}$ with $\alpha > 1$, we can construct a suitable 0-regular graph whose multiplication reduces this pole to simple poles, thereby enabling the systematic decomposition described in Section~\ref{sec:cross-ratio-identities}.

The 0-regular graphs can be understood as multiplicative, M\"obius-invariant factors: by definition they have vanishing net degree at each vertex and therefore preserve the regularity class when multiplied into a CHY integrand. For higher-order pole reduction, it is useful to view such factors as ``long cross-ratios'' associated with alternating even cycles in the colored graph.

A key practical point is that any 0-regular factor can be decomposed into a product of standard four-point cross-ratios. One constructive route is:
(i) pair incident solid and dashed edges locally at each vertex (possible because $\deg_s(v)=\deg_d(v)$);
(ii) follow these pairings to partition the edge set into edge-disjoint alternating even cycles; and
(iii) factorize each alternating $2m$-cycle into $(m-1)$ four-point cross-ratios via a telescoping identity.
A proof and an explicit factorization formula are given in Appendix~\ref{app:hamiltonian}.

\paragraph{Cycle-to-cross-ratio factorization (example).}
For an alternating six-cycle with vertices $(i,j,k,l,m,n)$ ordered as
\[
i\!-\!j\!-\!k\!-\!l\!-\!m\!-\!n\!-\!i,
\]
the associated long factor takes the form
\begin{equation}
\frac{z_{ij}z_{kl}z_{mn}}{z_{jk}z_{lm}z_{ni}}\,,
\end{equation}
and admits the identity (up to an overall sign from $z_{ab}=-z_{ba}$)
\begin{equation}
\frac{z_{ij}z_{kl}z_{mn}}{z_{jk}z_{lm}z_{ni}}
=
\frac{z_{ij}z_{kn}}{z_{jk}z_{ni}}
\times
\frac{z_{kl}z_{mn}}{z_{lm}z_{kn}}\,.
\label{eq:cross-ratio-decomposition}
\end{equation}
Each factor on the right-hand side is a standard four-point cross-ratio. More generally, an alternating $2m$-cycle factorizes into $(m-1)$ four-point cross-ratios by iterating this step.

\begin{figure}[h]
    \centering
    \includegraphics[width=1.0\linewidth]{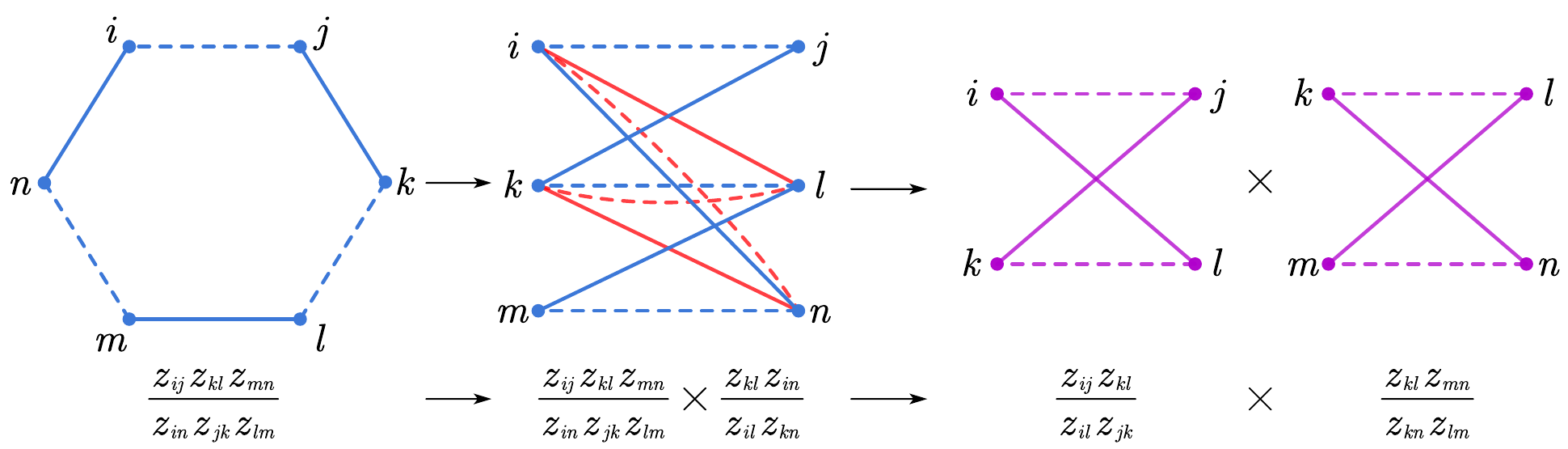}
    \caption{Example of decomposing a 0-regular factor via alternating-cycle extraction and telescoping factorization. A convenient visualization is obtained by a left/right representation of vertices; the resulting alternating even cycle can then be factorized into four-point cross-ratios by inserting intermediate diagonals (cf.\ Appendix~\ref{app:hamiltonian}).}
    \label{fig:decomposed-cross-ratios}
\end{figure}

\paragraph{Enhanced Decomposition Method for Higher-Order Pole Integrands.}
With the systematic decomposition of 0-regular graphs established, we can now develop a more efficient method for decomposing higher-order pole integrands that reduces computational complexity in our examples compared to the systematic approach outlined in Ref.~\cite{Cardona:2016gon}.

\textbf{Enhanced Algorithm:}
\begin{enumerate}
    \item \textbf{0-regular Graph Construction:} For a given higher-order pole integrand, apply the methods from Section~\ref{sec:generalized-chy} to construct an appropriate 0-regular graph that reduces the higher-order poles when multiplied with the original integrand.
    
    \item \textbf{Cross-ratio Decomposition:} Apply the algorithm described above to decompose this 0-regular graph into products of standard four-point cross-ratios, systematically identifying all minimal cross-ratio structures.
    
    \item \textbf{Cross-Ratio Identity Construction:} For each four-point cross-ratio identified in the decomposition, construct the corresponding cross-ratio identity using the methods from Section~\ref{sec:cross-ratio-identities}. This yields a set of multiplicative factors that can be applied to the original integrand.
\end{enumerate}

The key advantage of this enhanced approach is computational efficiency. Rather than applying multiple cross-ratio identities simultaneously (which leads to exponential growth in the number of terms), we apply them sequentially based on the systematic decomposition of the 0-regular graph.

\paragraph{Computational Advantages.}
Our preliminary analysis of two representative examples demonstrates that this enhanced decomposition method produces noticeably fewer intermediate terms compared to the symmetric approach mentioned in Section~\ref{sec:cross-ratio-identities}. Specifically:
\begin{itemize}
    \item The traditional symmetric approach applies all necessary cross-ratio identities simultaneously, leading to $\prod_{i} n_i$ terms where $n_i$ is the number of terms in each identity.
    \item Our enhanced method applies identities sequentially based on the 0-regular graph decomposition, typically reducing the total number of terms by factors of 2-5 depending on the complexity of the pole structure.
\end{itemize}

In the following sections, we will demonstrate this enhanced method through detailed examples of 6-point and 8-point CHY integrands with higher-order poles, showing explicit comparisons between the number of terms generated by each approach. These examples illustrate both the theoretical elegance and practical computational advantages of the graph-theoretic approach to pole decomposition in the CHY formalism.

\subsection{Examples: 6-Point CHY Integrand with Higher-Order Poles}

\begin{figure}[htbp]
    \centering
    \includegraphics[width=0.4\linewidth]{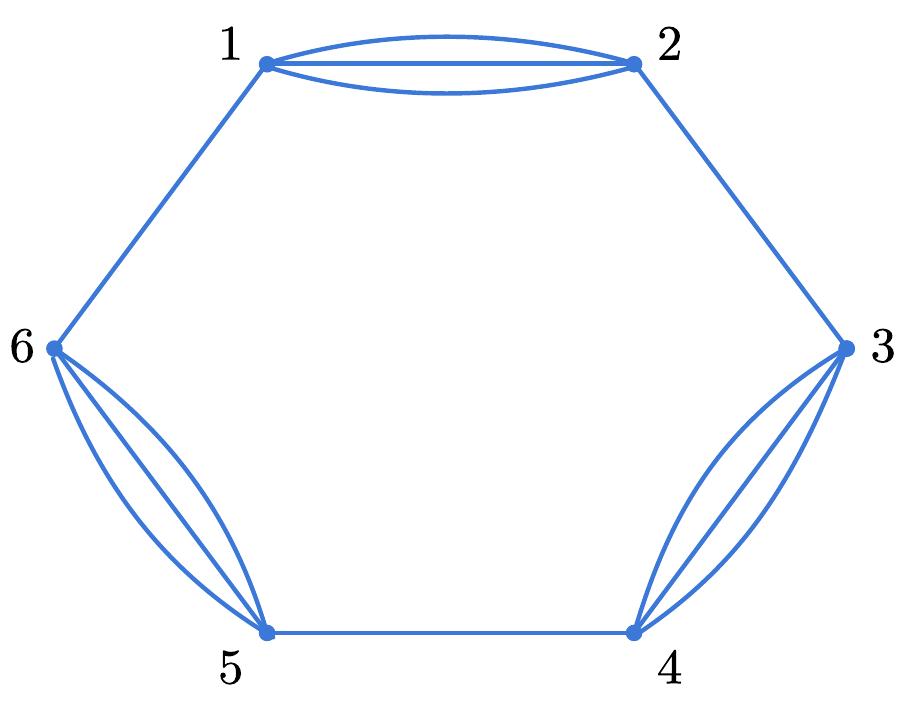}
    \caption{The 4-regular graph representation of the six-point CHY integrand $I_6$. Each vertex represents a particle, and the edge multiplicities encode the powers in the denominator of equation \eqref{eq:I6-integrand}. The triple edges connecting vertices $(1,2)$, $(3,4)$, and $(5,6)$ correspond to the factors $z_{12}^3$, $z_{34}^3$, and $z_{56}^3$ respectively, indicating the three double poles in the amplitude.}
    \label{fig:6pt-4regular}
\end{figure}

We demonstrate our enhanced decomposition method through a concrete six-point example that illustrates the computational advantages over traditional approaches. Consider the CHY integrand
\begin{equation}
I_6 = \frac{1}{z_{12}^3 z_{23} z_{34}^3 z_{45} z_{56}^3 z_{61}},
\label{eq:I6-integrand}
\end{equation}
whose 4-regular graph representation is shown in Figure \ref{fig:6pt-4regular}.

Using the pole index formula $\chi(A) = L[A] - 2(|A| - 1)$, we identify that this integrand contains exactly three double poles: $s_{12}$, $s_{34}$, and $s_{56}$ (where $\chi = 1$ for each). To systematically reduce these higher-order poles to simple poles, we employ the cross-ratio identity method with the six-point cross-ratio factor $\frac{z_{12}z_{34}z_{56}}{z_{16}z_{23}z_{45}}$.

\begin{figure}[htbp]
    \centering
    \includegraphics[width=0.9\linewidth]{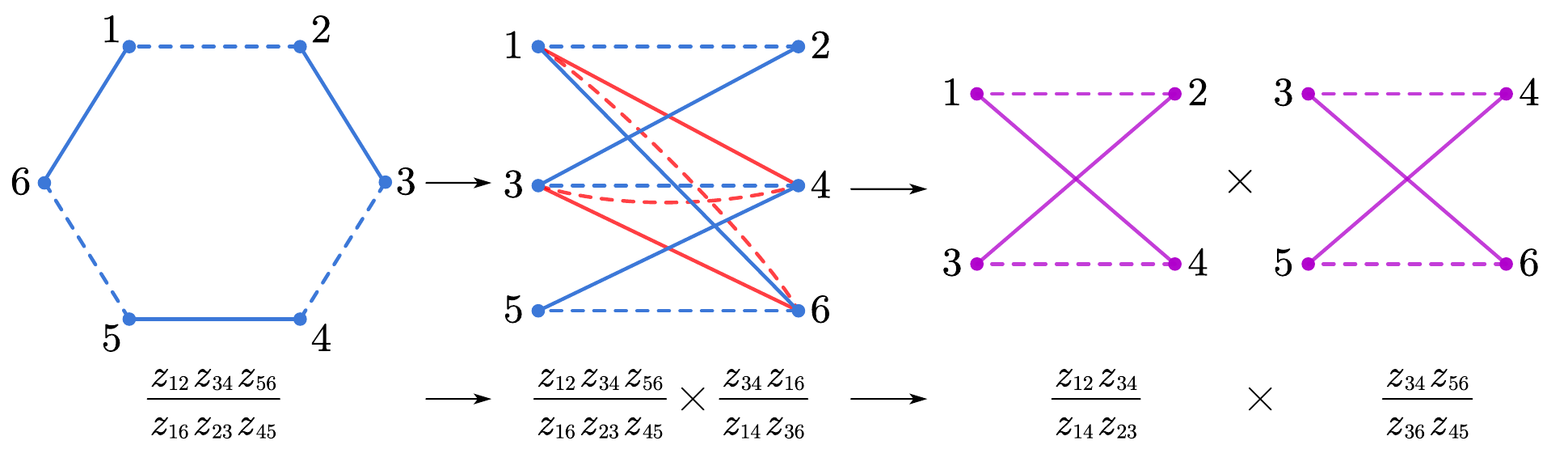}
    \caption{Systematic decomposition of the six-point cross-ratio factor $\frac{z_{12}z_{34}z_{56}}{z_{16}z_{23}z_{45}}$ into two four-point cross-ratios. The bipartite representation (center) reveals the natural factorization into $\frac{z_{12}z_{34}}{z_{14}z_{32}} \times \frac{z_{34}z_{56}}{z_{36}z_{54}}$, providing the foundation for our sequential application of cross-ratio identities.}
    \label{fig:crossratio-decomp}
\end{figure}

The key to our enhanced approach is the systematic decomposition of this six-point cross-ratio into elementary four-point factors, as illustrated in Figure \ref{fig:crossratio-decomp}.

Based on this decomposition, we strategically select two basic cross-ratio identities that minimize the generation of new higher-order poles:
\begin{align}
I_6[\{1,2\},1,4] &= -\frac{s_{23} z_{21} z_{34}}{s_{12} z_{14} z_{23}}-\frac{s_{25} z_{21} z_{54}}{s_{12} z_{14} z_{25}}-\frac{s_{26} z_{21} z_{64}}{s_{12} z_{14} z_{26}}, \label{eq:identity1}\\
I_6[\{5,6\},6,3] &= -\frac{s_{51} z_{13} z_{56}}{s_{56} z_{51} z_{63}}-\frac{s_{52} z_{23} z_{56}}{s_{56} z_{52} z_{63}}-\frac{s_{54} z_{43} z_{56}}{s_{56} z_{54} z_{63}}. \label{eq:identity2}
\end{align}

\begin{figure}[htbp]
    \centering
    \includegraphics[width=1\linewidth]{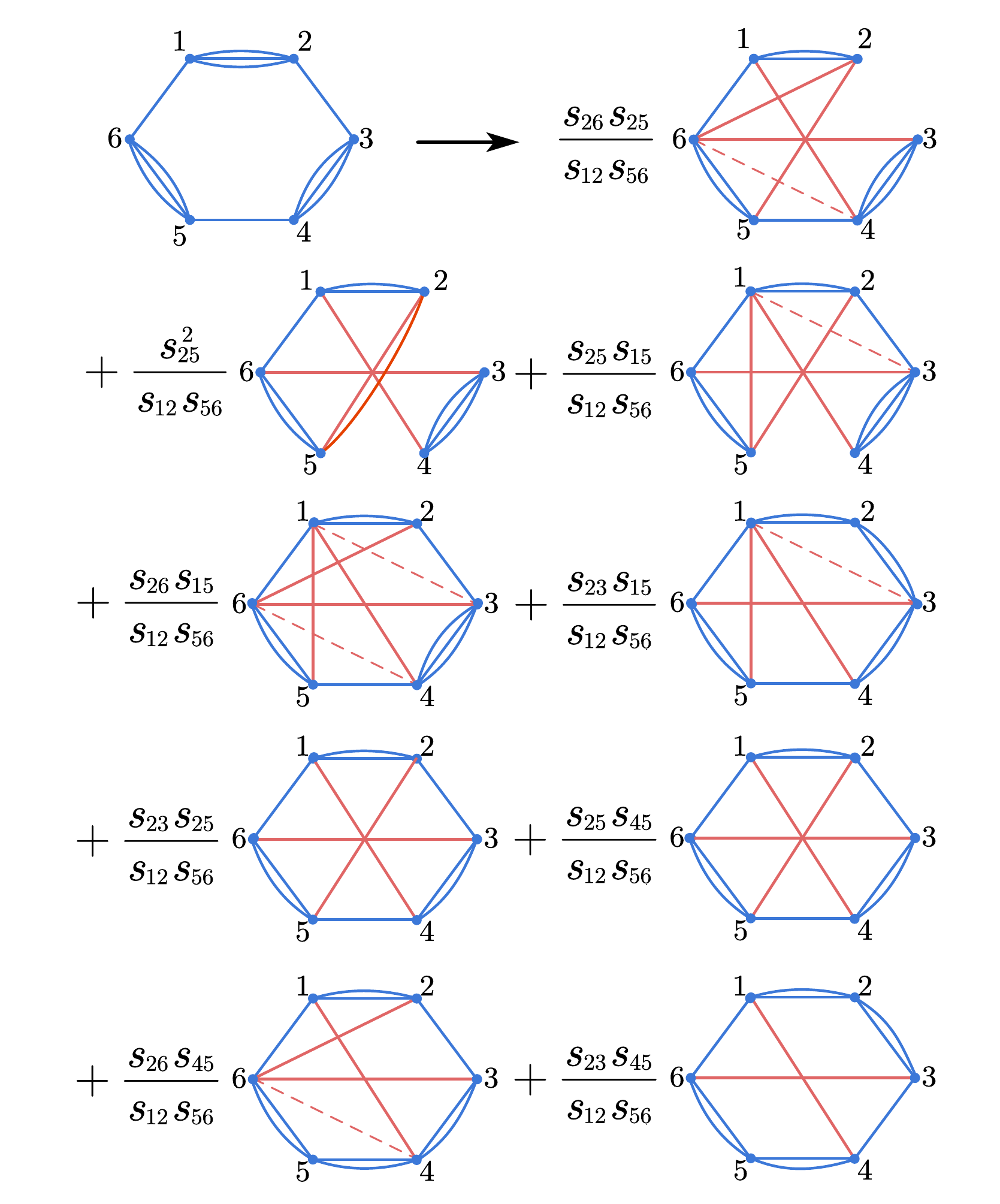}
    \caption{Complete decomposition tree showing how the original integrand $I_6$ with three double poles is systematically reduced through cross-ratio identities. The initial level shows the $3 \times 3 = 9$ terms from applying identities \eqref{eq:identity1} and \eqref{eq:identity2}. Five terms (shown in gray) already contain only simple poles, while four terms (shown in white) retain the double pole $s_{34}$, requiring further decomposition.}
    \label{fig:decomp-overview}
\end{figure}

Multiplying $I_6$ by these identities yields $3^2 = 9$ terms, whose pole structures are systematically tracked in Figure \ref{fig:decomp-overview}.

Among these nine terms, five immediately yield CHY integrands with only simple poles, while the remaining four terms
\begin{equation}
I_6\times\frac{s_{26}s_{52}}{s_{12} s_{56}}\frac{z_{12}z_{23}z_{46}z_{56}}{z_{14}z_{25}z_{26}z_{36}}, \quad I_6\times\frac{s_{25}s_{52}}{s_{12} s_{56}}\frac{z_{12}z_{23}z_{45}z_{56}}{z_{14}z_{25}^2z_{36}}, \quad \text{etc.}
\label{eq:intermediate-terms}
\end{equation}
still contain the double pole $s_{34}$. 

To eliminate this remaining higher-order pole, we apply the cross-ratio identity for subset $\{3,4\}$:
\begin{equation}
I_6[\{3,4\},3,1] = -\frac{s_{42} z_{21} z_{43}}{s_{34} z_{31} z_{42}}-\frac{s_{45} z_{51} z_{43}}{s_{34} z_{31} z_{45}}-\frac{s_{46} z_{61} z_{43}}{s_{34} z_{31} z_{46}}.
\label{eq:identity3}
\end{equation}

\begin{figure}[htbp]
    \centering
    \includegraphics[width=1\linewidth]{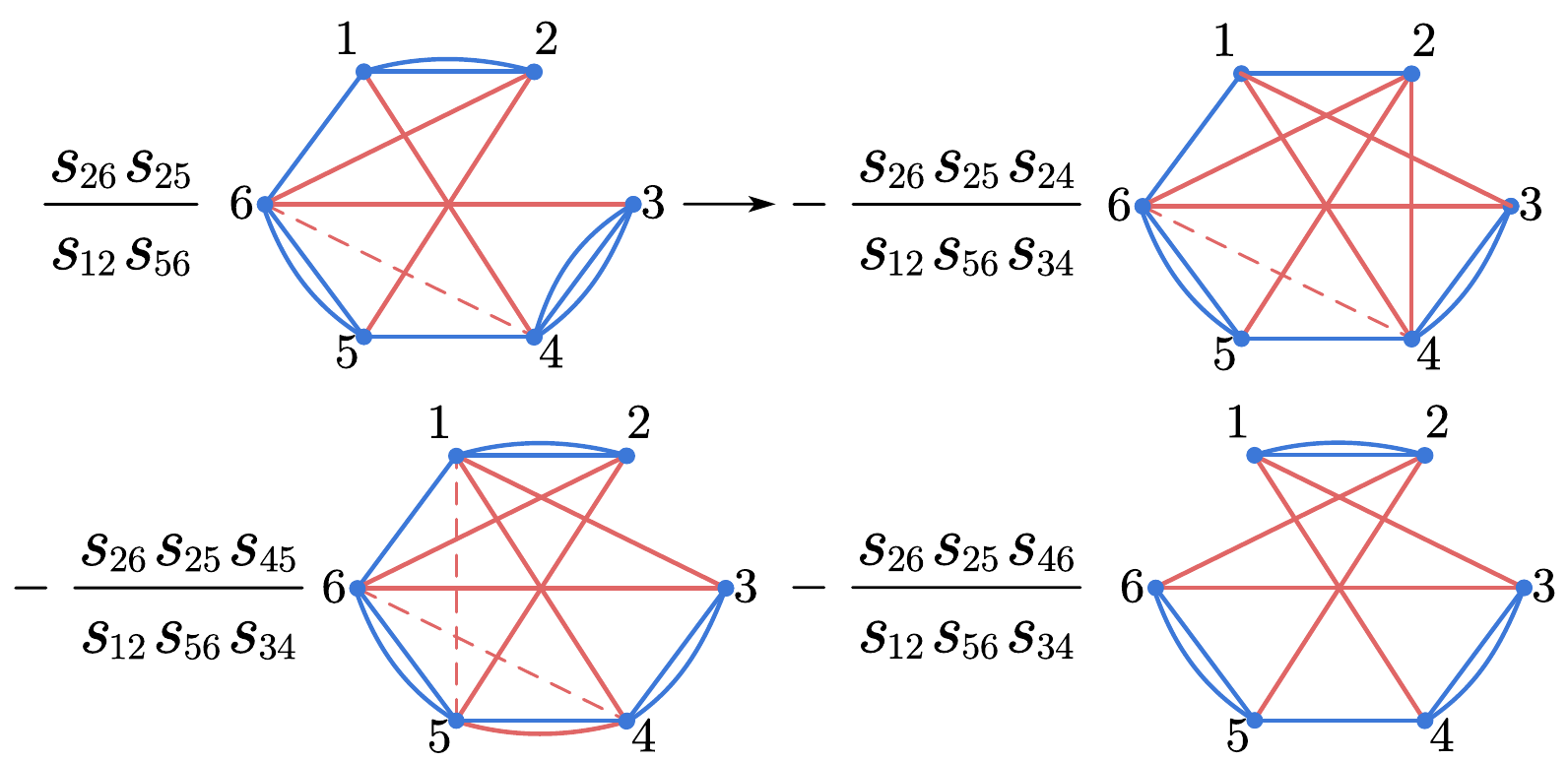}
    \caption{Decomposition of the initial intermediate term containing double pole $s_{34}$. Applying identity \eqref{eq:identity3} yields three CHY integrands with only simple poles, each corresponding to a distinct channel in the final amplitude. The cross-ratio factors effectively cancel the $z_{34}^3$ contribution, reducing the pole order from 2 to 1.}
    \label{fig:decomp1}
\end{figure}

\begin{figure}[htbp]
    \centering
    \includegraphics[width=1\linewidth]{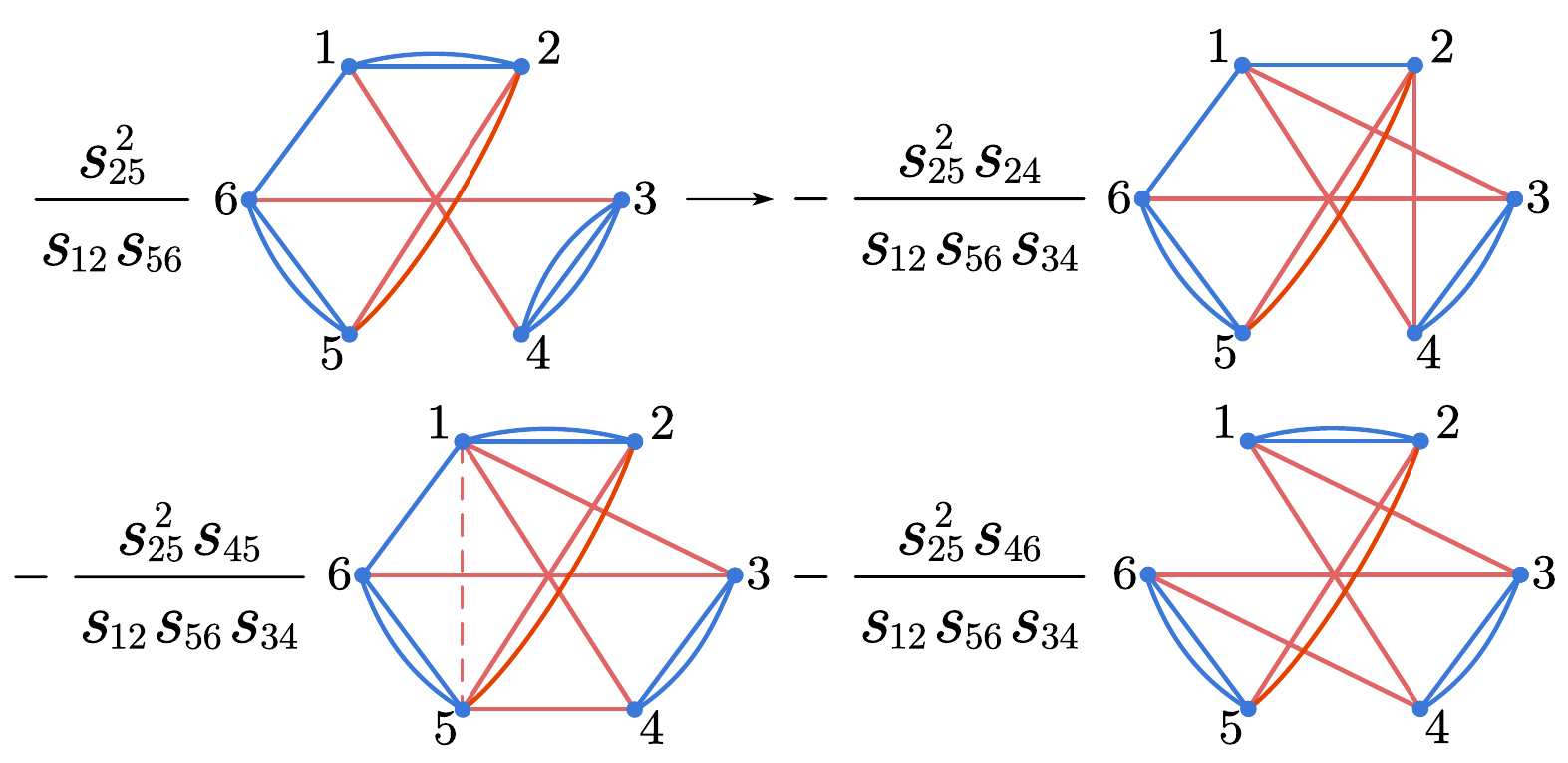}
    \caption{Decomposition of the second intermediate term. The systematic application of identity \eqref{eq:identity3} again produces three terms with simple poles, demonstrating the consistency of our approach across different intermediate configurations.}
    \label{fig:decomp2}
\end{figure}

\begin{figure}[htbp]
    \centering
    \includegraphics[width=1\linewidth]{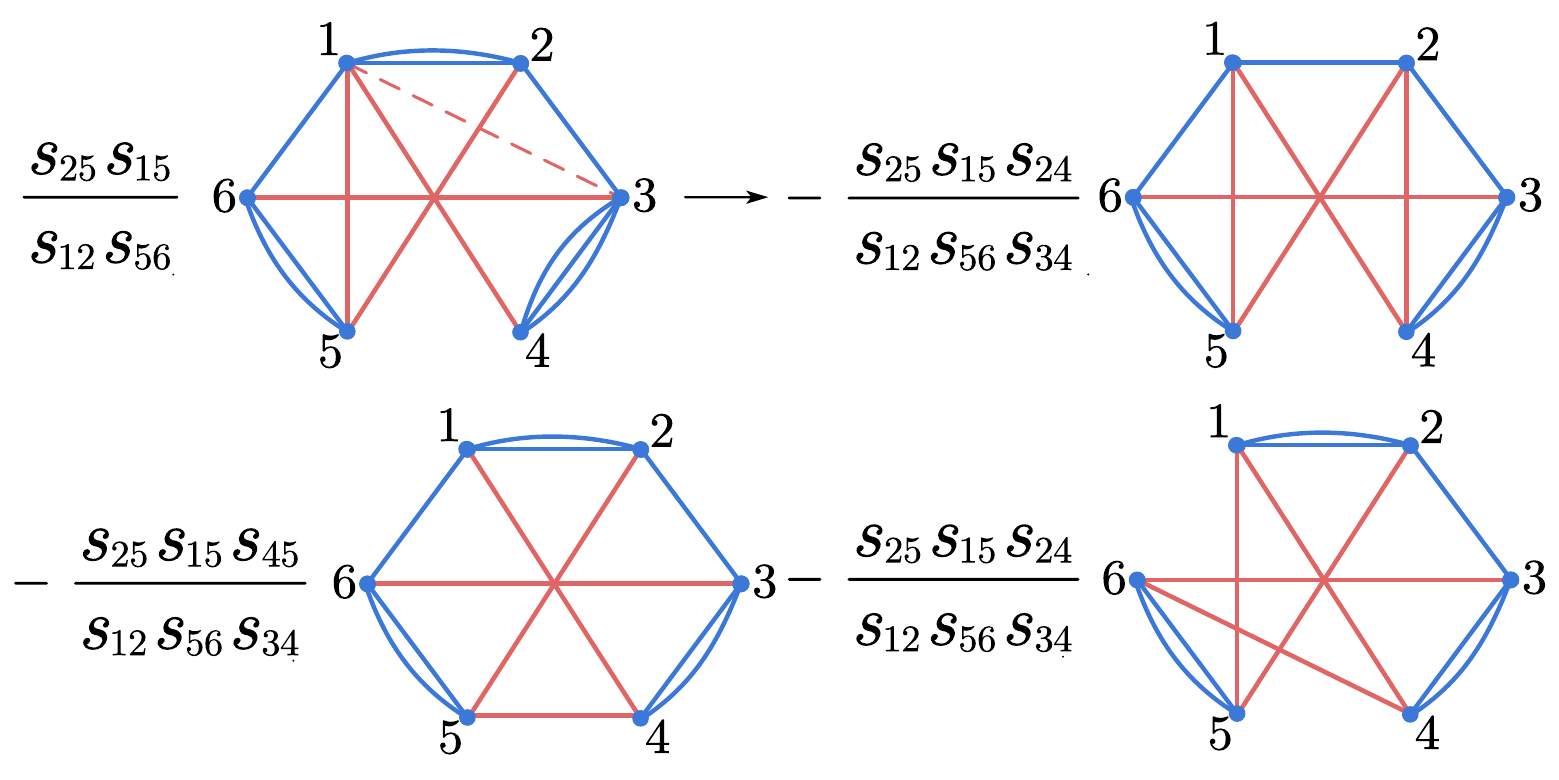}
    \caption{Decomposition of the third intermediate term. Note how the cross-ratio identity targets the $s_{34}$ pole while preserving the simple pole structure in other channels.}
    \label{fig:decomp3}
\end{figure}

\begin{figure}[htbp]
    \centering
    \includegraphics[width=1\linewidth]{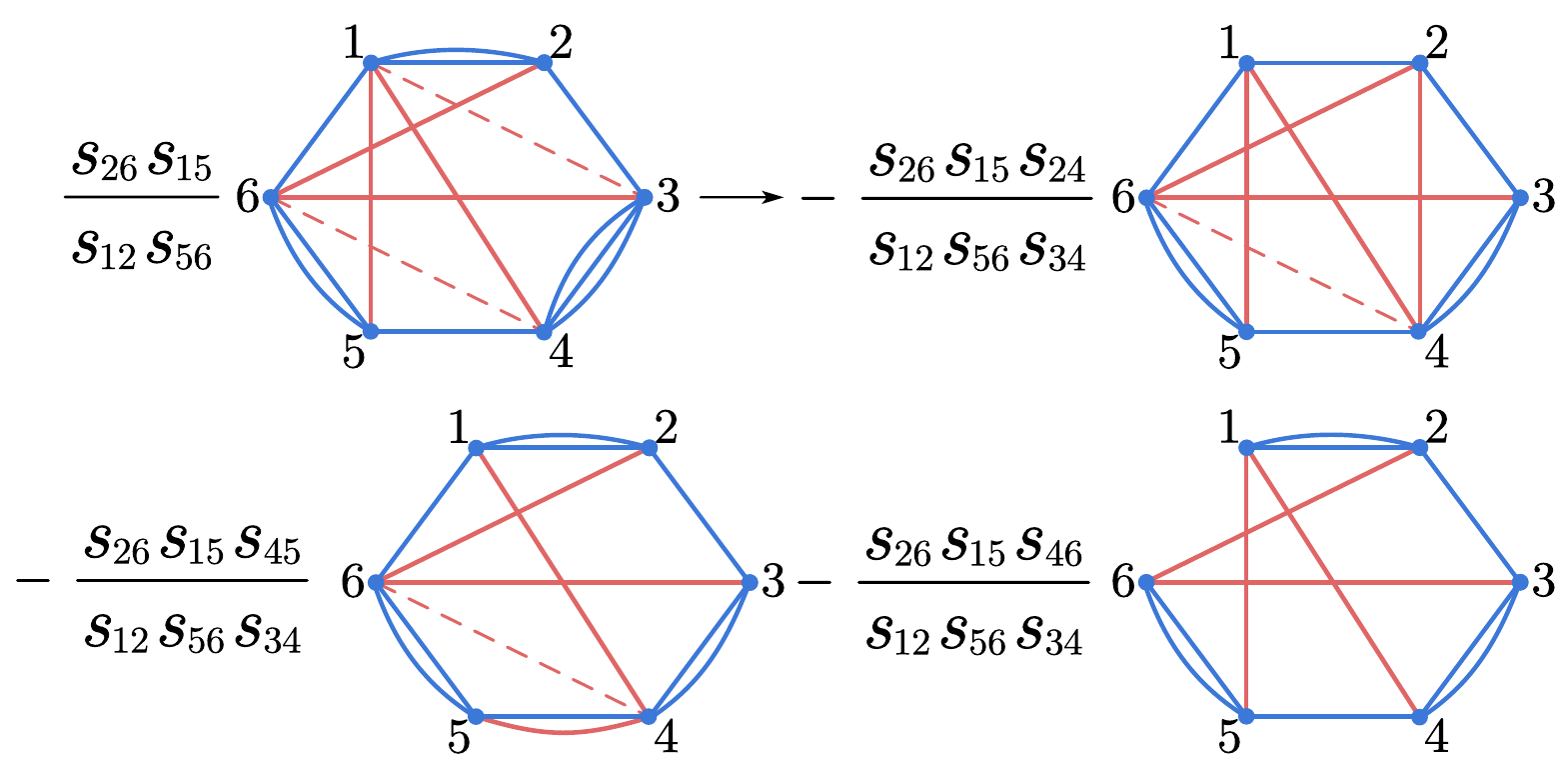}
    \caption{Decomposition of the fourth and final intermediate term. The complete decomposition process transforms one integrand with three double poles into 17 integrands with only simple poles.}
    \label{fig:decomp4}
\end{figure}

Figures \ref{fig:decomp1}--\ref{fig:decomp4} illustrate how each of the four intermediate terms is further decomposed into three terms with only simple poles.

Through this systematic process, the original CHY integrand $I_6$ with three double poles is decomposed into $5 + 4 \times 3 = 17$ terms, each containing only simple poles. The sum of these 17 terms reproduces the correct amplitude with all physical and spurious pole cancellations properly accounted for.

\clearpage
\subsection*{Computational Efficiency and Comparison}

Our enhanced decomposition method demonstrates clear computational advantages over traditional approaches. For this six-point example with three double poles ($s_{12}$, $s_{34}$, $s_{56}$):

\begin{center}
\begin{tabular}{|l|c|c|}
\hline
\textbf{Method} & \textbf{Total Terms} & \textbf{Decomposition Steps} \\
\hline
Systematic algorithm \cite{Cardona:2016gon} & 27 & 4 \\
\textbf{Our enhanced method} & \textbf{17} & \textbf{2} \\
\hline
\end{tabular}
\end{center}

This improvement becomes even more pronounced for higher-multiplicity amplitudes. A detailed eight-point example illustrating the scalability of our approach is presented in Appendix \ref{eg.8-Point-Higher-Order-Poles}, where the reduction from 265 to 153 terms illustrates the method's behavior for more complex cases. 

Through this three-round process, the eight-point CHY integrand $I_8$ with four double poles is systematically decomposed into $9 + 60 + 12 + 72 = 153$ terms, each containing only simple poles. The improvement over traditional methods is summarized in the following table:

\begin{center}
\begin{tabular}{|l|c|c|c|c|c|}
\hline
\textbf{Method} & \textbf{Round 1} & \textbf{Round 2} & \textbf{Round 3} & \textbf{Round 4} & \textbf{Total} \\
\hline
Systematic algorithm \cite{Cardona:2016gon} & 5 [5] & 25 [20] & 105 [20] & 265 [0] & 265 \\
\hline
\textbf{Our method} & 25 [16] & 89 [8] & 153 [0] & --- & \textbf{153} \\
\hline
\end{tabular}
\end{center}
\vspace{0.2cm}
\noindent Numbers in brackets [·] indicate terms with remaining higher-order poles at each stage.

The systematic nature of our decomposition, guided by the graph-theoretic structure of 0-regular graphs, suggests that this computational advantage can persist as we move toward even higher multiplicities relevant for precision calculations in quantum field theory.
Our enhanced method achieves this reduction by:
\begin{enumerate}
\item Systematically decomposing cross-ratio factors using the 0-regular graph structure
\item Applying identities sequentially to minimize intermediate term proliferation
\item Leveraging the hierarchical nature of pole constraints to optimize the decomposition order
\end{enumerate}

This eight-point example suggests that our graph-theoretic approach to higher-order pole decomposition can maintain computational advantages as complexity increases, making it a useful tool for the amplitude calculations considered here.

\clearpage

\section{Conclusion}
\label{sec:conclusion}

We presented an algorithmic approach to the inverse CHY construction problem: given a set of pole constraints, we construct rational CHY integrands whose colored-graph data realize them \cite{Cachazo:2013gna,Cachazo:2013iea,Cachazo:2013hca}.
The key object is the generalized pole degree $K(A)$, which can be read as a signed internal-edge count in the standard graph representation of CHY integrands \cite{Baadsgaard:2015voa,Baadsgaard:2015aha,Cardona:2016gon}.
In addition, we formulated the construction in a generalized ``regularity'' (net-degree) grading of colored graphs, where multiplication is graded and the degree-zero sector (0-regular graphs) forms a commutative group acting on a fixed regularity class (Sec.~\ref{sec:generalized-chy}).
 
Our workflow has three components.

\textbf{(i) Linearization and MILP.}
Using additivity under integrand multiplication and the elementary face recursion on the subset lattice, we express higher-channel $K(A)$ as linear functions of the two-particle variables $\{K(ij)\}$.
The inverse step can then be posed as a mixed-integer \emph{linear} feasibility problem for $\{K(ij)\}$ \cite{Schrijver:1998,Wolsey:1998}.

\textbf{(ii) Deterministic simplicial message passing.}
The same recursion induces a fixed dependency graph (the Hasse diagram of the subset lattice).
On this graph we implement deterministic constraint propagation in the factorial-rescaled variables $\widetilde K(A)=(|A|-2)!\,K(A)$, so each local update is integral.
In CoNN language, the forward pass evaluates the recursion upward, while a backward sweep propagates constraint residuals along the same fixed graph; this is a purely discrete procedure with no training and no learnable parameters \cite{ebli2020,Bodnar:2021,morris2019}.

\textbf{(iii) Applications and 0-regular factors.}
We demonstrated pick-pole selection and higher-order pole reduction on six- and eight-point examples.
In the graded regular-graph viewpoint, the reduction step can be implemented by multiplying by suitable 0-regular factors (degree-zero) and then decomposing them into products of standard four-point cross ratios.
This leads to a sequential use of cross-ratio identities which, in the examples studied, can reduce intermediate term proliferation compared to symmetric applications \cite{Feng:2016nsv,Feng:2019winn,Cardona:2016gon,Huang:2016FeynmanRules,Zhou:2017CrossRatio}.

Because all steps operate directly on integer edge data and preserve the recursion constraints, intermediate quantities remain exact in the integer-rescaled recursion variables.
This avoids workflows based on finite-field sampling with reconstruction or high-precision numerics with subsequent exactification \cite{Peraro:2016FiniteFields,Peraro:2019FiniteFlow,Klappert:2019FireFly,Ferguson:1999PSLQ,Laporta:2000DifferenceEquations,DeLaurentis:2019seampy}.
Equivalently, the overall computation can be viewed as a constraint-preserving algorithm on a fixed dependency graph.

Natural extensions include improving selection/branching heuristics for the MILP stage, studying scaling to higher multiplicity, and exploring broader CHY integrand ans\"atze where the same subset-lattice constraints apply.
Finally, the graded multiplication viewpoint is compatible with moduli-space formulations of CHY-type integrals and their combinatorial/topological organization \cite{Mizera:2019,Cachazo:CompatibleCycles:2019}.

\section*{Acknowledgements}
We acknowledge startup support from Ningxia University. The manuscript benefited from automated language-editing tools.

\clearpage
\appendix
\section{Proofs of Recursion Relations of $K(A)$}

\subsection*{Proof of the elementary face recursion}
\label{app:proof-face}
Fix $A$ with $|A|\ge 3$. Write $K(A)$ as a signed count of internal edges,
\(K(A)=\sum_{e\subset A}\sigma_e\), where $\sigma_e=+1$ for a solid edge and $\sigma_e=-1$ for a dashed edge.
For any $(|A|-1)$-subset $B\subset A$, the same edge $e=(i,j)$ contributes to $K(B)$ iff the removed
vertex is not $i$ or $j$. There are exactly $|A|-2$ such choices. Hence each edge is counted $|A|-2$
times in $\sum_{|B|=|A|-1}K(B)$, so
\[
\sum_{\substack{B\subset A\\|B|=|A|-1}}K(B)=(|A|-2)K(A),
\]
which implies \eqref{eq:Krec}.

\subsection*{Proof of the codimension-$d$ recursion}
\label{app:proof-general}
Fix $d$ with $1\le d\le |A|-2$. For an internal edge $e=(i,j)\subset A$, the number of subsets
$B\subset A$ of size $|A|-d$ that contain both $i$ and $j$ equals
\(\binom{|A|-2}{|A|-d-2}=\binom{|A|-2}{d}\).
Therefore each internal edge contributes the same multiplicity $\binom{|A|-2}{d}$ to
$\sum_{|B|=|A|-d}K(B)$, and summing with signs gives
\[
\sum_{\substack{B\subset A\\|B|=|A|-d}}K(B)=\binom{|A|-2}{d}K(A),
\]
which implies \eqref{eq:general-rec}.

\section{Solving Bootstrap Equations via Mixed-Integer Linear Programming}
\label{app:ilp}

The inverse CHY problem—constructing integrands from desired amplitude structures—presents unique mathematical challenges. Unlike the forward problem of evaluating known integrands, the inverse problem requires navigating a vast discrete space of possible graph configurations while satisfying intricate algebraic constraints. The combinatorial nature of this problem, combined with the requirement for exact integer solutions, places it firmly in the realm of NP-hard optimization problems \cite{Papadimitriou:1998, Korte:2012}.

This appendix describes how modern computational optimization techniques, specifically Integer Linear Programming (ILP), can be leveraged to solve the bootstrap equations systematically. The approach transforms the abstract mathematical problem into a concrete computational framework amenable to well-established algorithmic solutions.

\textbf{Mathematical Formulation.} The bootstrap construction requires finding integer values $\{K(s_{ij})\}$ that satisfy a system of linear constraints derived from physical requirements:
\begin{itemize}
\item \textbf{Equality constraints}: For desired poles, $K(s_A) = 2(|A| - 1)$
\item \textbf{Inequality constraints}: For unwanted poles, $K(s_A) < 2(|A| - 1)$  
\item \textbf{Regularity constraints}: For 4-regular graphs, $K_i = \sum_{j \neq i} K(s_{ij}) = 4$
\end{itemize}

This naturally maps to an Integer Linear Programming problem:
\begin{itemize}
\item \textbf{Variables}: $\vec{x} = (K(s_{12}), K(s_{13}), \ldots, K(s_{n-1,n}))^T \in \mathbb{Z}^{\binom{n}{2}}$
\item \textbf{Constraints}: $\mathbf{A}_{eq}\vec{x} = \vec{b}_{eq}$, $\mathbf{A}_{ineq}\vec{x} < \vec{b}_{ineq}$
\item \textbf{Feasibility}: Find any $\vec{x}$ satisfying all constraints (no objective function)
\end{itemize}

The key insight enabling this formulation is that all higher-order pole degrees $K(s_A)$ with $|A| > 2$ can be expressed as linear combinations of the two-particle variables $\{K(s_{ij})\}$ via the recursion relations (2.5). This linearization is crucial—without it, the constraints would involve nonlinear relationships that would make the problem much harder to solve.

\textbf{Computational Complexity.} The difficulty of this problem is substantial. For an $n$-particle amplitude:
\begin{itemize}
\item The number of variables scales as $\binom{n}{2} = O(n^2)$
\item The number of potential constraints scales as $2^n - 1$ (one for each subset)
\item The search space for integer solutions grows exponentially
\end{itemize}

Even for moderate values of $n$, brute-force enumeration becomes impossible. For $n=10$, there are already 45 variables and over 1000 potential constraints. The integer requirement transforms what would be a polynomial-time solvable linear program into an NP-hard problem \cite{Schrijver:1998, Wolsey:1998}.

\textbf{Algorithmic Approaches.} Modern ILP solvers employ sophisticated techniques to handle this complexity. The standard framework is Branch-and-Bound, enhanced with cutting plane methods \cite{Mitchell:2002, Bixby:2012}:

\begin{algorithm}[H]
\caption{Enhanced Branch-and-Cut Algorithm}
\begin{algorithmic}[1]
\STATE Solve LP relaxation using simplex or interior point methods
\IF{solution is integer}
    \RETURN solution
\ELSE
    \STATE Apply cutting plane generation (Gomory cuts, lift-and-project)
    \STATE Select branching variable using strong branching or pseudocost
    \STATE Create child nodes with disjoint constraints
    \STATE Apply preprocessing: bound tightening, constraint propagation
    \STATE Recursively solve subproblems with pruning based on bounds
    \STATE Use heuristics (feasibility pump, RINS) to find integer solutions
\ENDIF
\end{algorithmic}
\end{algorithm}

Recent advances in ILP solvers have improved performance through parallelization, machine learning-guided branching strategies, and symmetry detection \cite{Bengio:2021, Gasse:2019}. 

\textbf{Implementation Details.} For practical implementation, we utilize:

In \textbf{Mathematica} \cite{Wolfram:2024}:
\begin{itemize}
\item \texttt{k[i,j]}: Integer variable representation
\item \texttt{KValue[set]}: Recursive computation using memoization
\item \texttt{FindInstance}: Leverages internal ILP solver with symmetry detection
\end{itemize}

In \textbf{Python} \cite{Virtanen:2020}:
\begin{itemize}
\item \texttt{scipy.optimize.milp}: Interfaces with HiGHS solver \cite{Huangfu:2018}
\item \texttt{cvxpy}: Provides modeling language with automatic reformulation \cite{Diamond:2016}
\item \texttt{PuLP}: Connects to commercial solvers (Gurobi, CPLEX) \cite{Mitchell:2011}
\end{itemize}

\textbf{The 8-Particle Example.} We demonstrate the method on an 8-particle amplitude requiring poles at $s_{15}, s_{16}, s_{23}, s_{68}, s_{78}, s_{156}, s_{234}, s_{378}, s_{678}, s_{1256}$, and $s_{3478}$. This example, while manageable computationally, exhibits the full complexity of the problem:
\begin{itemize}
\item 28 integer variables $\{K(s_{ij})\}_{1 \leq i < j \leq 8}$
     includes 5 equalities:
\begin{align*}
K_{1,5}=2,K_{1,6}=2,K_{2,3}=2,K_{6,8}=2,K_{7,8}=2,
\end{align*}
     and 23 inequalities:
\begin{align*}
&K_{1,2}<2,K_{1,3}<2,K_{1,4}<2,K_{1,7}<2,K_{1,8}<2,K_{2,4}<2,K_{2,5}<2,K_{2,6}<2,\\
&K_{2,7}<2,K_{2,8}<2,K_{3,4}<2,K_{3,5}<2,K_{3,6}<2,K_{3,7}<2,K_{3,8}<2,K_{4,5}<2,\\
&K_{4,6}<2,K_{4,7}<2,K_{4,8}<2,K_{5,6}<2,K_{5,7}<2,K_{5,8}<2,K_{6,7}<2.
\end{align*}
\item 162 total constraints (20 equalities, 142 inequalities).
      In addition to the above 28 integer variables, it also includes 15 equalities:
\begin{align*}
&K_{1,2}+K_{1,6}+K_{2,6}=4,K_{1,5}+K_{1,6}+K_{5,6}=4,K_{2,3}+K_{2,4}+K_{3,4}=4,\\
&K_{3,7}+K_{3,8}+K_{7,8}=4,K_{6,7}+K_{6,8}+K_{7,8}=4,\\
&K_{1,2}+K_{1,5}+K_{1,6}+K_{2,5}+K_{2,6}+K_{5,6}=6,\\
&K_{3,4}+K_{3,7}+K_{3,8}+K_{4,7}+K_{4,8}+K_{7,8}=6,\\
&K_1=K_{1,2}+K_{1,3}+K_{1,4}+K_{1,5}+K_{1,6}+K_{1,7}+K_{1,8}=4,\\
&K_2=K_{1,2}+K_{2,3}+K_{2,4}+K_{2,5}+K_{2,6}+K_{2,7}+K_{2,8}=4,\\
&K_3=K_{1,3}+K_{2,3}+K_{3,4}+K_{3,5}+K_{3,6}+K_{3,7}+K_{3,8}=4,\\
&K_4=K_{1,4}+K_{2,4}+K_{3,4}+K_{4,5}+K_{4,6}+K_{4,7}+K_{4,8}=4,\\
&K_5=K_{1,5}+K_{2,5}+K_{3,5}+K_{4,5}+K_{5,6}+K_{5,7}+K_{5,8}=4,\\
&K_6=K_{1,6}+K_{2,6}+K_{3,6}+K_{4,6}+K_{5,6}+K_{6,7}+K_{6,8}=4,\\
&K_7=K_{1,7}+K_{2,7}+K_{3,7}+K_{4,7}+K_{5,7}+K_{6,7}+K_{7,8}=4,\\
&K_8=K_{1,8}+K_{2,8}+K_{3,8}+K_{4,8}+K_{5,8}+K_{6,8}+K_{7,8}=4.
\quad
\end{align*}
     and 51 inequalities, of three poles:
\begin{align*}
&K_{1,2}+K_{1,3}+K_{2,3}<4,K_{1,2}+K_{1,4}+K_{2,4}<4,K_{1,2}+K_{1,5}+K_{2,5}<4,\\
&K_{1,2}+K_{1,7}+K_{2,7}<4,K_{1,2}+K_{1,8}+K_{2,8}<4,K_{1,3}+K_{1,4}+K_{3,4}<4,\\
&K_{1,3}+K_{1,5}+K_{3,5}<4,K_{1,3}+K_{1,6}+K_{3,6}<4,K_{1,3}+K_{1,7}+K_{3,7}<4,\\
&K_{1,3}+K_{1,8}+K_{3,8}<4,K_{1,4}+K_{1,5}+K_{4,5}<4,K_{1,4}+K_{1,6}+K_{4,6}<4,\\
&K_{1,4}+K_{1,7}+K_{4,7}<4,K_{1,4}+K_{1,8}+K_{4,8}<4,K_{1,5}+K_{1,7}+K_{5,7}<4,\\
&K_{1,5}+K_{1,8}+K_{5,8}<4,K_{1,6}+K_{1,7}+K_{6,7}<4,K_{1,6}+K_{1,8}+K_{6,8}<4,\\
&K_{1,7}+K_{1,8}+K_{7,8}<4,K_{2,3}+K_{2,5}+K_{3,5}<4,K_{2,3}+K_{2,6}+K_{3,6}<4,\\
&K_{2,3}+K_{2,7}+K_{3,7}<4,K_{2,3}+K_{2,8}+K_{3,8}<4,K_{2,4}+K_{2,5}+K_{4,5}<4,\\
&K_{2,4}+K_{2,6}+K_{4,6}<4,K_{2,4}+K_{2,7}+K_{4,7}<4,K_{2,4}+K_{2,8}+K_{4,8}<4,\\
&K_{2,5}+K_{2,6}+K_{5,6}<4,K_{2,5}+K_{2,7}+K_{5,7}<4,K_{2,5}+K_{2,8}+K_{5,8}<4,\\
&K_{2,6}+K_{2,7}+K_{6,7}<4,K_{2,6}+K_{2,8}+K_{6,8}<4,K_{2,7}+K_{2,8}+K_{7,8}<4,\\
&K_{3,4}+K_{3,5}+K_{4,5}<4,K_{3,4}+K_{3,6}+K_{4,6}<4,K_{3,4}+K_{3,7}+K_{4,7}<4,\\
&K_{3,4}+K_{3,8}+K_{4,8}<4,K_{3,5}+K_{3,6}+K_{5,6}<4,K_{3,5}+K_{3,7}+K_{5,7}<4,\\
&K_{3,5}+K_{3,8}+K_{5,8}<4,K_{3,6}+K_{3,7}+K_{6,7}<4,K_{3,6}+K_{3,8}+K_{6,8}<4,\\
&K_{4,5}+K_{4,6}+K_{5,6}<4,K_{4,5}+K_{4,7}+K_{5,7}<4,K_{4,5}+K_{4,8}+K_{5,8}<4,\\
&K_{4,6}+K_{4,7}+K_{6,7}<4,K_{4,6}+K_{4,8}+K_{6,8}<4,K_{4,7}+K_{4,8}+K_{7,8}<4,\\
&K_{5,6}+K_{5,7}+K_{6,7}<4,K_{5,6}+K_{5,8}+K_{6,8}<4,K_{5,7}+K_{5,8}+K_{7,8}<4,
\quad
\end{align*}

     and 68 inequalities, of four poles:
\begin{align*}
&K_{1,2}+K_{1,3}+K_{1,4}+K_{2,3}+K_{2,4}+K_{3,4}<6,K_{1,2}+K_{1,3}+K_{1,5}+K_{2,3}+K_{2,5}+K_{3,5}<6,\\
&K_{1,2}+K_{1,3}+K_{1,6}+K_{2,3}+K_{2,6}+K_{3,6}<6,K_{1,2}+K_{1,3}+K_{1,7}+K_{2,3}+K_{2,7}+K_{3,7}<6,\\
&K_{1,2}+K_{1,3}+K_{1,8}+K_{2,3}+K_{2,8}+K_{3,8}<6,K_{1,2}+K_{1,4}+K_{1,5}+K_{2,4}+K_{2,5}+K_{4,5}<6,\\
&K_{1,2}+K_{1,4}+K_{1,6}+K_{2,4}+K_{2,6}+K_{4,6}<6,K_{1,2}+K_{1,4}+K_{1,7}+K_{2,4}+K_{2,7}+K_{4,7}<6,\\
&K_{1,2}+K_{1,4}+K_{1,8}+K_{2,4}+K_{2,8}+K_{4,8}<6,K_{1,2}+K_{1,5}+K_{1,7}+K_{2,5}+K_{2,7}+K_{5,7}<6,\\
&K_{1,2}+K_{1,5}+K_{1,8}+K_{2,5}+K_{2,8}+K_{5,8}<6,K_{1,2}+K_{1,6}+K_{1,7}+K_{2,6}+K_{2,7}+K_{6,7}<6,\\
&K_{1,2}+K_{1,6}+K_{1,8}+K_{2,6}+K_{2,8}+K_{6,8}<6,K_{1,2}+K_{1,7}+K_{1,8}+K_{2,7}+K_{2,8}+K_{7,8}<6,\\
&K_{1,3}+K_{1,4}+K_{1,5}+K_{3,4}+K_{3,5}+K_{4,5}<6,K_{1,3}+K_{1,4}+K_{1,6}+K_{3,4}+K_{3,6}+K_{4,6}<6,\\
&K_{1,3}+K_{1,4}+K_{1,7}+K_{3,4}+K_{3,7}+K_{4,7}<6,K_{1,3}+K_{1,4}+K_{1,8}+K_{3,4}+K_{3,8}+K_{4,8}<6,\\
&K_{1,3}+K_{1,5}+K_{1,6}+K_{3,5}+K_{3,6}+K_{5,6}<6,K_{1,3}+K_{1,5}+K_{1,7}+K_{3,5}+K_{3,7}+K_{5,7}<6,\\
&K_{1,3}+K_{1,5}+K_{1,8}+K_{3,5}+K_{3,8}+K_{5,8}<6,K_{1,3}+K_{1,6}+K_{1,7}+K_{3,6}+K_{3,7}+K_{6,7}<6,\\
&K_{1,3}+K_{1,6}+K_{1,8}+K_{3,6}+K_{3,8}+K_{6,8}<6,K_{1,3}+K_{1,7}+K_{1,8}+K_{3,7}+K_{3,8}+K_{7,8}<6,\\
&K_{1,4}+K_{1,5}+K_{1,6}+K_{4,5}+K_{4,6}+K_{5,6}<6,K_{1,4}+K_{1,5}+K_{1,7}+K_{4,5}+K_{4,7}+K_{5,7}<6,\\
&K_{1,4}+K_{1,5}+K_{1,8}+K_{4,5}+K_{4,8}+K_{5,8}<6,K_{1,4}+K_{1,6}+K_{1,7}+K_{4,6}+K_{4,7}+K_{6,7}<6,\\
&K_{1,4}+K_{1,6}+K_{1,8}+K_{4,6}+K_{4,8}+K_{6,8}<6,K_{1,4}+K_{1,7}+K_{1,8}+K_{4,7}+K_{4,8}+K_{7,8}<6,\\
&K_{1,5}+K_{1,6}+K_{1,7}+K_{5,6}+K_{5,7}+K_{6,7}<6,K_{1,5}+K_{1,6}+K_{1,8}+K_{5,6}+K_{5,8}+K_{6,8}<6,\\
&K_{1,5}+K_{1,7}+K_{1,8}+K_{5,7}+K_{5,8}+K_{7,8}<6,K_{1,6}+K_{1,7}+K_{1,8}+K_{6,7}+K_{6,8}+K_{7,8}<6,\\
&K_{2,3}+K_{2,4}+K_{2,5}+K_{3,4}+K_{3,5}+K_{4,5}<6,K_{2,3}+K_{2,4}+K_{2,6}+K_{3,4}+K_{3,6}+K_{4,6}<6,\\
&K_{2,3}+K_{2,4}+K_{2,7}+K_{3,4}+K_{3,7}+K_{4,7}<6,K_{2,3}+K_{2,4}+K_{2,8}+K_{3,4}+K_{3,8}+K_{4,8}<6,\\
&K_{2,3}+K_{2,5}+K_{2,6}+K_{3,5}+K_{3,6}+K_{5,6}<6,K_{2,3}+K_{2,5}+K_{2,7}+K_{3,5}+K_{3,7}+K_{5,7}<6,\\
&K_{2,3}+K_{2,5}+K_{2,8}+K_{3,5}+K_{3,8}+K_{5,8}<6,K_{2,3}+K_{2,6}+K_{2,7}+K_{3,6}+K_{3,7}+K_{6,7}<6,\\
&K_{2,3}+K_{2,6}+K_{2,8}+K_{3,6}+K_{3,8}+K_{6,8}<6,K_{2,3}+K_{2,7}+K_{2,8}+K_{3,7}+K_{3,8}+K_{7,8}<6,\\
&K_{2,4}+K_{2,5}+K_{2,6}+K_{4,5}+K_{4,6}+K_{5,6}<6,K_{2,4}+K_{2,5}+K_{2,7}+K_{4,5}+K_{4,7}+K_{5,7}<6,\\
&K_{2,4}+K_{2,5}+K_{2,8}+K_{4,5}+K_{4,8}+K_{5,8}<6,K_{2,4}+K_{2,6}+K_{2,7}+K_{4,6}+K_{4,7}+K_{6,7}<6,\\
&K_{2,4}+K_{2,6}+K_{2,8}+K_{4,6}+K_{4,8}+K_{6,8}<6,K_{2,4}+K_{2,7}+K_{2,8}+K_{4,7}+K_{4,8}+K_{7,8}<6,\\
&K_{2,5}+K_{2,6}+K_{2,7}+K_{5,6}+K_{5,7}+K_{6,7}<6,K_{2,5}+K_{2,6}+K_{2,8}+K_{5,6}+K_{5,8}+K_{6,8}<6,\\
&K_{2,5}+K_{2,7}+K_{2,8}+K_{5,7}+K_{5,8}+K_{7,8}<6,K_{2,6}+K_{2,7}+K_{2,8}+K_{6,7}+K_{6,8}+K_{7,8}<6,\\
&K_{3,4}+K_{3,5}+K_{3,6}+K_{4,5}+K_{4,6}+K_{5,6}<6,K_{3,4}+K_{3,5}+K_{3,7}+K_{4,5}+K_{4,7}+K_{5,7}<6,\\
&K_{3,4}+K_{3,5}+K_{3,8}+K_{4,5}+K_{4,8}+K_{5,8}<6,K_{3,4}+K_{3,6}+K_{3,7}+K_{4,6}+K_{4,7}+K_{6,7}<6,\\
&K_{3,4}+K_{3,6}+K_{3,8}+K_{4,6}+K_{4,8}+K_{6,8}<6,K_{3,5}+K_{3,6}+K_{3,7}+K_{5,6}+K_{5,7}+K_{6,7}<6,\\
&K_{3,5}+K_{3,6}+K_{3,8}+K_{5,6}+K_{5,8}+K_{6,8}<6,K_{3,5}+K_{3,7}+K_{3,8}+K_{5,7}+K_{5,8}+K_{7,8}<6,\\
&K_{3,6}+K_{3,7}+K_{3,8}+K_{6,7}+K_{6,8}+K_{7,8}<6,K_{4,5}+K_{4,6}+K_{4,7}+K_{5,6}+K_{5,7}+K_{6,7}<6,\\
&K_{4,5}+K_{4,6}+K_{4,8}+K_{5,6}+K_{5,8}+K_{6,8}<6,K_{4,5}+K_{4,7}+K_{4,8}+K_{5,7}+K_{5,8}+K_{7,8}<6,\\
&K_{4,6}+K_{4,7}+K_{4,8}+K_{6,7}+K_{6,8}+K_{7,8}<6,K_{5,6}+K_{5,7}+K_{5,8}+K_{6,7}+K_{6,8}+K_{7,8}<6,\quad
\end{align*}
\item Search space of approximately $10^{20}$ integer vectors
\end{itemize}

\textbf{The constraint system admits exactly four distinct integer solutions:}

\textbf{Solution 1:}
\begin{align*}
&K_{1,2} \to 1, K_{1,3} \to 0, K_{1,4} \to 0, K_{1,5} \to 2, K_{1,6} \to 2, K_{1,7} \to 0, K_{1,8} \to -1,\\
&K_{2,3} \to 2, K_{2,4} \to 1, K_{2,5} \to 0, K_{2,6} \to 1, K_{2,7} \to -1, K_{2,8} \to 0,\\
&K_{3,4} \to 1, K_{3,5} \to 0, K_{3,6} \to -1, K_{3,7} \to 1, K_{3,8} \to 1,\\
&K_{4,5} \to 1, K_{4,6} \to 0, K_{4,7} \to 1, K_{4,8} \to 0,\\
&K_{5,6} \to 0, K_{5,7} \to 1, K_{5,8} \to 0, K_{6,7} \to 0, K_{6,8} \to 2, K_{7,8} \to 2
\end{align*}

\textbf{Solution 2:}
\begin{align*}
&K_{1,2} \to 1, K_{1,3} \to 0, K_{1,4} \to 0, K_{1,5} \to 2, K_{1,6} \to 2, K_{1,7} \to 0, K_{1,8} \to -1,\\
&K_{2,3} \to 2, K_{2,4} \to 1, K_{2,5} \to 0, K_{2,6} \to 1, K_{2,7} \to 0, K_{2,8} \to -1,\\
&K_{3,4} \to 1, K_{3,5} \to 0, K_{3,6} \to -1, K_{3,7} \to 1, K_{3,8} \to 1,\\
&K_{4,5} \to 1, K_{4,6} \to 0, K_{4,7} \to 0, K_{4,8} \to 1,\\
&K_{5,6} \to 0, K_{5,7} \to 1, K_{5,8} \to 0, K_{6,7} \to 0, K_{6,8} \to 2, K_{7,8} \to 2
\end{align*}

\textbf{Solution 3:}
\begin{align*}
&K_{1,2} \to 1, K_{1,3} \to 0, K_{1,4} \to 0, K_{1,5} \to 2, K_{1,6} \to 2, K_{1,7} \to 1, K_{1,8} \to -2,\\
&K_{2,3} \to 2, K_{2,4} \to 1, K_{2,5} \to 0, K_{2,6} \to 1, K_{2,7} \to -1, K_{2,8} \to 0,\\
&K_{3,4} \to 1, K_{3,5} \to 0, K_{3,6} \to -1, K_{3,7} \to 1, K_{3,8} \to 1,\\
&K_{4,5} \to 1, K_{4,6} \to 0, K_{4,7} \to 1, K_{4,8} \to 0,\\
&K_{5,6} \to 0, K_{5,7} \to 0, K_{5,8} \to 1, K_{6,7} \to 0, K_{6,8} \to 2, K_{7,8} \to 2
\end{align*}

\textbf{Solution 4:}
\begin{align*}
&K_{1,2} \to 1, K_{1,3} \to 0, K_{1,4} \to 0, K_{1,5} \to 2, K_{1,6} \to 2, K_{1,7} \to 1, K_{1,8} \to -2,\\
&K_{2,3} \to 2, K_{2,4} \to 1, K_{2,5} \to 0, K_{2,6} \to 1, K_{2,7} \to 0, K_{2,8} \to -1,\\
&K_{3,4} \to 1, K_{3,5} \to 0, K_{3,6} \to -1, K_{3,7} \to 1, K_{3,8} \to 1,\\
&K_{4,5} \to 1, K_{4,6} \to 0, K_{4,7} \to 0, K_{4,8} \to 1,\\
&K_{5,6} \to 0, K_{5,7} \to 0, K_{5,8} \to 1, K_{6,7} \to 0, K_{6,8} \to 2, K_{7,8} \to 2
\end{align*}

All solutions satisfy the regularity condition $K_i = 4$ for $i = 1, \ldots, 8$. Each solution yields a distinct CHY integrand via:
\begin{equation}
I_n = \prod_{1 \leq i < j \leq n} (z_i - z_j)^{-K(s_{ij})}
\end{equation}

For instance, 
Solution 1 produces:
\begin{equation}
I_8^{[1]} = \frac{z_{18}z_{27}z_{36}}{z_{12}z_{15}^2z_{16}^2z_{23}^2z_{24}z_{26}z_{34}z_{37}z_{38}z_{45}z_{47}z_{57}z_{68}^2z_{78}^2}
\end{equation}
Solution 2 produces:
\begin{equation}
I_8^{[2]} = \frac{z_{18}z_{28}z_{36}}{z_{12}z_{15}^2z_{16}^2z_{23}^2z_{24}z_{26}z_{34}z_{37}z_{38}z_{45}z_{48}z_{57}z_{68}^2z_{78}^2}
\end{equation}
Solution 3 produces:
\begin{equation}
I_8^{[3]} = \frac{z_{18}^2z_{27}z_{36}}{z_{12}z_{15}^2z_{16}^2z_{17}z_{23}^2z_{24}z_{26}z_{34}z_{37}z_{38}z_{45}z_{47}z_{58}z_{68}^2z_{78}^2}
\end{equation}
Solution 4 produces:
\begin{equation}
I_8^{[4]} = \frac{z_{18}^2z_{28}z_{36}}{z_{12}z_{15}^2z_{16}^2z_{17}z_{23}^2z_{24}z_{26}z_{34}z_{37}z_{38}z_{45}z_{48}z_{58}z_{68}^2z_{78}^2}
\end{equation}

The existence of exactly four solutions is non-trivial. It demonstrates that the pole structure alone does not uniquely determine the CHY integrand, indicating a degeneracy in the map from amplitudes to worldsheet representations. Related structures appear in \cite{Mizera:2019, feng:2022, Cachazo:CompatibleCycles:2019}; we leave a systematic investigation of any connection to the space of consistent quantum field theories within the CHY framework for future work.

\section{Alternating Cycle Decomposition for 0-Regular Graphs}
\label{app:hamiltonian}

This appendix provides a constructive justification for the decomposition used in
Section~\ref{Decompose the 0-regular}: any 0-regular multiplicative factor can be written as a product
of standard four-point cross-ratios. We only use elementary graph-theoretic notions \cite{Bondy:1976,Diestel:2017}.

\paragraph{0-regular factors.}
We represent a multiplicative factor by a colored multigraph
$G=(V,E_s\cup E_d)$, where each solid edge $(i,j)\in E_s$ corresponds to a denominator factor
$z_{ij}^{-1}$ and each dashed edge $(i,j)\in E_d$ corresponds to a numerator factor $z_{ij}^{+1}$.
Up to an overall sign from $z_{ij}=-z_{ji}$, the associated monomial is
\begin{equation}
\mathcal{P}_G(z)\;:=\;\frac{\prod\limits_{(i,j)\in E_d} z_{ij}}{\prod\limits_{(i,j)\in E_s} z_{ij}}\,.
\end{equation}
The graph is \textbf{0-regular} if for every vertex $v\in V$,
\begin{equation}
\deg_s(v)=\deg_d(v),
\end{equation}
i.e.\ the net degree $\deg_{\rm net}(v)=\deg_s(v)-\deg_d(v)$ vanishes.

\begin{lemma}[Alternating cycle decomposition]
\label{lem:alt-cycle-decomp}
Let $G=(V,E_s\cup E_d)$ be a 0-regular multigraph. Then $E_s\cup E_d$ can be partitioned into
edge-disjoint closed alternating cycles (each cycle alternates between solid and dashed edges).
\end{lemma}

\begin{proof}
At each vertex $v$, choose an arbitrary bijection between the incident solid edges and the incident dashed edges.
This is possible because $\deg_s(v)=\deg_d(v)$. Use these bijections to define a local ``next-edge'' rule:
whenever a walk arrives at $v$ along a solid edge, it exits along the paired dashed edge, and vice versa.

Start from any unused edge and iterate this rule. Because the number of edges is finite and the update is deterministic,
the walk eventually repeats an edge and therefore closes into an alternating cycle. Removing the edges of this cycle
from $G$ preserves 0-regularity on the remaining multigraph (degrees of each color are reduced equally at each visited
vertex). Repeating the procedure exhausts all edges and yields a partition into edge-disjoint alternating cycles.
\end{proof}

Each alternating cycle $C$ contributes a factor $\mathcal{P}_C(z)$, and since the cycles are edge-disjoint we have
\begin{equation}
\mathcal{P}_G(z)=\prod_C \mathcal{P}_C(z).
\end{equation}

\begin{lemma}[Even-cycle telescoping into four-point cross-ratios]
\label{lem:cycle-to-4pt}
Let $C=(v_1,v_2,\ldots,v_{2m})$ be an alternating cycle with $2m\ge4$ distinct vertices and
$v_{2m+1}\equiv v_1$. Define its associated ``long'' factor by
\begin{equation}
\mathcal{P}_C(z)\;:=\;\frac{\prod\limits_{r=1}^{m} z_{v_{2r-1}\,v_{2r}}}{\prod\limits_{r=1}^{m} z_{v_{2r}\,v_{2r+1}}}\,.
\label{eq:long-cycle-factor}
\end{equation}
Then (up to an overall sign) $\mathcal{P}_C(z)$ factorizes into $(m-1)$ four-point cross-ratios as
\begin{equation}
\mathcal{P}_C(z)
=
\prod_{r=2}^{m}
\frac{
z_{v_{2r-3}\,v_{2r-2}}\; z_{v_{2r-1}\,v_{2m}}
}{
z_{v_{2r-2}\,v_{2r-1}}\; z_{v_{2r-3}\,v_{2m}}
}\,.
\label{eq:cycle-telescope}
\end{equation}
\end{lemma}

\begin{proof}
Multiply the right-hand side of \eqref{eq:cycle-telescope}. The factors involving $z_{(\cdot)\,v_{2m}}$
telescope: every intermediate $z_{v_{2r-1}\,v_{2m}}$ cancels between numerator and denominator, leaving only
$z_{v_{2m-1}\,v_{2m}}$ in the numerator and $z_{v_1\,v_{2m}}$ in the denominator. The remaining uncanceled factors
reproduce exactly the product of consecutive edge factors in \eqref{eq:long-cycle-factor}. This proves the identity.
\end{proof}

\begin{theorem}[0-regular decomposition into four-point cross-ratios]
\label{thm:0regular-decomposition}
For any 0-regular multigraph $G$, the associated factor $\mathcal{P}_G(z)$ can be written (up to an overall sign)
as a product of standard four-point cross-ratios.
\end{theorem}

\begin{proof}
By Lemma~\ref{lem:alt-cycle-decomp}, $\mathcal{P}_G$ factorizes into a product over alternating cycles.
Each cycle factor then decomposes into four-point cross-ratios by Lemma~\ref{lem:cycle-to-4pt}.
\end{proof}

\begin{algorithm}[H]
\caption{Decomposition of a 0-regular factor into four-point cross-ratios}
\label{alg:0regular-decomposition}
\begin{algorithmic}[1]
\REQUIRE 0-regular multigraph $G=(V,E_s\cup E_d)$
\ENSURE List $\mathcal{L}$ of four-point cross-ratios whose product equals $\mathcal{P}_G(z)$ (up to sign)
\STATE Choose, for each vertex $v$, a bijection pairing incident solid and dashed edges
\STATE Mark all edges unused; set $\mathcal{L}\leftarrow\emptyset$
\WHILE{there exists an unused edge}
    \STATE Extract an alternating cycle $C=(v_1,\ldots,v_{2m})$ by following the local pairings
    \FOR{$r=2$ \TO $m$}
        \STATE Append $\displaystyle
        \frac{
        z_{v_{2r-3}\,v_{2r-2}}\; z_{v_{2r-1}\,v_{2m}}
        }{
        z_{v_{2r-2}\,v_{2r-1}}\; z_{v_{2r-3}\,v_{2m}}
        }$ to $\mathcal{L}$
    \ENDFOR
    \STATE Mark the edges of $C$ used
\ENDWHILE
\RETURN $\mathcal{L}$
\end{algorithmic}
\end{algorithm}

\section{Discrete AD-Based Neural Network Examples on 8-Point CHY Integrands}
\label{AD-8-Point CHY Integrands}

To demonstrate the scalability and effectiveness of our discrete automatic differentiation framework, we present a comprehensive example using an 8-point CHY integrand. This example highlights the three-layer network structure and shows how our neural-inspired pick-pole algorithm isolates desired pole structures while maintaining constraint satisfaction.

Consider the 8-point CHY integrand:
\begin{equation}
I_8 = \frac{z_{18}^2z_{28}z_{36}}{z_{12}z_{15}^2z_{16}^2z_{17} z_{23}^2 z_{24}z_{26}z_{34}z_{37}z_{38}z_{45}z_{48}z_{58}z_{68}^2 z_{78}^2},
\end{equation}

which generates seven distinct Feynman diagrams:
\begin{equation}
\begin{split}
&\frac{1}{s_{16}s_{23}s_{78}s_{156}s_{234}} + \frac{1}{s_{15}s_{23}s_{68}s_{234}s_{678}} + \frac{1}{s_{15}s_{23}s_{78}s_{234}s_{678}} + \frac{1}{s_{16}s_{78}s_{126}s_{378}s_{1256}}\\
&+ \frac{1}{s_{15}s_{78}s_{156}s_{378}s_{1256}} + \frac{1}{s_{16}s_{78}s_{156}s_{378}s_{1256}} + \frac{1}{s_{15}s_{23}s_{78}s_{156}s_{234}}.
\end{split}
\end{equation}

The amplitude contains simple poles at $s_{15}$, $s_{16}$, $s_{23}$, $s_{68}$, $s_{78}$, $s_{156}$, $s_{234}$, $s_{378}$, $s_{678}$, and $s_{1256}$. Our objective is to isolate all contributions containing the pole $s_{15}$ by systematically removing all incompatible poles.

\begin{figure}[H]
    \centering
    \includegraphics[width=1\linewidth]{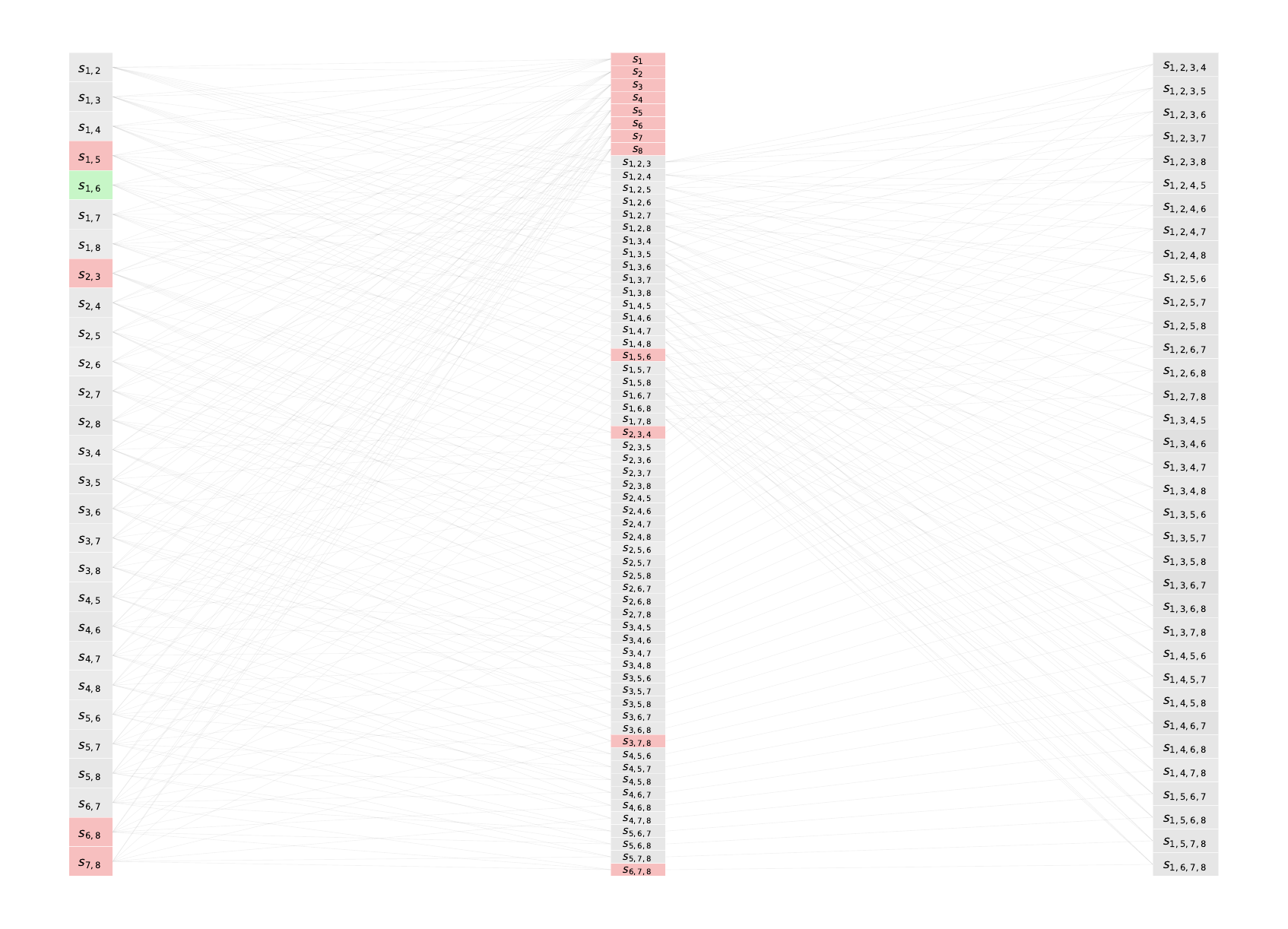}
    \caption{Three-layer hierarchical network structure for 8-point CHY integrand pole selection. The network comprises: (Layer 1) two-particle poles forming the base layer, (Layer 2) single-particle and three-particle poles in the intermediate layer, and (Layer 3) four-particle poles at the top. Gray edges encode the dependency relationships from the recursion relation $K(s_A) = \frac{1}{(|A|-2)}\sum_{B\subset A,|B|=|A|-1}K(s_B)$. Red nodes represent fixed poles that must be preserved during the algorithm, including the target pole $s_{15}$ and all compatible poles necessary for a consistent amplitude structure.}
    \label{eg-8-01}
\end{figure}

We construct a three-layer hierarchical network where the dependency relationships are governed by the recursion relation:
\begin{equation}
K(s_A)=\frac{1}{(|A|-2)}\sum_{B\subset A,|B|=|A|-1}K(s_B).
\end{equation}

As illustrated in Figure~\ref{eg-8-01}, this creates a complex network structure where two-particle and single-particle poles form the foundational layer, three-particle poles constitute the intermediate layer, and four-particle poles occupy the top layer. We fix all poles compatible with $s_{15}$ (marked in red) to ensure they remain unchanged throughout the discrete automatic differentiation process.

\subsection*{Removing Incompatible Pole $s_{16}$}

Our initial step involves removing the incompatible pole $s_{16}$ by decrementing its K-value:
\begin{equation}
K^*(s_{16}) = K(s_{16}) - 1 = 1.
\end{equation}

\begin{figure}[H]
    \centering
    \includegraphics[width=1\linewidth]{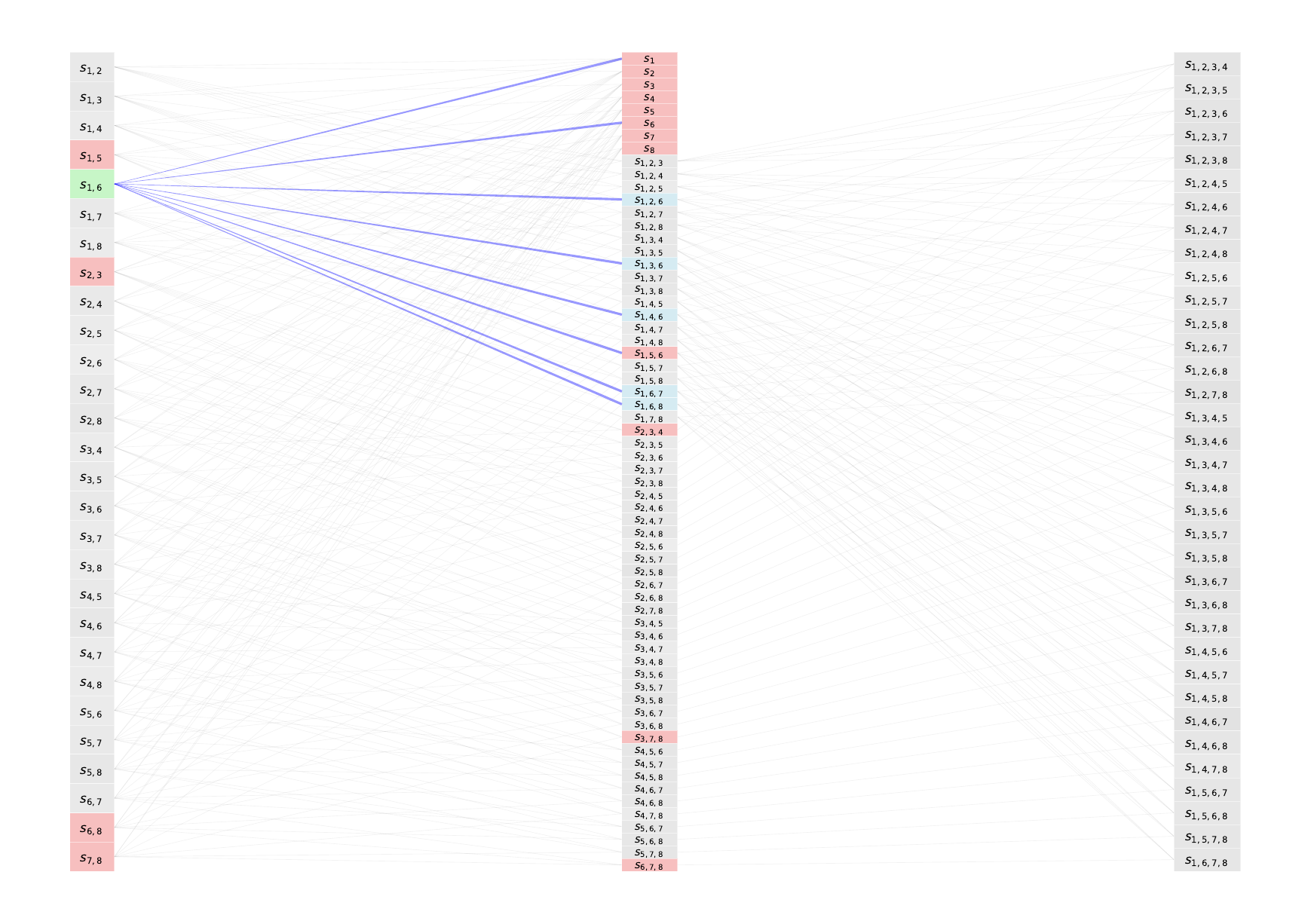}
    \caption{Initial upward propagation phase in the discrete automatic differentiation algorithm. The modification $\Delta K = -1$ at pole $s_{16}$ (highlighted in green) triggers upward propagation throughout the network hierarchy. Blue arrows trace the propagation pathways to all three-particle supersets and single-particle subsets containing pole $s_{16}$. The algorithm systematically decrements the K-values of $s_{16}$'s three-particle supersets: $s_{126}$, $s_{136}$, $s_{146}$, $s_{156}$, $s_{167}$, and $s_{168}$ by 1 unit each, following the recursion constraints. Critical fixed constraints are encountered at $s_{156}$, $s_1$, and $s_6$ (red nodes), which triggers the backward propagation mechanism to maintain network consistency.}
    \label{eg-8-02}
\end{figure}

The upward propagation phase, shown in Figure~\ref{eg-8-02}, demonstrates how the initial modification cascades through the network hierarchy. The algorithm encounters fixed constraints at nodes $s_{156}$, $s_1$, and $s_6$, which cannot be modified due to physical requirements. This triggers the backward propagation mechanism.

\begin{figure}[H]
    \centering
    \includegraphics[width=1\linewidth]{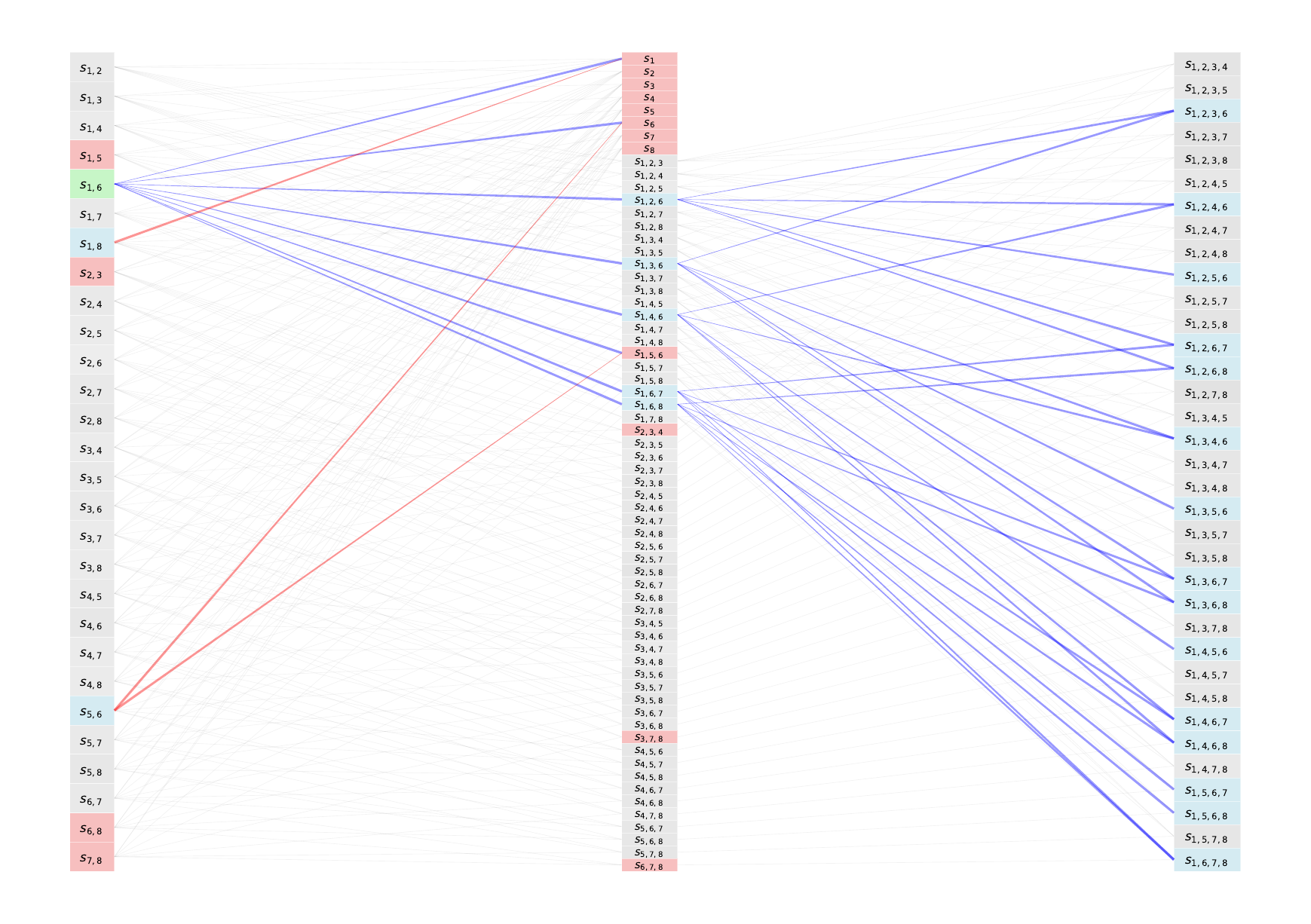}
    \caption{Backward propagation with simultaneous multi-layer updates. When upward propagation encounters fixed constraints, the discrete automatic differentiation algorithm activates backward propagation (red arrows) to redistribute changes while preserving all recursion relations. The algorithm strategically selects $s_{56}$ (intersection of dependent subsets from fixed point $s_{156}$ and superset of fixed node $s_6$) and $s_{18}$ (superset of fixed point $s_1$), incrementing their K-values by +1. Simultaneously, the algorithm propagates from the second to third layer (blue arrows), updating four-point supersets by -0.5 to maintain consistency across all hierarchical levels. These -0.5 updates are in the $K$ normalization; in $\widetilde K$ they correspond to integer updates.}
    \label{eg-8-03}
\end {figure}

The backward propagation phase, illustrated in Figure~\ref{eg-8-03}, shows the algorithm's sophisticated constraint handling. The system compensates for fixed constraints by selecting appropriate alternative poles and adjusting their values. Concurrently, changes propagate to the third layer, maintaining global consistency.

\begin{figure}[H]
    \centering
    \includegraphics[width=1\linewidth]{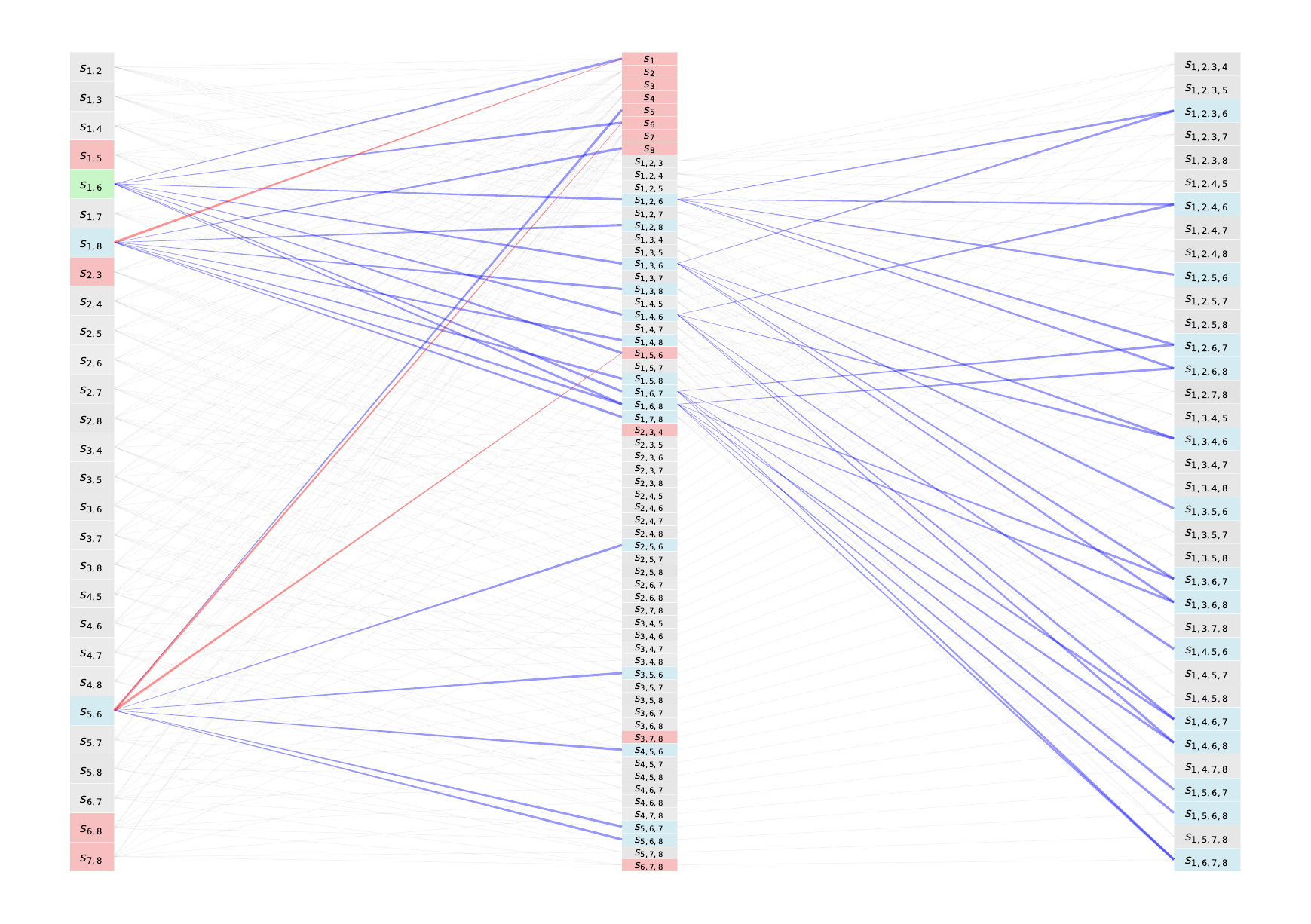}
    \caption{Second upward propagation cascade from compensatory modifications. The changes introduced at $s_{18}$ and $s_{56}$ during backward propagation now trigger their own upward propagation sequences (blue arrows). Each pole's three-particle supersets receive +1 increments: $s_{18}$ affects $s_{128}$, $s_{138}$, $s_{148}$, $s_{158}$, $s_{168}$, and $s_{178}$, while $s_{56}$ affects $s_{156}$, $s_{256}$, $s_{356}$, $s_{456}$, $s_{567}$, and $s_{568}$. The propagation encounters additional fixed points $s_5$ and $s_8$, necessitating another backward propagation cycle to resolve the emerging constraint conflicts.}
    \label{eg-8-04}
\end {figure}

As shown in Figure~\ref{eg-8-04}, the compensatory changes from the previous step now generate their own propagation effects. The algorithm encounters additional fixed points, demonstrating the iterative nature of the constraint resolution process.

\begin{figure}[H]
    \centering
    \includegraphics[width=1\linewidth]{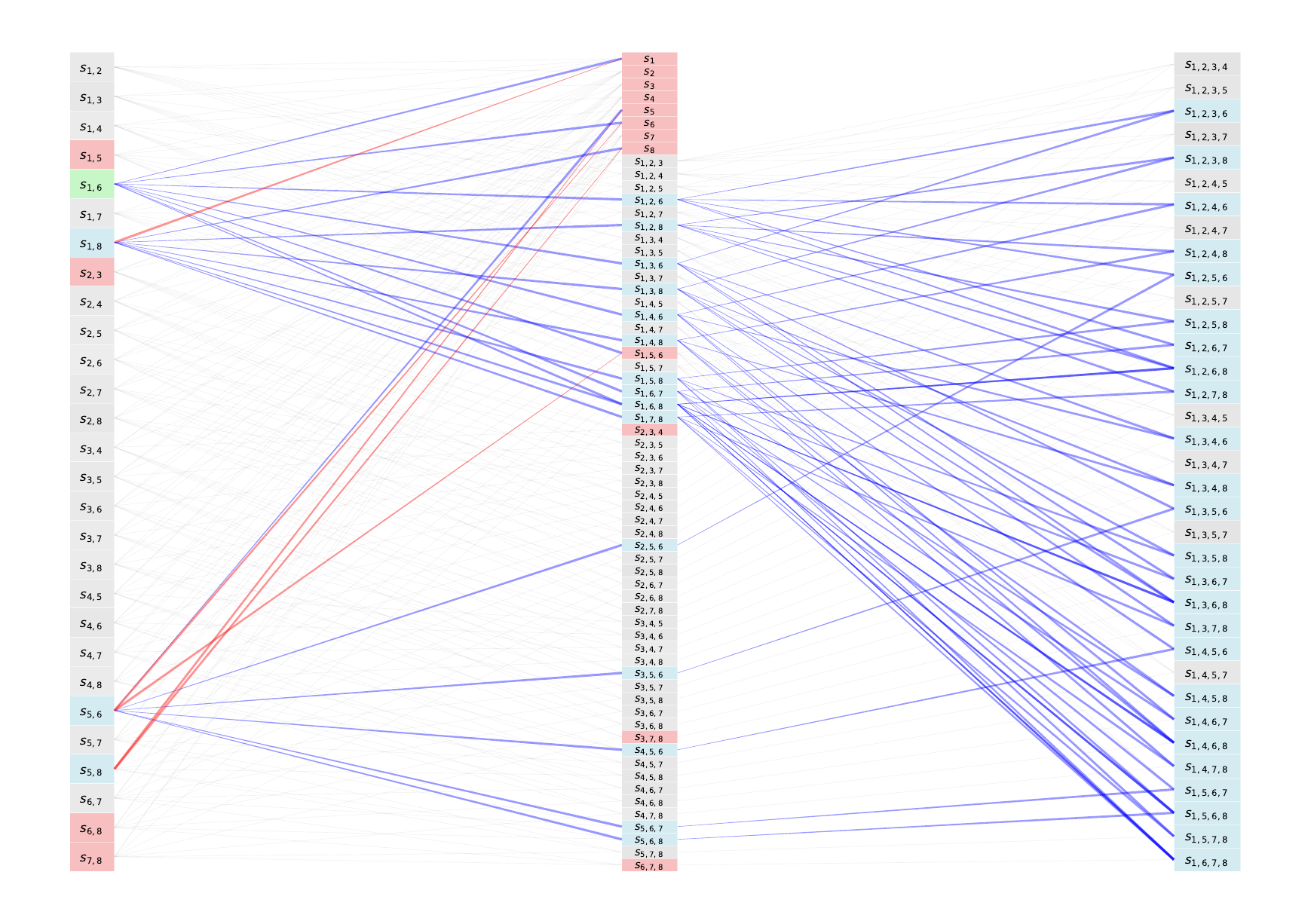}
    \caption{Second backward propagation cycle with inter-layer constraint resolution. The encounter with fixed points $s_5$ and $s_8$ activates another backward propagation phase (red arrows from second to initial layer). The algorithm identifies $s_{58}$ as the intersection of dependent supersets from fixed nodes $s_5$ and $s_8$, decrementing its K-value by -1 to compensate for the constrained propagation. Simultaneously, upward propagation continues to the third layer (blue arrows), where four-point supersets of the affected three-particle poles receive +0.5 increments, ensuring that all recursion relations remain exactly satisfied throughout the multi-layer network. These +0.5 updates are in the $K$ normalization; in $\widetilde K$ they correspond to integer updates.}
    \label{eg-8-05}
\end {figure}

The second backward propagation cycle, depicted in Figure~\ref{eg-8-05}, shows how the algorithm systematically resolves cascading constraints. The selection of $s_{58}$ as a compensatory adjustment illustrates a workable solution within the constraint space.

\begin{figure}[H]
    \centering
    \includegraphics[width=1\linewidth]{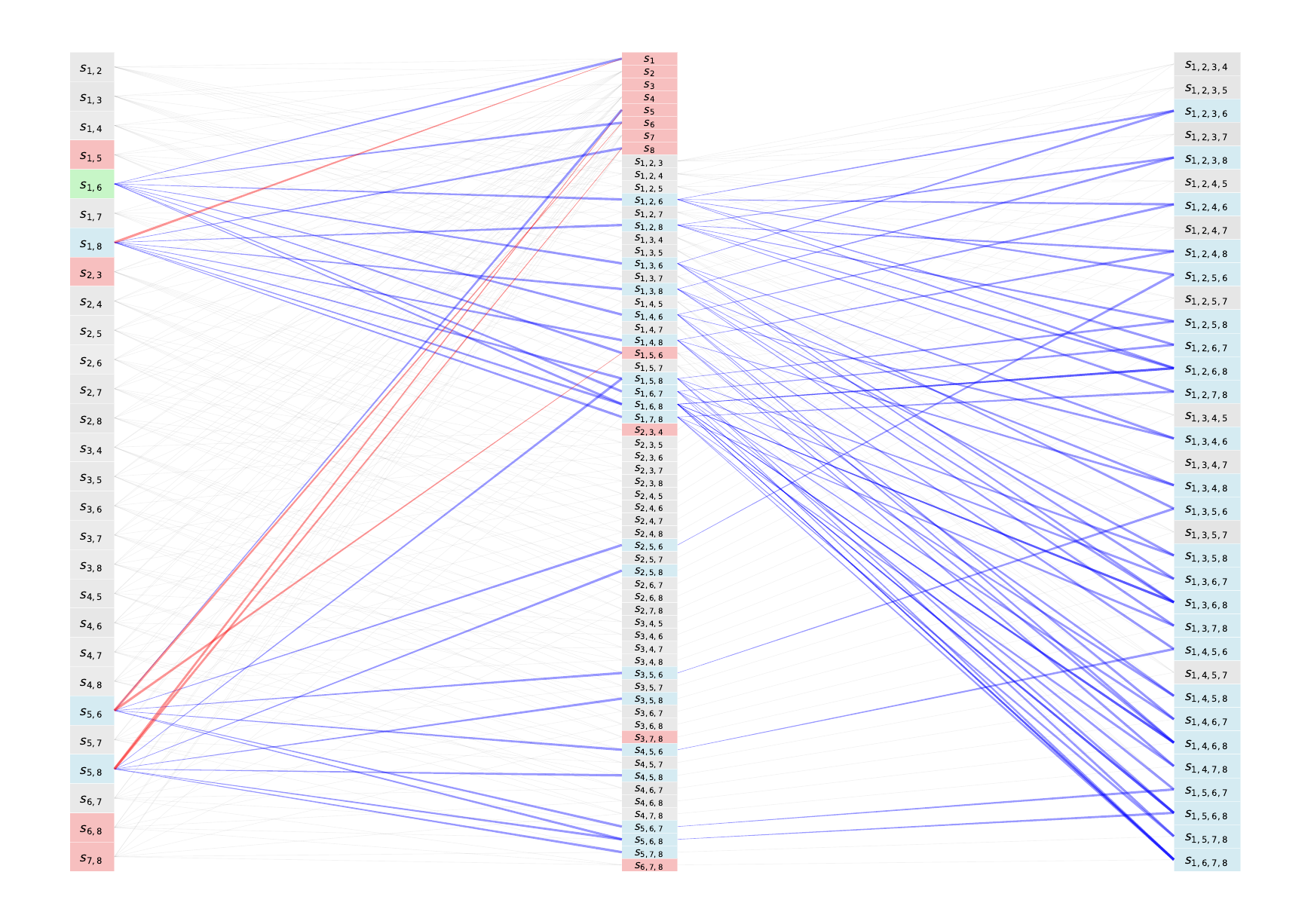}
    \caption{Final upward propagation completing the constraint resolution. The modification at $s_{58}$ triggers the final upward propagation phase (blue arrows), decrementing its three-particle supersets $s_{158}$, $s_{258}$, $s_{358}$, $s_{458}$, $s_{568}$, $s_{578}$, and $s_{678}$ by -1 each. This propagation proceeds without encountering additional fixed constraints, indicating successful resolution of all constraint conflicts. }
    \label{eg-8-06}
\end {figure}

\begin{figure}[H]
    \centering
    \includegraphics[width=1\linewidth]{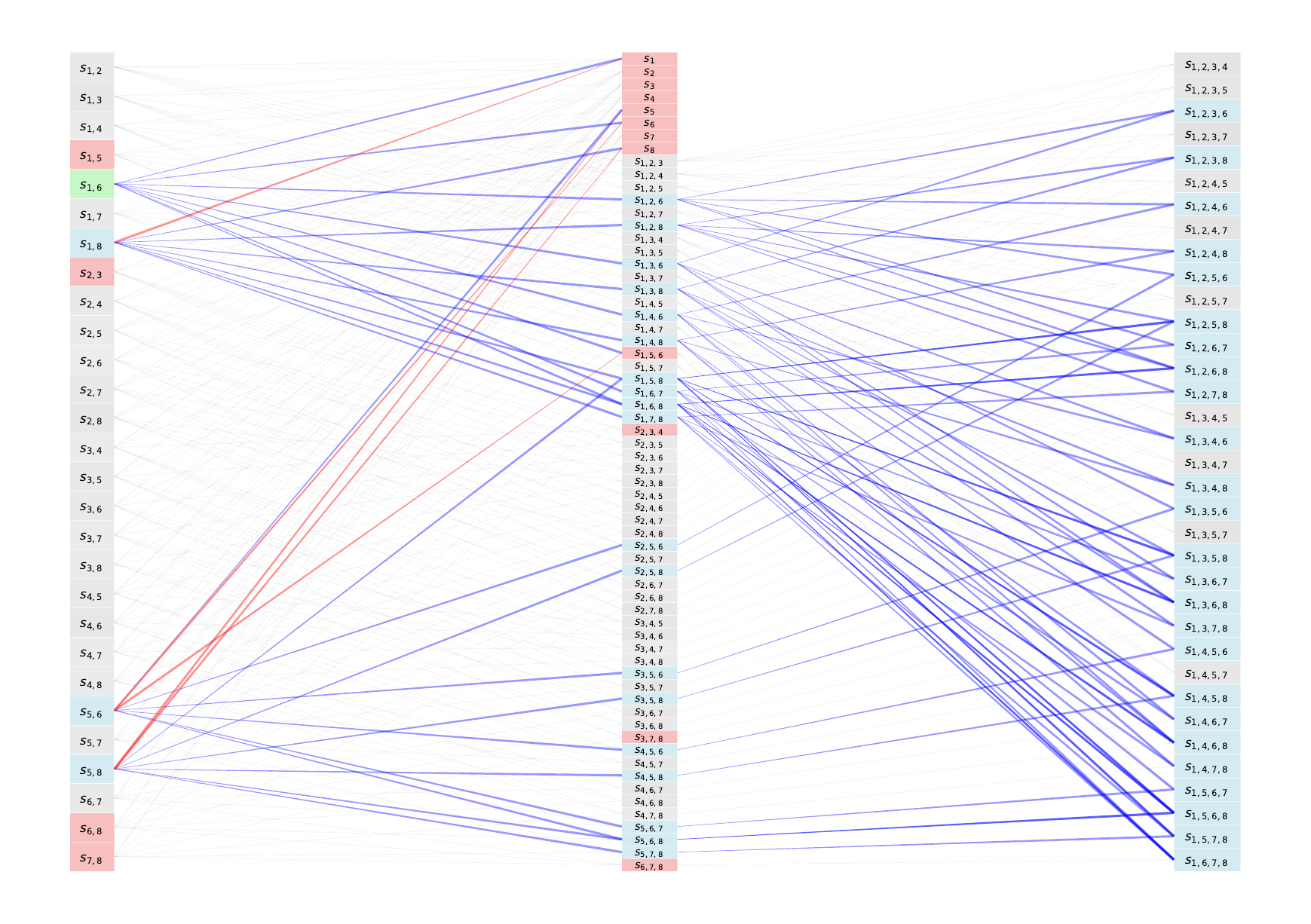}
    \caption{The algorithm seamlessly continues to update the third layer, where four-point supersets receive corresponding -0.5 decrements to maintain the recursive relationships across all network levels. This systematic propagation through all network layers keeps each recursion relation $K(s_A) = \frac{1}{(|A|-2)}\sum_{B\subset A,|B|=|A|-1}K(s_B)$ satisfied. The -0.5 updates are shown in the $K$ normalization; in $\widetilde K$ they correspond to integer updates. The successful completion of this propagation without further constraint violations confirms that the removal of incompatible pole $s_{16}$ has been achieved while preserving all required poles and maintaining network consistency.}
    \label{eg-8-07}
\end{figure}

The final propagation stages, shown in Figures~\ref{eg-8-06} and~\ref{eg-8-07}, demonstrate the algorithm's convergence. The absence of additional constraint violations indicates successful completion of the pole removal process.

Upon completion of the discrete automatic differentiation process, the algorithm removes the incompatible pole $s_{16}$ while preserving the imposed constraints in this example. The modified CHY integrand becomes:

\begin{equation}
I_8^* = \frac{z_{18}z_{28}z_{36}}{z_{12}z_{15}^2z_{16}z_{17} z_{23}^2 z_{24}z_{26}z_{34}z_{37}z_{38}z_{45}z_{48}z_{56}z_{68}^2 z_{78}^2},
\end{equation}

which produces the refined amplitude:
\begin{equation}
\frac{1}{s_{15}s_{23}s_{68}s_{234}s_{678}} + \frac{1}{s_{15}s_{23}s_{78}s_{234}s_{678}} + \frac{1}{s_{15}s_{78}s_{156}s_{378}s_{1256}} + \frac{1}{s_{15}s_{23}s_{78}s_{156}s_{234}}.
\end{equation}

This result isolates the terms containing the desired pole $s_{15}$ in this example, illustrating the behavior of our neural-inspired approach for higher-multiplicity CHY integrands.

The 8-point example highlights several key features of our discrete automatic differentiation framework:

\begin{itemize}
\item \textbf{Three-layer architecture}: Unlike the 6-point case with its two-layer structure, the 8-point example requires a three-layer network, demonstrating scalability to higher multiplicities.

\item \textbf{Cascading constraint resolution}: The algorithm successfully handles multiple rounds of backward propagation, showing robustness in complex constraint environments.

\item \textbf{Exact constraint satisfaction}: Throughout the entire process, all recursion relations remain exactly satisfied, with no approximations or numerical errors.

\end{itemize}

This 8-point demonstration is consistent with the practical applicability of our approach to realistic amplitude calculations and provides a starting point for extensions to higher multiplicities and more complex pole structures. The systematic nature of the constraint resolution process suggests that the method may scale effectively as we move toward higher multiplicities relevant to current amplitude computation capabilities.
\clearpage

\section{Examples: 8-Point CHY Integrand with Higher-Order Poles}
\label{eg.8-Point-Higher-Order-Poles}
\label{app:8point-higher}

We now demonstrate the scalability of our enhanced decomposition method through a more complex eight-point example. This case illustrates how the systematic approach based on 0-regular graph decomposition can maintain computational advantages in this example as the complexity increases.

\begin{figure}[htbp]
    \centering
    \includegraphics[width=0.4\linewidth]{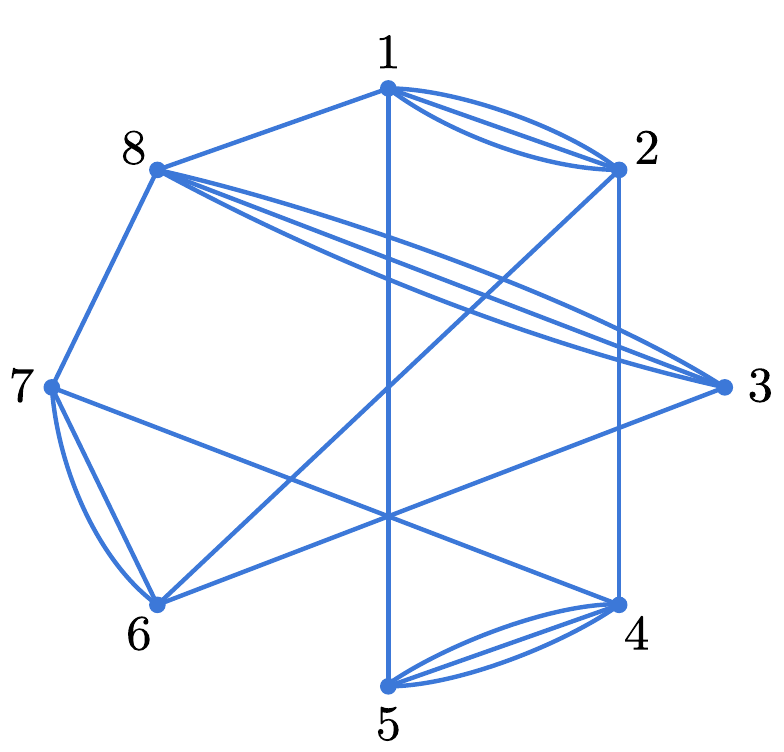}
    \caption{The 4-regular graph representation of the eight-point CHY integrand $I_8$. The edge multiplicities directly encode the denominator powers in equation \eqref{eq:I8-integrand}. The triple edges at $(1,2)$, $(3,8)$, and $(4,5)$ indicate three double poles, while the more complex edge structure suggests an additional higher-order pole that merits further analysis.}
    \label{fig:8pt-4regular}
\end{figure}

Consider the eight-point CHY integrand
\begin{equation}
I_8 = \frac{1}{z_{12}^3 z_{15} z_{26} z_{36} z_{38}^3 z_{45}^3 z_{47} z_{67}^2 z_{78}},
\label{eq:I8-integrand}
\end{equation}
whose 4-regular graph representation is shown in Figure \ref{fig:8pt-4regular}.

Using the pole index formula, we identify all subsets with $\chi(A) > 0$:
\begin{equation}
\{1,2\}, \{3,8\}, \{4,5\}, \{1,2,4,5\}
\end{equation}
where $\chi(A) = 1$ for each, indicating four constraints that must be addressed. The presence of the four-particle pole $s_{1245}$ makes this example particularly challenging, as it introduces correlations between the decomposition steps.

In the original method \cite{Cardona:2016gon}, since the CHY integrand \(I_8\) has four higher-order poles \(s_{12}\), \(s_{38}\), \(s_{45}\), and \(s_{1245}\), one multiplies \(I_8\) by four basic cross-ratio identities, one for each of the subsets \(\{1,2\}\), \(\{3,8\}\), \(\{4,5\}\), and \(\{1,2,4,5\}\).
Since the identities for \(\{1,2\}\), \(\{3,8\}\), and \(\{4,5\}\) contain 5 terms and the identity for \(\{1,2,4,5\}\) contains 9 terms, multiplying these four identities decomposes the original CHY integrand into \(5\times5\times5\times9=1125\) terms.

\begin{figure}[htbp]
    \centering
    \includegraphics[width=1.0\linewidth]{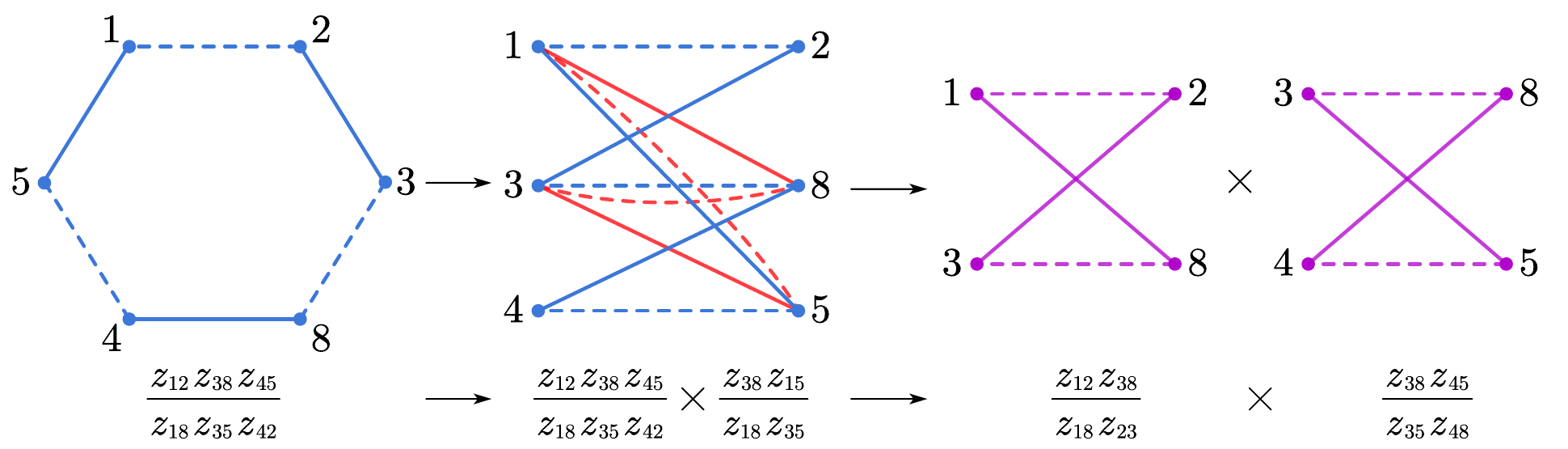}
    \caption{Decomposition of the six-point cross-ratio factor for the eight-point integrand. The bipartite representation reveals the natural factorization into $\frac{z_{12}z_{38}}{z_{18}z_{32}} \times \frac{z_{38}z_{45}}{z_{35}z_{48}}$, providing the foundation for our two-stage decomposition strategy. This decomposition is specifically designed to address the three two-particle double poles while minimizing interference with the four-particle pole $s_{1245}$.}
    \label{fig:8pt-crossratio}
\end{figure}

To reduce these higher-order poles and redundant terms, we construct a six-point cross-ratio factor $\frac{z_{12}z_{38}z_{45}}{z_{18}z_{35}z_{42}}$ targeting the three two-particle double poles. Figure \ref{fig:8pt-crossratio} shows its systematic decomposition.

\subsection*{Round 1: Initial Decomposition}

Based on the cross-ratio decomposition, we select two strategic identities:
\begin{align}
I_8[\{1,2\},1,8] &= -\frac{s_{23}z_{12}z_{38}}{s_{12}z_{18}z_{23}} - \frac{s_{24}z_{12}z_{48}}{s_{12}z_{18}z_{24}} - \frac{s_{25}z_{12}z_{58}}{s_{12}z_{18}z_{25}} - \frac{s_{26}z_{12}z_{68}}{s_{12}z_{18}z_{26}} - \frac{s_{27}z_{12}z_{78}}{s_{12}z_{18}z_{27}}, \label{eq:8pt-id1}\\
I_8[\{4,5\},5,3] &= -\frac{s_{41}z_{13}z_{45}}{s_{45}z_{14}z_{35}} - \frac{s_{42}z_{23}z_{45}}{s_{45}z_{24}z_{35}} - \frac{s_{46}z_{36}z_{45}}{s_{45}z_{35}z_{46}} - \frac{s_{47}z_{37}z_{45}}{s_{45}z_{35}z_{47}} - \frac{s_{48}z_{38}z_{45}}{s_{45}z_{35}z_{48}}. \label{eq:8pt-id2}
\end{align}

\begin{figure}[htbp]
    \centering
    \includegraphics[width=1\linewidth]{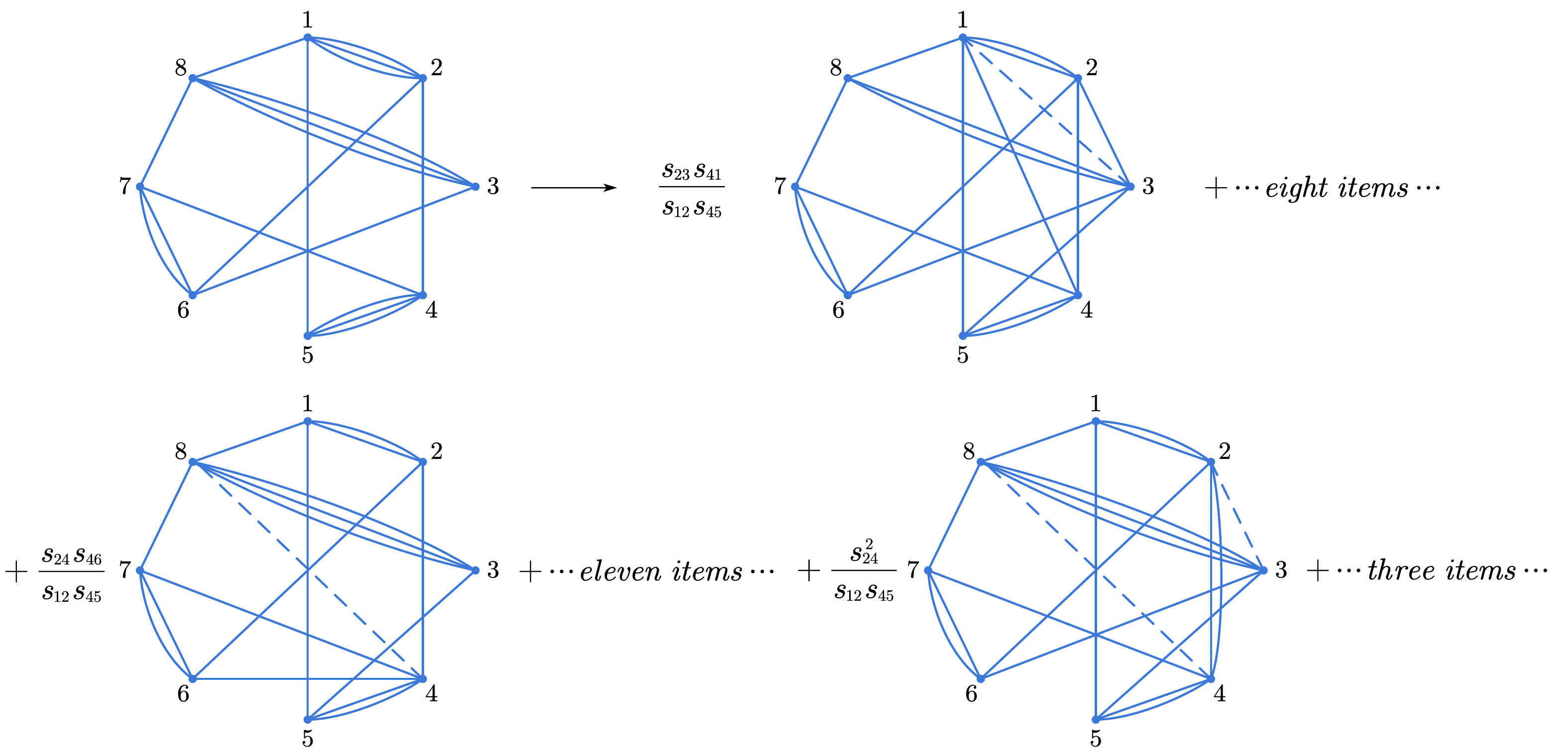}
    \caption{Round 1 decomposition results: The initial application of identities \eqref{eq:8pt-id1} and \eqref{eq:8pt-id2} produces 25 terms organized in a $5 \times 5$ grid. The initial 9 terms (top-left, shown in gray) already contain only simple poles. The central 12 terms retain the double pole $s_{38}$, while the final 4 terms (bottom-right) contain both $s_{38}$ and $s_{1245}$ as double poles, requiring the most extensive further treatment.}
    \label{fig:8pt-round1}
\end{figure}

Multiplying $I_8$ by these identities produces $5^2 = 25$ terms, whose structure is shown in Figure \ref{fig:8pt-round1}.

The decomposition yields three categories:
\begin{itemize}
\item \textbf{9 terms with only simple poles} 
\item \textbf{12 terms with double pole $s_{38}$} 
\item \textbf{4 terms with double poles $s_{38}$ and $s_{1245}$}
\end{itemize}

\subsection*{Round 2: Eliminating the $s_{38}$ Double Pole}

For the 12 terms containing double pole $s_{38}$, we apply the identity:
\begin{equation}
I_8[\{3,8\},3,4] = -\frac{s_{81}z_{14}z_{38}}{s_{38}z_{18}z_{34}} - \frac{s_{82}z_{24}z_{38}}{s_{38}z_{28}z_{34}} - \frac{s_{85}z_{45}z_{38}}{s_{45}z_{34}z_{58}} - \frac{s_{86}z_{46}z_{38}}{s_{38}z_{34}z_{68}} - \frac{s_{87}z_{47}z_{38}}{s_{38}z_{34}z_{78}}.
\label{eq:8pt-id3}
\end{equation}

\begin{figure}[htbp]
    \centering
    \includegraphics[width=1\linewidth]{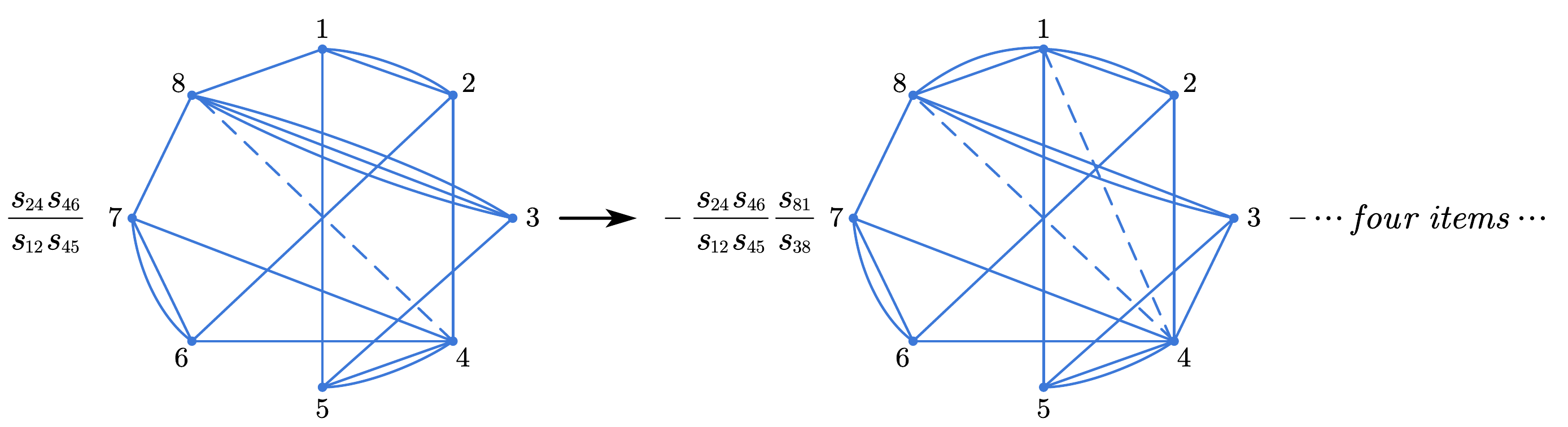}
    \caption{Round 2 decomposition of a term with double pole $s_{38}$. The application of identity \eqref{eq:8pt-id3} to the term $\frac{s_{24}s_{46}}{s_{12}s_{45}}\frac{z_{84}}{z_{21}^2 z_{42} z_{51} z_{53} z_{54}^2 z_{62} z_{64} z_{74} z_{76}^2 z_{81} z_{83}^3 z_{87}}$ produces 5 terms with only simple poles. The cross-ratio factors in the identity precisely cancel the $z_{83}^3$ contribution, reducing the pole order as required.}
    \label{fig:8pt-round2a}
\end{figure}

Figure \ref{fig:8pt-round2a} illustrates this decomposition for one representative term.

This process transforms the 12 terms with double pole $s_{38}$ into $12 \times 5 = 60$ terms with only simple poles.

\begin{figure}[htbp]
    \centering
    \includegraphics[width=1\linewidth]{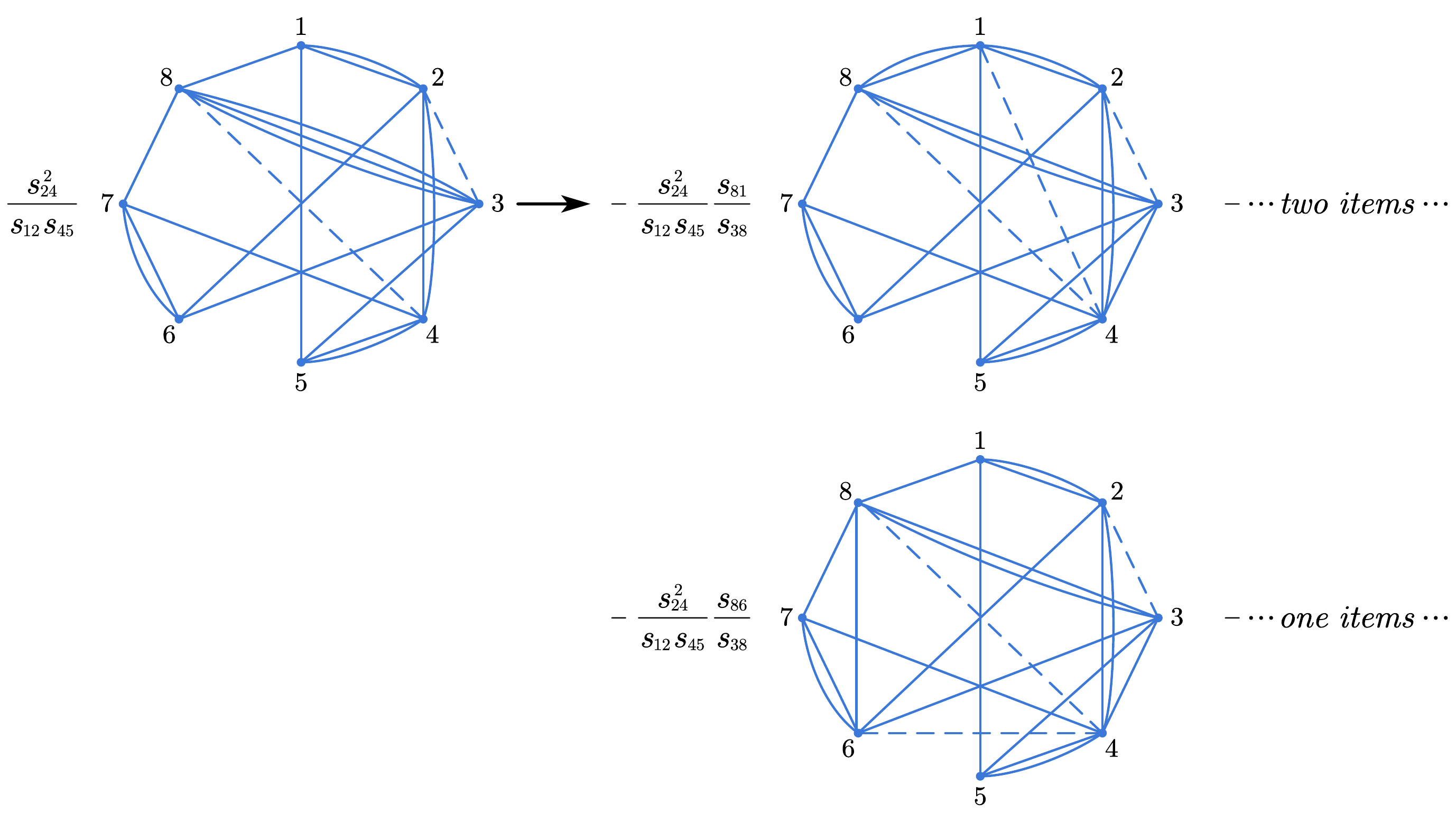}
    \caption{Round 2 decomposition of a term with double poles $s_{38}$ and $s_{1245}$. After applying identity \eqref{eq:8pt-id3}, the resulting 5 terms split into two categories: 3 terms (shown in gray) with only simple poles, and 2 terms (shown in white) that still contain the double pole $s_{1245}$, necessitating a third round of decomposition.}
    \label{fig:8pt-round2b}
\end{figure}

For the 4 terms containing both $s_{38}$ and $s_{1245}$ double poles, applying identity \eqref{eq:8pt-id3} yields 20 terms, of which 12 have only simple poles while 8 retain the double pole $s_{1245}$. Figure \ref{fig:8pt-round2b} shows a representative example.

\subsection*{Round 3: Final Decomposition of $s_{1245}$}

The remaining 8 terms with double pole $s_{1245}$ require the application of a nine-term identity:
\begin{align}
I_8[\{1,2,4,5\},1,3] = &-\frac{s_{26}z_{12}z_{36}}{s_{1245}z_{13}z_{26}} - \frac{s_{46}z_{14}z_{36}}{s_{1245}z_{13}z_{46}} - \frac{s_{56}z_{15}z_{36}}{s_{1245}z_{13}z_{56}} - \frac{s_{27}z_{12}z_{37}}{s_{1245}z_{13}z_{27}} \nonumber\\
&-\frac{s_{28}z_{12}z_{38}}{s_{1245}z_{13}z_{28}} - \frac{s_{47}z_{14}z_{37}}{s_{1245}z_{13}z_{47}} - \frac{s_{48}z_{14}z_{38}}{s_{1245}z_{13}z_{48}} - \frac{s_{57}z_{15}z_{37}}{s_{1245}z_{13}z_{57}} - \frac{s_{58}z_{15}z_{38}}{s_{1245}z_{13}z_{58}}.
\label{eq:8pt-id4}
\end{align}

\begin{figure}[htbp]
    \centering
    \includegraphics[width=1\linewidth]{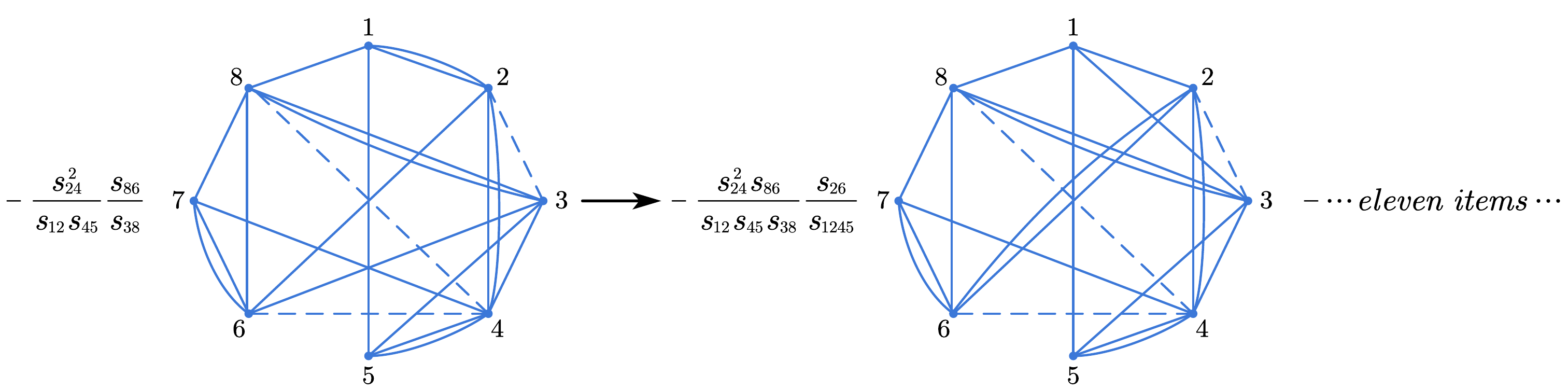}
    \caption{Round 3 final decomposition. The nine-term identity \eqref{eq:8pt-id4} applied to a term with double pole $s_{1245}$ produces 9 CHY integrands with only simple poles. This completes the systematic reduction of all higher-order poles in the original eight-point integrand. The complexity of this final step (9 terms) reflects the four-particle nature of the $s_{1245}$ pole.}
    \label{fig:8pt-round3}
\end{figure}

Figure \ref{fig:8pt-round3} demonstrates the final decomposition step.

\subsection*{Summary and Computational Efficiency}

Through this three-round process, the eight-point CHY integrand $I_8$ with four double poles is systematically decomposed into $9 + 60 + 12 + 72 = 153$ terms, each containing only simple poles. The  improvement over traditional methods is summarized in the following table:

\begin{center}
\begin{tabular}{|l|c|c|c|c|c|}
\hline
\textbf{Method} & \textbf{Round 1} & \textbf{Round 2} & \textbf{Round 3} & \textbf{Round 4} & \textbf{Total} \\
\hline
Systematic algorithm \cite{Cardona:2016gon} & 5 [5] & 25 [20] & 105 [20] & 265 [0] & 265 \\
\hline
\textbf{Our method} & 25 [16] & 89 [8] & 153 [0] & --- & \textbf{153} \\
\hline
\end{tabular}
\end{center}
\vspace{0.2cm}
\noindent Numbers in brackets [·] indicate terms with remaining higher-order poles at each stage.

Our enhanced method achieves a reduction factor by:
\begin{enumerate}
\item Systematically decomposing cross-ratio factors using the 0-regular graph structure
\item Applying identities sequentially to minimize intermediate term proliferation
\item Leveraging the hierarchical nature of pole constraints to optimize the decomposition order
\end{enumerate}

This eight-point example suggests that our graph-theoretic approach to higher-order pole decomposition can maintain computational advantages as complexity increases, making it a useful tool for the amplitude calculations considered here.

\bibliographystyle{JHEP} 
\bibliography{reference}

\end{document}